\documentclass[11pt,a4paper]{article}
\usepackage{jheppub}
\usepackage{amsmath,amssymb,amsfonts,bm,amscd}
\usepackage{mathtools}
\usepackage{graphicx}
\usepackage{multirow}
\usepackage{verbatim}
\usepackage{appendix}
\usepackage{fancybox}
\usepackage{array}

\newcommand{\bea}{\begin{eqnarray}}
\newcommand{\eea}{\end{eqnarray}}
\newcommand{\be}{\begin{equation}}
\newcommand{\ee}{\end{equation}}

\newcommand{\unknot}{{\,\raisebox{-0.1cm}{\includegraphics[width=0.4cm]{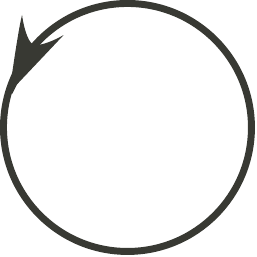}}\,}}

\newcommand{\Z}{{\mathbb Z}}
\newcommand{\R}{{\mathbb R}}
\newcommand{\C}{{\mathbb C}}

\def\Tr{{\rm Tr \,}}

\def\k{\kappa}

\def\frak{\mathfrak}

\def\tilde{\widetilde}
\def\hat{\widehat}
\def\bar{\overline}

\def\CA{{\mathcal A}}

\def\CC{{\mathcal C}}
\def\CE{{\mathcal E}}

\def\CG{{\mathcal G}}
\def\CH{{\mathcal H}}
\def\CI{{\mathcal I}}

\def\CL{{\mathcal L}}
\def\CM{{\mathcal M}}
\def\CN{{\mathcal N}}
\def\CO{{\mathcal O}}

\def\CQ{{\mathcal Q}}

\def\CS{{\mathcal S}}
\def\CT{{\mathcal T}}

\def\CW{{\mathcal W}}

\def\CZ{{\mathcal Z}}

\newcommand{\cp}{{\mathbb{C}}{\mathbf{P}}}

\renewcommand{\bar}{\overline}
\renewcommand{\hat}{\widehat}

\renewcommand{\d}{\partial}

\def\Coker{{\mathrm{Coker}\,}}

\def\deg{{\mathrm{deg}}}

\def\Tr{{\mathrm{Tr}}}

\title{Vertex algebras and 4-manifold invariants}
\author[a]{Mykola Dedushenko,}
\author[a]{Sergei Gukov}
\author[b]{and Pavel Putrov}
\affiliation[a]{Walter Burke Institute for Theoretical Physics, California Institute of Technology, \\ Pasadena, CA 91125, USA}
\affiliation[b]{School of Natural Sciences, Institute for Advanced Study,\\ Princeton, NJ 08540, USA}

\abstract{We propose a way of computing 4-manifold invariants, old and new, as chiral correlation functions in
half-twisted 2d $\CN=(0,2)$ theories that arise from compactification of fivebranes.
Such formulation gives a new interpretation of some known statements about Seiberg-Witten invariants,
such as the basic class condition, and gives a prediction for structural properties of the multi-monopole invariants
and their non-abelian generalizations.

\vspace{2cm}

\vspace{2cm}

\texttt{CALT-TH-2017-008}
}

\begin{document}
\maketitle



\section{Introduction and motivation}

There are at least three parallel tracks that lead to the study of $\bar Q_+$-cohomology and chiral
correlation functions in a certain class of 2d $\CN=(0,2)$ theories.

\subsection{Unorthodox invariants of smooth 4-manifolds}

Searching for new invariants of smooth 4-manifolds, that potentially could go beyond the Seiberg-Witten and Donaldson invariants,
it was proposed in \cite{Gadde:2013sca}
to consider a two-dimensional quantum field theory $T[M_4,G]$ as a rather unusual invariant of smooth structures on a 4-manifold $M_4$.
Specifically, $T[M_4,G]$ is a 2d $\CN=(0,2)$ superconformal theory that, apart from $M_4$, also depends on a choice of a root system $G$
and is invariant under the Kirby moves.

Luckily, conformal field theories in two dimensions exhibit rich mathematical structure which, on the one hand,
is rich enough to (potentially) describe the wild world of smooth 4-manifolds and, on the other hand, is rigorous enough
to hope for a precise mathematical definition of the invariant $T[M_4,G]$.
In fact, for many practical purposes and applications in this paper, a mathematically inclined reader can
think of $T[M_4,G]$ as a functor that assigns a vertex operator algebra (VOA) to a smooth 4-manifold $M_4$ (and a ``gauge'' group $G$).
Composing it with other functors that assign various quantities to 2d conformal theories, one can obtain more
conventional invariants of smooth 4-manifolds:
\be
M_4 \quad \leadsto \quad T[M_4;G] \quad \leadsto \quad Z_{T[M_4;G]} = \text{4-manifold invariant}.
\ee
Here, $Z$ can be any invariant of a 2d conformal theory with $\CN=(0,2)$ supersymmetry,
{\it e.g.}, its elliptic genus, chiral ring, moduli space of marginal couplings, or central charge.
Since 2d theory $T[M_4;G]$ is systematically determined by $M_4$ and invariant under the Kirby moves,
all such invariants lead to various 4-manifold invariants; some are simple and some are quite powerful.
In particular, it was conjectured in \cite{Gadde:2013sca} that chiral ring of the theory $T[M_4;G]$ for $G=SU(2)$
or, equivalently, its $\bar Q_+$-cohomology knows about Donaldson invariants of $M_4$.

One of the main goals in this paper is to present some evidence to this conjecture and to build a bridge between
VOA$_G [M_4]$ and more traditional 4-manifold invariants.
Conjecturally, at least for manifolds with $b_2^+ > 1$, one can trade $SU(2)$ Donaldson invariants
for Seiberg-Witten invariants defined in a simpler gauge theory, with gauge group $G=U(1)$,
which would be too trivial if not for an extra ingredient, the additional spinor fields.
We wish to study how these invariants are realized in 2d theory $T[M_4;G]$ with $G=SU(2)$ and $G=U(1)$, respectively.
In particular, we shall see that in 2d realization of Seiberg-Witten invariants, much like in gauge theory on $M_4$,
the non-trivial information about the 4-manifold comes from an extra ingredient, a particular vertex operator $\CS (z)$.

The functor that associates a vertex operator algebra to a smooth 4-manifold is a natural generalization
of the pioneering work by Nakajima \cite{Nakajima} where connection between gauge theory on ALE spaces and Kac-Moody algebras was observed
(see also \cite{Nakajima:1995ka} for a visionary exposition).
These results alone suffice to determine VOA$_G [M_4]$ for many negative-definite 4-manifolds by analyzing
its implications for the Kirby moves; in physics, this might be called ``bootstrapping'' $T[M_4]$ with ALE spaces and Kirby moves.

A separate line of development that motivates the study of VOA$_G [M_4]$ has to do with categorification of quantum group invariants.

\subsection{4d TQFT}

\begin{figure}[ht]
\centering
\includegraphics[trim={0 0.5in 0 0.5in},clip,width=5.0in]{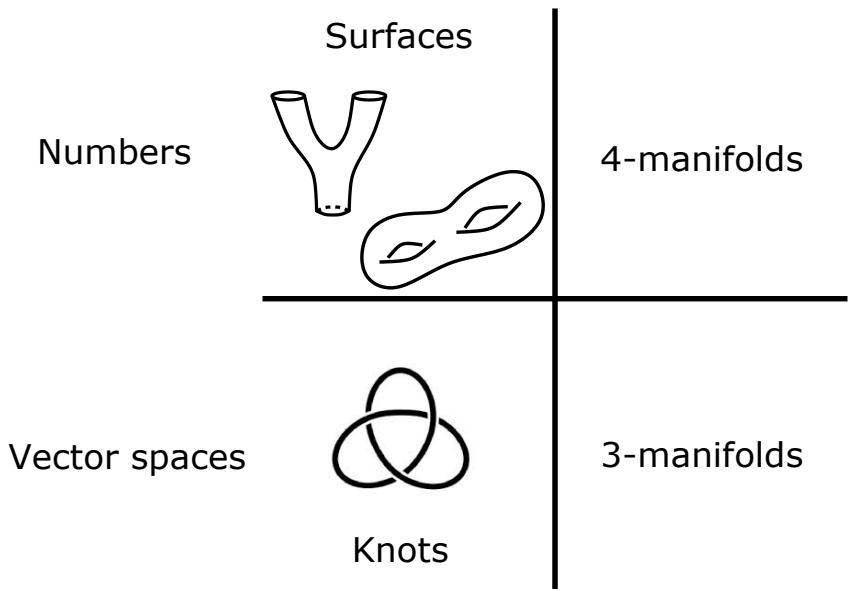}
\caption{Various corners of a 4d TQFT with topological ``surface operators'' or ``foams''.
It assigns vector spaces to knots and 3-manifolds,
maps between these homological invariants to the corresponding cobordisms,
and numerical invariants to closed 4-manifolds and embedded surfaces.}
\label{fig:TQFT}
\end{figure}

In 1994, Crane and Frenkel envisioned a 4d TQFT that categorifies quantum group invariants of knots and 3-manifolds.
Their paper \cite{MR1295461} was way ahead of its time since, even twenty years later,
the proposed 4d TQFT is slowly being built, one brick at a time.
Much like any 4d TQFT that obeys Atiyah-Segal axioms, it should assign numerical invariants to closed 4-manifolds and vector spaces to 3-manifolds.
Moreover, if the 4d TQFT in question admits 2-dimensional topological defects --- that we shall call either ``foams'' or ``surface operators'' ---
then, it should also assign numbers to closed surfaces $D \subset M_4$ and vector spaces to knots and links, {\it cf.} Figure~\ref{fig:TQFT}.
A non-trivial requirement is that Witten-Reshetikhin-Turaev (WRT) invariants arise as graded Euler characteristics of vector spaces assigned to knots and 3-manifolds.

The first major piece of the desired structure came with the construction of Khovanov-Rozansky homology \cite{Khovanov,KR1,KR2}
that belongs to the lower left corner in Figure~\ref{fig:TQFT}. This corner is by far the most developed
element of the sought after 2d-4d TQFT on $D \subset M_4$, and even that only for $M_4 = \R^4$ and $D = \R \times K$.
Its physical interpretation, proposed in \cite{Gukov:2004hz}, led to many new predictions and connections between various areas,
which include knot contact homology \cite{Aganagic:2013jpa}, gauge theory \cite{Gukov:2007ck,Witten:2011zz},
and algebras of interfaces \cite{Carqueville:2011zea,Chun:2015gda}, just to name a few.
(A more complete account of these connections can be found, {\it e.g.}, in \cite{Chun:2015gda,Nawata:2015wya}.)

Homological invariants of knots and links were soon generalized to invariants of cobordisms
and closed surfaces \cite{MR2113903,MR2174270,MR2213759,MR2171235,Rasmussen,MR3100886}.
This generalization, illustrated in the top left corner of Figure~\ref{fig:TQFT},
corresponds to passing from $D = \R \times K$ to more general surfaces, while keeping $M_4 = \R^4$.
Apart from a lone exception \cite{Chun:2015gda}, the 2d TQFT on a ``foam'' $D$ has not been studied in the physics literature.

The situation is roughly reversed as it comes to generalization in a different direction,
namely to homological invariants of non-trivial 3-manifolds.
It was recently studied in the physics literature \cite{Gukov:2016gkn,Gukov:2016njj},
while in math the lower right corner of Figure~\ref{fig:TQFT}
remains a largely unexplored territory.
If the homological invariants of 3-manifolds proposed in {\it loc. cit.} are functorial (perhaps under certain conditions),
then one must be able to replace $\R \times M_3$ by a more general cobordism $M_4$, finally taking us to
the upper right corner of Figure~\ref{fig:TQFT}:
\begin{equation}
\begin{array}{ccc}
\R & \times & M_3 \\
 & \cup &  \\
\R & \times & K \\
\end{array}
\qquad \leadsto \qquad
\begin{array}{c}
M_4 \\
\cup \\
D \\
\end{array}
\label{cobordisms}
\end{equation}
Now comes the key point. The physical setup that extends homological invariants of knots and 3-manifolds to their cobordisms is, in fact,
precisely what defines the $\bar Q_+$-cohomology of 2d $\CN=(0,2)$ theory $T[M_4,G]$ or, equivalently, the vertex operator algebra VOA$_G [M_4]$.

{}From this perspective, the goal of this paper is to bring the categorification program closer its original motivation
and move from homological invariants of 3-manifolds with knots to the corresponding invariants of 4-manifolds with embedded surfaces,
{\it i.e.}, from the lower half to the upper half of Figure~\ref{fig:TQFT}.

\subsection{$G_2$ perspective}

Another motivation for this work comes from M-theory compactifications on 7-manifolds with special holonomy group $G_2$.
Such compactifications lead to $\CN=1$ supersymmetric physics in the remaining four space-time dimensions,
which is interesting for a number of reasons. If supersymmetry is a part of Nature, it may very well be the geometry
of the world we live in. Moreover, 4d $\CN=1$ physics is very interesting on its own, exhibiting a wide spectrum of
physical phenomena --- dynamical SUSY breaking, confinement and other phases, rich landscape of superconformal points, {\it etc.} --- many
of which offer enormous potential for future discoveries, perhaps through $G_2$ compactifications.

One way to explore the physics of such compactifications is with the help of extended objects, which in the context
of M-theory basically limits us to either M2-branes or M5-branes. Since the dimension of M2-brane world-volume is
barely enough for it to wrap a supersymmetric cycle in a $G_2$-manifold, fivebranes quickly take the center stage.
In particular, one can try to probe the physics of $G_2$ compactifications by studying M5-branes supported
on coassociative 4-manifolds. In 4d $\CN=1$ theory, such objects look like half-BPS ``cosmic strings''
that preserve 2d $\CN=(0,2)$ supersymmetry on their world-sheet.
A natural question, then, is: What degrees of freedom does such a string carry?

As the reader might have anticipated by now, the answer is $T[M_4; G]$ for $G = U(N)$,
where $N$ is the number of fivebranes. Indeed, a neighborhood of every coassociative 4-manifold $M_4$
in the ambient $G_2$ space looks like a bundle of self-dual 2-forms, $\Lambda^{2,+} (M_4)$,
\begin{equation}
 \begin{array}{cccc}
  \text{space-time:} & \quad \R^4 & ~\times & \Lambda^{2,+} (M_4) \\
   & \quad \cup &  & \cup \\
N\,\text{M5-branes:} & \quad \R^2 & ~\times & M_4 \\
 \end{array}
\label{Msetup}
\end{equation}
and the partial topological twist of 6d $(2,0)$ theory on the fivebrane world-volume is precisely
what defines 2d $\CN=(0,2)$ theory $T[M_4; G]$ in two of its remaining space-time directions.
These directions can also be ``twisted'', so that $\R^2$ can be replaced by a more general
world-sheet $\Sigma$ of a cosmic string, with $\R^4$ accordingly replaced by $T^* \Sigma$.
Since the resulting theory on a Riemann surface $\Sigma$ is not fully topological, it is often
called {\it half-twisted A-model} and the corresponding twist is often called {\it holomorphic} (as opposed to topological).\\

The paper is organized as follows.
In sections \ref{sec:fluxes} and \ref{sec:Kondo}, we discuss 2d $\CN=(0,2)$ theory $T[M_4; G]$ and its impurity vertex operators.
In particular, we propose a framework in which gauge theoretic 4-manifold invariants (not necessarily abelian!)
are given by correlation functions of two vertex operators, $\CS (z)$ and $\CS_+ (z)$,
and we find an intriguing relation between these two vertex operators.

This framework leads to new predictions, which we verify in section~\ref{sec:NfSW}
using traditional gauge theory methods in the case of multi-monopole generalization of Seiberg-Witten theory.\footnote{A mathematically
inclined reader may want to skip directly to section~\ref{sec:NfSW}.}
Mathematically, this theory is interesting in its own right and formulating it on arbitrary 4-manifolds requires extra care.
We address these challenges using equivariant techniques extensively used in physics in recent years,
starting with \cite{Moore:1997dj} and culminating in the formulation of the Nekrasov instanton partition function \cite{Nekrasov:2002qd}
on 4-manifolds with $U(1) \times U(1)$ symmetry.
We use similar methods to carefully define and study multi-monopole invariants of arbitrary 4-manifolds and
comment on the structure of the corresponding Floer theory for 3-manifolds.

Once we reproduce the structure of 2d chiral correlators in the abelian gauge theory, in section~\ref{sec:non-abelian}
we comment on non-abelian generalizations. In particular, as an illustration, we show how correlation
functions of $\CS (z)$ and $\CS_+ (z)$ lead to new predictions for invariants of 4-manifolds in
a non-abelian gauge theory. It would be interesting to verify these predictions by direct gauge theoretic
techniques in a way similar to the analysis of section~\ref{sec:NfSW} and explore more general theories.
Returning to our original motivation, in section~\ref{sec:non-abelian} we also discuss the structure of the 4d TQFT illustrated in Figure~\ref{fig:TQFT}
and propose a simple criterion that can help to identify new opportunities for constructing 4-manifold invariants
via 4d $\CN=2$ theories (possibly, non-Lagrangian).

Since most of the time the choice of $G$ is clear from the context, to reduce clutter we often omit it
and refer to $T[M_4; G]$ simply as $T[M_4]$.


\section{Flux vacua of $T {[ M_4 ]}$}
\label{sec:fluxes}

Many applications of string theory and M-theory, from AdS/CFT to building semi-realistic models of particle physics,
involve dimensional reduction (a.k.a. compactification) on a non-trivial manifold in the presence of background fluxes.
Generically, such background fluxes break supersymmetry, unless geometric moduli of the compactification manifold obey certain conditions.
Sometimes, these conditions can be interpreted as equations for a critical point of a function $\CW$ that, in turn,
often admits a simple physical interpretation in the low-dimensional effective theory.

The six-dimensional $(2,0)$ theory on the fivebrane world-volume, which can be thought of as a younger sister of M-theory~\cite{Losev:1997hx},
also admits flux compactifications. Thanks to non-gravitational physics of such flux vacua, they can serve as simpler
examples for a much richer landscape of fluxes in M-theory.
In particular, in order to preserve supersymmetry, background values of fields in 6d fivebrane theory must obey
certain conditions; equivalently, one can interpret these conditions as constraints on the geometry of the compactification
manifold $M$ when fluxes are non-zero.
For example, in compactification on a 4-manifold $M_4$, the relevant condition is the anti-self-duality equation
for the 2-form flux $F$ on $M_4$,
\be
F^+ \; = \; 0.
\label{ASDeq}
\ee
One of the goals in the present section is to describe such SUSY flux vacua in 2d $\CN=(0,2)$ theory $T[M_4]$.

\subsection{Theory $T {[ M_4 ]}$}

Before we incorporate fluxes and vertex operators, our first task is to describe 2d theory $T[M_4]$ itself.
Defined as a reduction of 6d $(2,0)$ theory on a 4-manifold $M_4$ with a partial topological twist along $M_4$,
the resulting 2d theory $T[M_4]$ carries $\CN=(0,2)$ supersymmetry. In other words, its right-moving sector
has $\CN=2$ supersymmetry and the left-moving sector is basically the vertex operator algebra VOA$_G [M_4]$.

In general, deriving $T[M_4]$ is a rather non-trivial task, which so far has been achieved only for particular
types of 4-manifolds (see {\it e.g.}, \cite{Gadde:2013sca} for simply connected $M_4$ with definite intersection form).
Luckily, in the special case of a single fivebrane, that is for $G=U(1)$ which we need in this paper,
the 2d $\CN=(0,2)$ theory $T[M_4]$ can be derived for {\it any} $M_4$ using the standard rules of the Kaluza-Klein
reduction.\footnote{This analysis was done jointly with S.~Schafer-Nameki and J.~Wong, whom we wish to thank for many enjoyable discussions on this topic.}.
In this special case, the 6d fivebrane theory is simply a free theory of a $(2,0)$ tensor multiplet,
whose bosonic fields include a self-dual 2-form gauge field $B$ and five real scalars that transform in the vector
representation of $SO(5)_R$ R-symmetry group. In the simply-connected case such reduction was considered in \cite{Verlinde:1995mz,Ganor:1996xg}.
Unless explicitly noted otherwise, we shall assume throughout the paper that the homology $H_*(M_4,\Z)$ has no torsion.
In principle, it is not hard to relax this assumption, as we shall illustrate in \ref{sec:example-plumbed} where
a class of examples with torsion is considered.

Under the partial topological twist\footnote{described in detail, {\it e.g.}, in \cite[sec.5]{Gukov:2016gkn}}
that corresponds to embedding the fivebrane in a $G_2$-manifold \eqref{Msetup}, three out of five scalar
fields combine into components of a self-dual 2-forms on $M_4$,
whose Kaluza-Klein modes contribute $b_2^+$ real scalars to the spectrum of $T[M_4]$.
By McLean's theorem, these modes can be identified with the moduli of the coassociative 4-manifold $M_4$,
{\it i.e.}, with SUSY-preserving displacements of M5-brane inside the $G_2$-manifold \eqref{Msetup}.
Since $\CN=(0,2)$ supersymmetry requires right-moving scalars to be complex-valued,
these $b_2^+$ real scalar fields must be in pairs with other Kaluza-Klein (KK) modes.
And, indeed, there are precisely $b_2^+$ right-moving compact scalars that come from Kaluza-Klein modes of
the self-dual 2-form gauge field $B$. Its field strength $H=dB$ satisfies the self-duality equation $H = * H$
and, therefore, after reduction on $M_4$ gives rise to $b_2^+$ right-moving compact scalars $X_R^i$
and $b_2^-$ left-moving compact scalars $X_L^i$, as well as $b_1=b_3$ vector fields $A^k$:
\be
H \; = \; \sum_{j=1}^{b_2^-} \partial X_L^j \wedge \omega_j^{-}
\; + \; \sum_{i=1}^{b_2^+} \bar \partial X_R^i \wedge \omega_i^{+}+\sum_{k=1}^{b_1} dA^k \wedge \omega^1_k+\sum_{k=1}^{b_1} n^k \wedge \omega^3_k,
\label{HviaXLR}
\ee
where $\omega_i^\pm$, $\omega^1_k$, $\omega^3_k$ are generators of $H^{2,\pm} (M_4,\R)$, $H^{1}(M_4,\R)$,
and $H^{3}(M_4,\R)$, respectively,
and $n^k\in \Z$ denote fluxes of 2d gauge fields $A^k$. Finally, the only remaining part of the bosonic KK spectrum
are the two real scalars of the 6d theory not affected by the topological twist.
They parametrize transverse displacements of the fivebrane \eqref{Msetup} inside $\R^4$
and can be naturally combined into a complex scalar of a standard 2d $\CN=(0,2)$ chiral multiplet.
In what follows, we refer to it as the ``center of mass'' multiplet $\Phi_0$.
The complete Kaluza-Klein spectrum of $T[M_4]$,
including fermions, is summarized in Table~\ref{tab:TMfields}.

\begin{table}[h]
\centering
\begin{tabular}{l c c c}
\hline\hline
\multicolumn{2}{c}{Field} & Chirality & Description
\\ [0.5ex]
\hline
$X_R^i$ & $i=1, \ldots, b_2^+$ & right-moving & compact real bosons \\
$\sigma^i$ & $i=1, \ldots, b_2^+$ & non-chiral & non-compact real bosons \\
$\psi_+^i$ & $i=1, \ldots, b_2^+$ & right-moving & complex Weyl fermions \\[2ex]
\multicolumn{2}{l}{$\phi_0$, $\bar \phi_0$} & non-chiral & non-compact complex boson \\
\multicolumn{2}{l}{$\chi_+$} & right-moving & complex Weyl fermion \\[2ex]
$X_L^j$ & $j=1, \ldots, b_2^-$ & left-moving & compact real bosons \\[2ex]
$A^k$ & $k=1, \ldots, b_1$ & non-chiral & vector fields \\
$\gamma^k_-$ & $k=1, \ldots, b_1$ & left-moving & complex Weyl fermions \\[2ex]
\hline
\end{tabular}
\caption{The field content of 2d theory $T[M_4]$ for $G = U(1)$.}
\label{tab:TMfields}
\end{table}

To complete the description of $T[M_4]$ we also need to specify the operator product expansion between compact bosons.
Note, this is not necessary for non-compact bosons since they can always be rescaled.
Suppose that the fields $X_{L,R}$ are normalized such that if $X_{L,R}$ is considered an element of $H^{2,\pm}(M_4)\equiv P_\pm H^2(M_4,\R)$
with $P_{\pm}=(1\pm \ast)/2$, the compactness is realized by periodicity with respect to
elements of $H^2(M_4,\Z) \subset H^2(M_4,\R)$. Much as the non-abelian version of the theory \cite{Gadde:2013sca},
our $T[M_4]$ here depends in a crucial way on the intersection form on $M_4$:
\begin{equation}
	\begin{array}{cccc}
		Q: & H_2(M_4,\Z)\otimes H_2(M_4,\Z) & \longrightarrow & \Z \\
		& \mu \otimes \nu &\longmapsto & \#(\mu \cap \nu ).
	\end{array}
	\label{Q-intersection}
\end{equation}
Its inverse is a bilinear form on the dual lattice:
\begin{equation}
	\begin{array}{cccc}
		Q^{-1}: & H^2(M_4,\Z)\otimes H^2(M_4,\Z) & \longrightarrow & \Z \\
		& [\mu] \otimes [\nu] &\longmapsto & \int_{M_4}\mu \wedge \nu,
	\end{array}	
\end{equation}
which can be extended to a real-valued bilinear form on $H^2(M_4,\R)\cong \CH^2(M_4)$. On the other hand, the OPE of fields $X_{L,R}$ is determined by the metric on $H^2(M_4,\R)$:
\begin{equation}
	\begin{array}{cccc}
		G: & H^2(M_4,\R)\otimes H^2(M_4,\R) & \longrightarrow & \R \\
		& \mu \otimes \nu &\longmapsto & \int_{M_4} \mu \wedge *\nu.
	\end{array}	
\end{equation}
The intersection form $Q$ depends only on the topology of $M_4$,
whereas the Hodge $\ast$ operator, acting on $H^2(M_4,\R)$, and therefore $G$ and $P_\pm$ all depend on the conformal structure of $M_4$. 

Note that $G$ and $Q^{-1}$ coincide on the self-dual subspace but differ by a sign on anti-self-dual classes. What appears in the OPE is $G^{-1}$, and so we can write:
\begin{equation}
	\d X_L(z)\otimes \d X_L(0)\;\sim\; -\frac{(P_-\otimes P_-) (Q)}{z^2},
\end{equation}
\begin{equation}
	\d X_R(\bar z)\otimes \d X_R(0)\;\sim\; \frac{(P_+\otimes P_+) (Q)}{\bar{z}^2},
\end{equation}
where $Q$ from (\ref{Q-intersection}) is understood as an element of $H^2(M_4,\Z)^{\otimes 2}\subset 	H^{2}(M_4,\R)^{\otimes 2}$, so that $(P_\pm\otimes P_\pm)(Q) \in H^{2,\pm}(M_2)^{\otimes 2}$.

Normally, non-abelian 2d $\CN=(0,2)$ gauge theories exhibit rich structure of phases \cite{Gadde:2014ppa,Gadde:2016khg}:
conformal, confining, dynamical SUSY breaking, {\it etc.}
In part for this reason, the infra-red theory $T[M_4]$ is usually very different from the result of a naive Kaluza-Klein reduction.
However, in the abelian case, that is for a single fivebrane, the theory is free and, therefore, a careful
Kaluza-Klein reduction is expected to correctly capture the physics of $T[M_4]$.
To verify that this is indeed the case, we can write down the central charges.
Combining the contributions of all left-moving fields in Table~\ref{tab:TMfields}, we find
\be
c_L \; = \;
\underbrace{2 + b_2^+ (M_4)}_{\text{non-compact} \atop \text{non-chiral bosons}}
+ \underbrace{~b_2^- (M_4)~}_{\text{compact} \atop \text{chiral bosons}}
-\underbrace{~3b_1 (M_4)~}_{\text{vectors}}+\underbrace{~b_1 (M_4)~}_{\text{Weyl} \atop \text{fermions}}
\; = \; \chi (M_4),
\label{cleft}
\ee
and, similarly, in the right-moving sector:
\be
c_R \; = \;
\underbrace{2 + b_2^+ (M_4)}_{\text{non-compact} \atop \text{non-chiral bosons}}
+ \underbrace{\frac{1}{2}(2 + 2 b_2^+ (M_4))}_{\text{right-moving} \atop \text{real fermions}}
+ \underbrace{~b_2^+ (M_4)~}_{\text{compact} \atop \text{chiral bosons}}
-\underbrace{~3b_1 (M_4)~}_{\text{vectors}}
\; = \; \frac{3}{2} (\chi + \sigma),
\label{cright}
\ee
where $\chi$ is the Euler characteristic of the 4-manifold $M_4$ and $\sigma$ is its signature.
These expressions agree with the infra-red central charges of $T[M_4]$ computed from the anomaly
polynomial of a single fivebrane \cite{Witten:1996hc,Harvey:1998bx}:
\be
A_{M5} \; = \; \frac{1}{48} \Big[ p_2 (N) - p_2 (T) + \frac{1}{4} (p_1 (T) - p_1 (N))^2 \Big],
\label{AM5}
\ee
where $T$ and $N$ denote, respectively, the tangent and normal bundle of the fivebrane world-volume
in the eleven-dimensional space-time. In our setup \eqref{Msetup}, we have $T = T\Sigma \oplus TM_4$
and $N = R \oplus \Lambda^{2,+} (M_4)$, where $R$ is the $U(1)_R$ R-symmetry bundle.
Substituting these into \eqref{AM5} and integrating over $M_4$ as in \cite{Alday:2009qq},
we obtain the standard form of the anomaly polynomial in a 2d $\CN=(0,2)$ theory:
\be
A_{2d} \; = \; \frac{c_R}{6} c_1 (R)^2 + \frac{c_L - c_R}{24} p_1 (T\Sigma),
\label{A2d}
\ee
with the values of $(c_L,c_R)$ in \eqref{cleft} and \eqref{cright}.
This gives a further justification to the result of the Kaluza-Klein reduction.

Still, there is something very peculiar about the fields listed in Table~\ref{tab:TMfields}.
While $(\phi_0, \chi_+)$ combine into a standard 2d $\CN = (0,2)$ chiral superfield
$\Phi_0 = \phi_0 + \theta^+ \chi_+ - i \theta^+ \bar \theta^+ \bar D \phi_0$,
the other fields form somewhat unusual representations of 2d $\CN=(0,2)$ supersymmetry algebra.
For example, the left-moving compact scalars $X_L^j$ live in a trivial representation of the supersymmetry algebra.
This is consistent because they satisfy $\bar \partial X_L^j = 0$, so one indeed can impose $Q_+ X_L^j = \bar Q_+ X_L^j = 0$.\footnote{This can be contrasted with the Fermi multiplets that contain only left-moving on-shell degrees of freedom, yet are acted on non-trivially by right-moving supersymmetries.}
These are not off-shell multiplets, however.

Similarly, the right-moving compact scalars $X_R^i$ belong to peculiar multiplets $\Phi_i = (\phi^i, \psi_+^i)$
whose lowest components are complex scalar fields $\phi^i := \sigma^i + i X_R^i$.
The latter has the unusual property that its imaginary part is a chiral (right-moving) scalar,
while the real part is an ordinary scalar.
We can still treat $\Phi_i$ as an off-shell $\CN = (0,2)$ chiral multiplet if we regard
the chirality condition of $X_R^i$ as part of the equations of motion.\\

At this point, it is instructive to put our theories $T[M_4]$ in the context of general 2d supersymmetric sigma-models
used, {\it e.g.}, as world-sheet theories in string compactifications. Such sigma-models describe maps
\be
\phi: \quad \Sigma \to X
\label{SmaptoX}
\ee
into a target space $X$, which in theories with $\CN=(2,2)$ or $\CN=(0,2)$ supersymmetry must be K\"ahler (if there is no $B$-field) \cite{Hull:1985jv}.
In the case of 2d $\CN=(0,2)$ sigma-models that are directly relevant to us here, the geometric data also
involves a holomorphic bundle $\CE$ over $X$, such that:
\begin{align}
c_1 (\CE) \; &= \; c_1 (TX) \quad \text{mod}~ 2 , \cr
c_2 (\CE) \; &= \; c_2 (TX) \,.
\end{align}
These anomaly cancellation conditions are trivially satisfied for our theories $T[M_4]$
which, for simply-connected $M_4$, have trivial bundle $\CE$ and
\be
X \; = \; \C \times (\C^*)^{b_2^+}.
\label{XforTM}
\ee
Actually, our theories are slightly more subtle than traditional sigma-models with this target due to asymmetry between left and right-moving bosons (see Table~\ref{tab:TMfields}). We will return to this class of theories shortly;
for a moment, though, let us continue with the general sigma-model based on a K\"ahler target space $X$.

Following the standard conventions,
we introduce local complex coordinates $\phi^i$, $i = 1, \ldots, \dim_{\C} (X)$,
their complex conjugate $\phi^{\bar i} : = \bar{\phi_i}$,
and denote by  $\psi_+^i$ and $\psi_+^{\bar i}$
projections of the right-moving fermions --- which couple to the tangent bundle $TX$ ---
into $K_{\Sigma}^{1/2} \otimes \phi^* (T^{1,0} X)$ and $K_{\Sigma}^{1/2} \otimes \phi^* (T^{0,1} X)$, respectively:
\be
\psi_+^i \; \in \; \Gamma \Big( K_{\Sigma}^{1/2} \otimes \phi^* (T^{1,0} X) \Big)
\qquad , \qquad
\psi_+^{\bar i} \; \in \; \Gamma \Big( K_{\Sigma}^{1/2} \otimes \phi^* (T^{0,1} X) \Big).
\ee
Here, $K_{\Sigma}$ is the canonical line bundle of $\Sigma$, the bundle of one-forms of type $(1,0)$.

The half-twisted model is obtained by modifying the Lorentz transformations of the fields
by the $U(1)_R$ R-symmetry, under which spin-$\tfrac{1}{2}$ fields $\psi_+^i$ have charge $+1$
and $\psi_+^{\bar i}$ have charge $-1$.
As a result, in the half-twisted model they transform as spin-0 and spin-1 fields:
\be
\psi_+^i \; \in \; \Gamma \Big( \phi^* (T^{1,0} X) \Big)
\qquad , \qquad
\psi_+^{\bar i} \; \in \; \Gamma \Big( K_{\Sigma} \otimes \phi^* (T^{0,1} X) \Big).
\ee
Their zero modes contribute to the ``ghost number anomaly'' which,
according to the Hirzebruch-Riemann-Roch theorem, on a genus-$g$ surface $\Sigma$
is given by (see, {\it e.g.}, \cite{Katz:2004nn,Witten:2005px,Melnikov:2012hk}):
\begin{multline}
\label{ghostanomaly}
~~~~~~~~~~~~\# (\psi_+^i~\text{zero-modes}) - \# (\psi_+^{\bar i}~\text{zero-modes}) \; = \\
= \; \dim_{\C} (X) \cdot (1-g) - \int_{\Sigma} \phi^* \left( c_1 (X) \right).~~~~~~~~~~~~
\end{multline}
Note, to ensure that the twisted model has non-anomalous Lorentz symmetry, one should twist
by a non-anomalous combination of global currents. For theories $T[M_4]$, this is not an issue
since the $U(1)_R$ symmetry that acts on the fermions $\psi_+$ as described above is non-anomalous.
Indeed, from \eqref{XforTM} it is clear that $c_1 (X)=0$ and, moreover,
the ghost number anomaly \eqref{ghostanomaly} is
\be
\Delta R_{T[M_4]} \; = \; (1-g)(1+b_2^+) \; = \; (1-g) \frac{\chi + \sigma}{2}.
\label{RTM4a}
\ee
This calculation can be easily generalized to non-simply connected 4-manifolds.

Note, another consistency check of the R-charge assignments in the theory $T[M_4]$
is the relation to the right-moving central charge, which follows from 2d $\CN = (0,2)$ superconformal algebra:
\be
c_R \; = \; 3 \Tr \gamma^3 R^2.
\label{cRRR}
\ee
Since among the fields in Table~\ref{tab:TMfields}, only complex Weyl fermions carry non-trivial R-charge, namely $R=1$,
the right-hand side is equal to $3 (1 + b_2^+)$. This agrees with the value of central charge $c_R$ in \eqref{cright}.\\

Before we turn the page, let is make a few brief remarks on the 2d $\CN=(0,2)$ theory $T[M_4]$ with $G = U(N)$ or $SU(N)$.
In particular, the chiral ring of such theory for $G=SU(2)$ or, equivalently, the ground ring of its half-twisted model
is expected \cite{Gadde:2013sca} to contain information about all Donaldson invariants of $M_4$.
While it does not require any additional impurity operators that are crucial for Seiberg-Witten invariants,
one serious drawback of the higher-rank version is that it can not be obtained by a simple Kaluza-Klein reduction.
Nevertheless, as in the abelian case \eqref{AM5}, one can derive the central charges $(c_L, c_R)$ from the anomaly polynomial
of $N$ fivebranes \cite{Harvey:1998bx,Intriligator:2000eq,Yi:2001bz}:
\be
A_{N \, \text{fivebranes}} \; = \; N A_{M5} + \frac{N(N^2-1)}{24} p_2 (N).
\ee
Integrating over $M_4$ and comparing with \eqref{A2d} gives, {\it cf.} \cite{Alday:2009qq}:
\begin{eqnarray}
c_L & = & \chi N + (2\chi + 3 \sigma) (N^3-N), \\
c_R & = & \frac{3}{2} (\chi + \sigma) N + (2 \chi + 3 \sigma) (N^3-N). \nonumber
\end{eqnarray}
Now it is easy to verify that, for $N>1$, these central charges of $T[M_4]$ do not agree with the naive Kaluza-Klein spectrum;
in particular, one needs to use alternative routes to find $T[M_4]$ (some of such routes were offered in \cite{Gadde:2013sca}).
Moreover, unitarity of $T[M_4]$ requires $c_L \ge 0$ and $c_R \ge 0$.
In the abelian case, these bounds automatically follow from positivity of the Betti numbers.
In the non-abelian case, though, they appear to imply a non-trivial condition on $c (M_4) = 2 \chi + 3 \sigma$,
which must be non-negative for the large-$N$ limit to preserve superconformal symmetry.

It is interesting to compare the large-$N$ limit of $T[M_4; U(N)]$ with a particular class of supergravity
solutions constructed in \cite{Gauntlett:2000ng}. The latter exists when $M_4$ admits a conformally half-flat Einstein metric,
which is a relatively strong condition. In particular, the virtual dimension of the moduli space of half-flat
structures on $M_4$ is \cite{Itohmoduli}:
\be
\text{expected}~\dim \CM_{\text{half-flat}}
\; = \; 26 \chi_h (M_4) - 7 c (M_4),
\ee
where, for balance, instead of $\chi$ and $\sigma$ we use
\begin{eqnarray}
\chi_h (M_4) & = & \frac{\chi (M_4) + \sigma (M_4)}{4}, \label{chihc} \\
c (M_4) & = & 2 \chi (M_4) + 3 \sigma (M_4) \qquad (= K_{M_4}^2 \quad \text{when $M_4$ is a complex surface}). \nonumber
\end{eqnarray}
It would be interesting to explore the relation between the space $\CM_{\text{half-flat}}$
and the conformal manifold of 2d $\CN=(0,2)$ theory $T[M_4]$ in the large-$N$ limit.

\subsubsection{Example: $M_4=S^2\times S^2$}
\label{sec:example-S2xS2}

The discussion so far has been fairly general. To make it a bit more concrete, let us consider an
example of a 4-manifold which, on the one hand, is basic enough to show up a number of times through our study
and, on the other hand, plays an important role in classification of smooth structures:\footnote{In classification
of smooth structures, taking a connected sum with $S^2 \times S^2$ plays the role of
a ``nilpotent'' operation which, if repeated sufficient number of times, can relate any pair of smooth structures.
In particular, according to Wall's stable $h$-cobordism theorem,
{\it if $M_4$ and $M_4'$ are $h$-cobordant simply-connected smooth 4-manifolds, then there exists an integer $n \ge 0$,
such that $M_4 \, \# \, n (S^2 \times S^2)$ is diffeomorphic to $M_4' \, \# \, n (S^2 \times S^2)$.}}
\be
M_4 \; = \; S^2\times S^2.
\ee
This is a simply-connected 4-manifold with $b_2^+=b_2^-=1$.
Its intersection form is represented by the following matrix:
\begin{equation}
	Q^{-1}=Q=\left(
	\begin{array}{cc}
		0 & 1 \\
		1 & 0
	\end{array}
	\right).
\end{equation}
Let $\omega^\pm \in H^2(M_4,\R)\cong \R^2$ be eigenvectors of $\ast$:
\begin{equation}
	\ast \omega^\pm =\pm \omega^\pm,
\end{equation}
that are normalized such that
\begin{equation}
	\int_{S^2\times S^2} \omega^\pm \wedge \omega^\pm =\pm 1.
\end{equation}
They are related to the basis elements $e_I$ of $H^2(M_4,\Z)$ as follows:
\begin{equation}
	\omega^\pm = \frac{e_1}{R} \pm \frac{e_2 R}{2},
\end{equation}
where $R$ depends on the conformal structure of the $S^2\times S^2$ and describes the ratio
of the sizes of the two $S^2$'s. The linear combination $X_L^1 \, \omega^- + X_R^1 \, \omega^+ \in H^2(M_4,\R)$
is periodic with respect to shifts by $e_{1,2}$. Therefore $X_L^1+X_R^1$ can be understood as a non-chiral compact boson valued in a circle with radius $R$. The T-duality $R\leftrightarrow 2/R$ corresponds to the exchange of two $S^2$'s in $M_4=S^2\times S^2$.

It follows that the full theory $T[M_4,U(1)]$ can be described as a 2d $\CN=(0,2)$ sigma-model
with target space $X=\C\times \C^*$ parametrized by \textit{non-chiral} bosons.
The same result follows from the twisted compactification on $S^2$ of the 4d theory $T[S^2,U(1)]$,
the sigma-model with the Nahm one-monopole moduli space $S^1 \times \R^3$ \cite{Assel:2016lad}.
The result of such compactification is a $\CN=(0,4)$ sigma-model with the target space $S^1\times \R^3$
and trivial bundle of left-moving fermions \cite{Putrov:2015jpa}.

Our next class of examples is more delicate and involves theories akin to the ones which play a role in classification of topological phases of matter.

\subsubsection{Examples with torsion in homology}
\label{sec:example-plumbed}

\begin{figure}[ht]
\centering
 \includegraphics[scale=3]{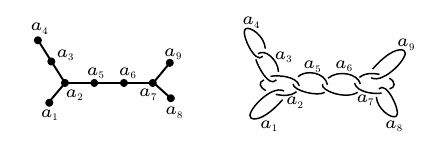}
\caption{An example of a plumbing graph $\Gamma$ (left) and the corresponding link $\CL(\Gamma)$ of framed unknots in $S^3$ (right). The associated 3-manifold $M_3(\Gamma)$ can be constructed as a Dehn surgery on $\CL(\Gamma)$.}
\label{fig:plumbing-example}
\end{figure}

Consider first a 3-manifold $M_3(\Gamma)$ associated to a (genus zero) plumbing graph $\Gamma$.
There are various ways to define $M_3$ for a given $\Gamma$.
One of them is to say that $M_3$ is obtained by a Dehn surgery on the corresponding link $\CL(\Gamma)$
of unknots, illustrated in Figure~\ref{fig:plumbing-example}.
For example, Seifert fibrations over $S^2$ correspond to star-shaped plumbing graphs.
Denote by $L$ the number of vertices of $\Gamma$.
It is equal to the number of components of the link $\CL(\Gamma)$.
We will also need $L\times L$ linking matrix of $\CL(\Gamma)$ which, somewhat suggestively, will be called $Q$:
\begin{equation}
 Q_{v_1,v_2}=\left\{
\begin{array}{ll}
 1,& v_1,v_2\text{ connected}, \\
 a_v, & v_1=v_2=v, \\
0, & \text{otherwise}.
\end{array}
\right.\qquad v_i \in \text{Vertices of }\Gamma \;\cong\;\{1,\ldots,L\}
\label{linking}
\end{equation}
We will assume that $Q$ is non-degenerate, unless explicitly noted otherwise. A plumbed 3-manifold $M_3$ can also be viewed as a link of a singularity. The plumbing graph then plays the role of the resolution graph. The resolution is a smooth 4-manifold $M_4'$ bounded by $M_3$ which has matrix $Q$ as the intersection form. In \cite{Gadde:2013sca} it was argued that $T[M_3,U(1)]$ is (an appropriate $\CN=2$ version of) $U(1)^L$ Chern-Simons theory with level matrix $Q$. Note that the 2d theory $T[M_4',U(1)]$ described in the previous section indeed naturally couples to 3d theory $T[M_3,U(1)]$ via gauging $U(1)^L$ global symmetry of $T[M_4',U(1)]$ in the 3d bulk. The $U(1)^L$ global symmetry acts as translations in the target space of compact bosons. This defines a gauge-invariant boundary condition for $U(1)^L$ Chern-Simons theory. Note that a 3d Wilson line of charge $h\in \Z^L/Q\Z^L$ ending on a boundary must couple to a boundary field of appropriate charge $\lambda\in H_2(M_4',\Z)\cong \Z^L$, such that $\lambda=h\mod Q\Z^L$, {\it e.g.}, a vertex operator $e^{2\pi i \lambda\cdot(X_L+X_R)}$.

The theory $T[M_4]$ for $M_4=M_3\times S^1$ then can be easily obtained by the dimensional reduction of 3d theory $T[M_3]$ to two dimensions.
Note that $Q$ is the intersection form for $M_4'$ described earlier, but not for $M_4$. Moreover, $b_2(M_4)=0$ when $Q$ is non-degenerate.
The non-trivial sector of the theory is given by compactification of abelian Chern-Simons theory on a circle.
This is a $(U(1)^L/U(1)^L)_Q$ gauged WZW with $U(1)^L$ WZW fields $e^{i\varphi_i}=\exp i\int_{S^1} A_i$.
Equivalently, the effective 2d theory, considered on surface $\Sigma$, can be described as a BF theory with level matrix $Q$:
\begin{equation}
	\frac{1}{2}\sum_{ij}Q_{ij}\int_{\Sigma\times S^1} A_idA_j= \sum_{ij}Q_{ij}\int_\Sigma \varphi_i dA_j,
	\label{BF-2d}
\end{equation}
which, in turn, can be described as a discrete gauge theory (of Dijkraaf-Witten type with trivial cocycle) with gauge group $\Coker Q$.
Note that one could also obtain such desciption by first compactifying the M5-brane on $S^1$
and then compactifying the 5d $U(1)$ gauge theory describing dynamics of the D4-brane on $M_3$. From this point of view,
\begin{equation}
	\varphi_i=\int_{e_i\subset M_3} A^{\text{5d}},
	\label{5d-red}
\end{equation}
where $e_i$ is the corresponding generator of $\Z^L$ in $\Z^L/Q\Z^L \cong H_1(M_3)$ and $A^{\text{5d}}$ is the 5d $U(1)$ gauge field.

The fields $A_j$ on the right-hand side of (\ref{BF-2d}) then play the role of Lagrange multipliers
imposing the constraint that $e_i\in H_1(M_3,\Z)$ is a torsion element
(contrary to a free generator, which would produce an ordinary free field via (\ref{5d-red})). Integrating out $A_j$ (together with summation over fluxes through $\Sigma$)
on the right-hand side of (\ref{BF-2d}) localizes on the configurations of the form:
\begin{equation}
	\varphi_i=\text{const along $\Sigma$}, \qquad \sum_iQ_{ij}\varphi_i =0\mod 2\pi.
\end{equation}
The fact that $T[M_4]$ is topological for such class of 4-manifolds is consistent with the fact that $c_L=c_R=0$ which follows from $\chi=\sigma=0$.

\subsection{Holomorphic differentials and fluxes}

After describing general features of the theory $T[M_4]$, let us study its flux vacua, in particular those that preserve supersymmetry.
We already mentioned in \eqref{ASDeq} that one of the conditions to preserve supersymmetry is $F^+=0$.
Because we are mostly working\footnote{Part of the reason for this assumption is to avoid wall crossing phenomena, whose 2d interpretation in the theory $T[M_4]$ we hope to address elsewhere.} with 4-manifolds that have $b_2^+>1$, one can always pick a generic metric on $M_4$ for which the space of harmonic anti-self-dual forms $\mathcal{H}^{2,-}$ has no integral points, thus there are no solutions to $F^+=0$ with $F/2\pi$ representing an integral cohomology class.

This suggests that generic $T[M_4]$ does not have any supersymmetric flux vacua, just like string compactifications with generic values of moduli.
Adjusting moduli to special values that allow supersymmetric flux vacua corresponds to choosing non-generic metrics on $M_4$ that allow solutions to $F^+=0$. In order to learn how one can circumvent this condition, we need to understand better its two-dimensional interpretation in $T[M_4]$.
Recall \cite{Witten:1997sc}, that in twisted compactification of the 6d $(2,0)$ theory on a Riemann surface $\Sigma$, four-dimensional gauge fields originate from the 6d self-dual two-form gauge field, and they correspond to normalizable holomorphic differentials on $\Sigma$. If $\Sigma$ is closed and has genus $g$, the space of such differentials has complex dimension $g$, and we pick a basis $h_I,\, I=1,\dots g$. The equation that relates 6d and 4d gauge field strengths is:
\be
\label{HviaF}
H=\sum_{I=1}^g F^{I-}\wedge h_I + F^{I+}\wedge \overline{h}_I,
\ee
and it is a close analogue of \eqref{HviaXLR}. From the 4d point of view, the supersymmetry equation $F^+=0$ simply follows from the SUSY variation of gaugino.
Now we can see the two-dimensional interpretation of fluxes by expanding $F^{I\pm}$ in the basis of (anti-)self-dual harmonic 2-forms $\omega^\pm_j$ and comparing \eqref{HviaF} to \eqref{HviaXLR}.

{}From the 2d point of view, $\partial X_L^j$ becomes a non-trivial linear combination of basic holomorphic differentials $h_I$ and $\bar\partial X_R^i$ becomes a linear combination of anti-holomorphic differentials $\overline{h}_I$. This means that scalar fields $X_L^j$ and $X_R^i$ are only locally defined, while globally they might have monodromies. Their derivatives $\partial X_L^j$ and $\bar\partial X_R^i$ are, of course, globally defined holomorphic and anti-holomorphic differentials.

We can say more about two-dimensional flux vacua if we recall that there are $\CN=(0,2)$ chiral multiplets $\Phi_i=(\phi^i,\psi_+^i)$ with $\phi^i=\sigma^i+i X_R^i$. Supersymmetry variation of $\psi_+^i$ implies another BPS equation:
\be
\label{BPS2d}
\bar\partial\left(\sigma^i + i X_R^i\right)=0,
\ee
which means that $\sigma^i + i X_R^i$ is locally holomorphic on $\Sigma$. In the absence of topologically nontrivial fluxes, we would say that this is also true globally and use it (as well as chirality of $X_R^i$) to argue that $\sigma^i$ and $X_R^i$ have to be constant separately. In the presence of fluxes, however, we already know that $\bar\partial X_R^i$ becomes a non-trivial anti-holomorphic differential, so this equation implies the same for $\bar\partial\sigma^i$. In other words, in the presence of fluxes, $\sigma^i$ is only locally defined too. Globally, it has monodromies around cycles of $\Sigma$.

While the fact that $X_L^j$ and $X_R^i$ are only locally defined is natural --- after all, we can think of them as originating from the terms $\sum_{j=1}^{b_2^-} X_L^j \omega_j^- + \sum_{i=1}^{b_2^+} X_R^i \omega_i^+$ in the $B$-field, and $B$, being a two-form gauge fields, is certainly only locally defined --- the fact that $\sigma^i$ are also only locally defined on $\Sigma$ comes as a surprise. The fields $\sigma^i$ parametrize deformations of coassociative 4-cycles in a $G_2$-manifold that locally looks like $\Lambda^{2,+}(M_4)$. The point where all $\sigma^i=0$ corresponds to the choice of $M_4\subset \Lambda^{2,+}(M_4)$ as a zero section of $\Lambda^{2,+}(M_4)$. Non-zero $\sigma^i$ describe deformations of this $M_4\subset\Lambda^{2,+}(M_4)$ that preserve its calibrated property. Therefore, vacuum expectation values of $\sigma^i$ describe the choice of $M_4$ in a $G_2$ manifold, and so describe the geometry of the fivebrane world-volume.

The fact that for supersymmetric flux vacua $\sigma^i$ has monodromies on $\Sigma$ means that, as we go around cycles of $\Sigma$, $\sigma^i$ get shifted by constants. This means that the M5-brane worldvolume is not really a compact manifold $M_4\times\Sigma$, as was assumed from the very beginning, but rather its non-compact covering: every time we go around a cycle on $\Sigma$, we end up on a different copy of $M_4$ inside of $\Lambda^{2,+}(M_4)$. We interpret this contradiction with the initial assumptions as a statement that the theory $T[M_4]$ (which was \emph{defined} by twisted compactification on $M_4$) does not have supersymmetric flux vacua.

This problem can be avoided by introducing our next ingredient, the impurity vertex operators in 2d theory $T[M_4]$,
which correspond to extra matter fields in the 4d theory on $M_4$ and contribute to the right-hand side of the BPS equation $F^+=0$.


\section{Seiberg-Witten invariants and the Kondo problem}
\label{sec:Kondo}

The classical version of the Kondo problem has to do with the structure of a ground state
in a system of free fermions coupled to a localized magnetic impurity.
In particular, it explains a peculiar rise in resistivity (observed in 1930s) that some metals exhibit at low temperature.
The model, proposed by Jun Kondo in 1964, is based on the so-called s-d Hamiltonian,
\be
H \; = \; \psi_{\alpha}^{\dagger} \left( - \frac{\nabla^2}{2m} - \epsilon_F \right) \psi_{\alpha}
+ J \delta (\vec x) \vec \CS \cdot \frac{\vec \tau_{\alpha \beta}}{2} \psi_{\alpha}^{\dagger} \psi_{\beta},
\label{Kondo}
\ee
which describes the conduction band (s-band) electrons $\psi_{\alpha}$ of a non-magnetic metal
interacting with a magnetic impurity (unfilled d-level) represented here by the spin-$\tfrac{1}{2}$ operator $\vec \CS$.
Here, $\alpha = \uparrow$ or $\downarrow$ is the spin index, $\epsilon_F$ is the Fermi energy, and $J$ is the Kondo coupling.
In the renormalization group approach, the Kondo problem is reduced to the effective theory of
massless left-moving Dirac fermions in 1+1 dimensions, a close cousin of $T[M_4]$, interacting with
a localized impurity.\footnote{The s-wave reduction and the doubling trick lead to the effective 1+1 model with the Hamiltonian:
\be
H_{\rm eff} = \frac{i}{2\pi} \int_{-\infty}^{+ \infty} dr \, \psi^{\dagger}_L \frac{d}{dr} \psi_L
+ \lambda \, \vec \CS \cdot \psi^{\dagger}_L (0) \frac{\vec \tau}{2} \psi_L (0).
\ee}

When the Kondo interaction is ferromagnetic, $J < 0$, the effective coupling flows to zero
and at low energies the impurity completely decouples from the conduction electrons.
On the other hand, when the interaction is anti-ferromagnetic ($J>0$),
the effective Kondo coupling grows logarithmically at low temperatures, so that
the impurity is completely screened by forming a singlet with the electrons around it.
In the description where impurity is represented by auxiliary ``slave'' fermions,
the formation of a Kondo singlet manifests itself as a condensation of a $0+1$ dimensional charged scalar,
made of an electron and a slave fermion.

The physics of this simple model and its generalizations is closely related to 2d and 3d realization
of Seiberg-Witten invariants discussed below, where $T[M_4]$ or $T[M_3]$ play the role of conduction electrons
and interesting physics comes from interaction with a certain ``impurity'' operator $\CS$.
Similar generalizations of the Kondo model, where free electrons are replaced by an interacting CFT,
describe, {\it e.g.}, impurities in Luttinger liquids.
Another natural generalization involves a version of the model \eqref{Kondo} where fermions carry an extra index $i = 1, \ldots, N_f$.
This ``multi-channel'' Kondo model exhibits interesting phases:
in the underscreened regime ($N_f < 2S_{\text{imp}}$) the ground state has non-trivial degeneracy and
the physics is similar to that of a single-channel Kondo model, whereas
in the overscreened case ($N_f > 2S_{\text{imp}}$) the RG flow leads to a non-trivial fixed point
and the ground state is not described by standard Fermi-liquid theory.

\subsection{Electric and magnetic impurities}

As our first example of a supersymmetric Kondo problem, let us consider a simple three-dimensional theory:
\be
\text{3d $\CN=2$ super-Chern-Simons theory $U(1)_p$}.
\label{superCS}
\ee
This theory by itself is rather boring; it has only one supersymmetric ground state on a 2-sphere $S^2$
and, correspondingly, the $Q$-cohomology $\CH_{\text{BPS}}$ is one-dimensional.\footnote{And
it has $p$ isolated supersymmetric vacua on a 2-torus $T^2$. There are several ways to see this,
{\it e.g.}, from the supersymmetric index $\Tr (-1)^F = |p|$ or by reducing on one circle at a time.
Indeed, on a circle it gives a 2d $\CN=(2,2)$ theory of a single twisted chiral multiplet
with a twisted superpotential $\tilde W = p \sigma^2 / 2$. The vacua of such theory are the critical points,
{\it i.e.}, solutions to:
\be
\exp \left( \frac{\partial \tilde W}{\partial \sigma} \right) \; = \; 1,
\ee
and, in our case at hand, it is easy to see that there are $p$ of them.}

Now, let us see how this changes when we introduce an impurity localized in space \cite{Gukov:2016gkn}:
a pair of chiral multiplets $\phi$ and $\tilde \phi$ with $U(1)$ gauge charges $+1$ and $-1$, respectively.
The Lagrangian of this combined 1d-3d system is similar to that of a 3d $\CN=2$ SQED, except the chiral
multiplets $\phi$ and $\tilde \phi$ are restricted to a 0+1 dimensional world-line of the impurity.
Then, in the presence of the impurity, the space of BPS states is simply the space of states
of a harmonic oscillator, which are excitations of the ``meson'' $\phi \tilde \phi$.
We give it a name:
\be
\CT^+ \quad =
{\,\raisebox{-1.5cm}{\includegraphics[width=3.5cm]{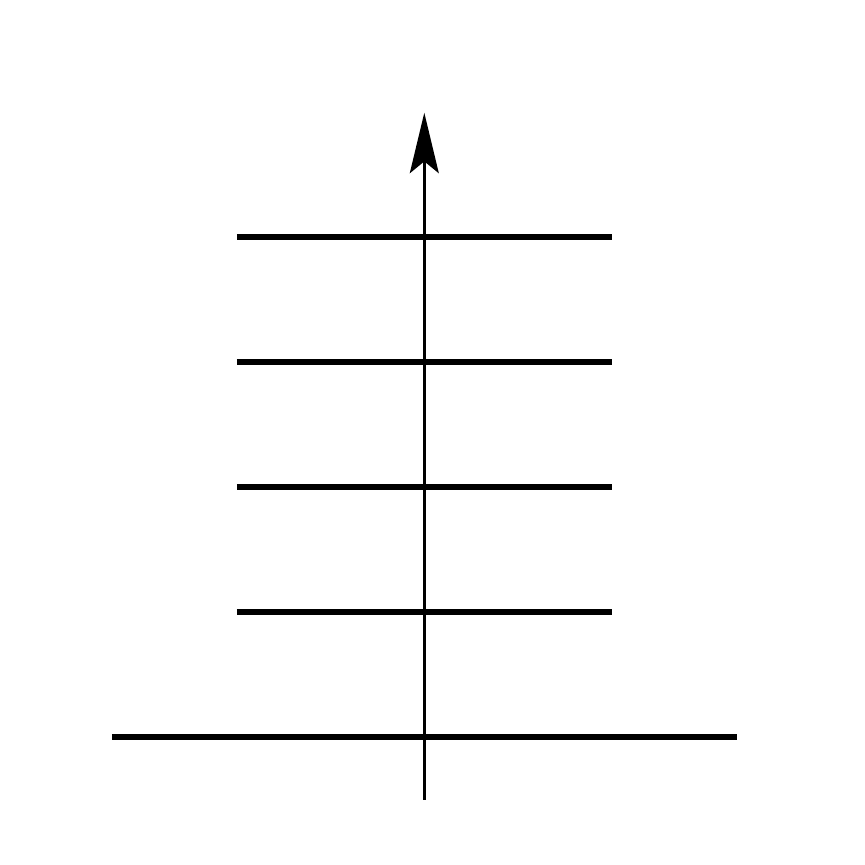}}\,}
\label{Tplus}
\ee
What we just produced is the Heegaard Floer homology $HF^+ (M_3)$ of a particular simple 3-manifold $M_3$.\\

This simple example can be generalized in a number of ways, in particular, allowing to describe
homological invariants of an arbitrary 3-manifold (possibly with knots) as a space of supersymmetric ground states
in a 3d $\CN=2$ theory with a certain half-BPS impurity localized in space, see, {\it e.g.}, \cite{Gukov:2017kmk}.
The prototypical example of a 3d $\CN=2$ theory $T[M_3]$ is the three-dimensional supersymmetric quantum electrodynamics (SQED).
In fact, for many 3-manifolds the theory $T[M_3]$ admits duality frames where its UV Lagrangian
involves only abelian gauge group coupled to matter fields. In such duality frames, half-BPS impurities
are basically electric or magnetic impurities of 3d $\CN=2$ SQED studied, {\it e.g.}, in \cite{Hook:2013yda,Tong:2013iqa}.

\begin{figure}[ht]
\centering
\includegraphics[trim={0 0.5in 0 0.5in},clip,width=4.0in]{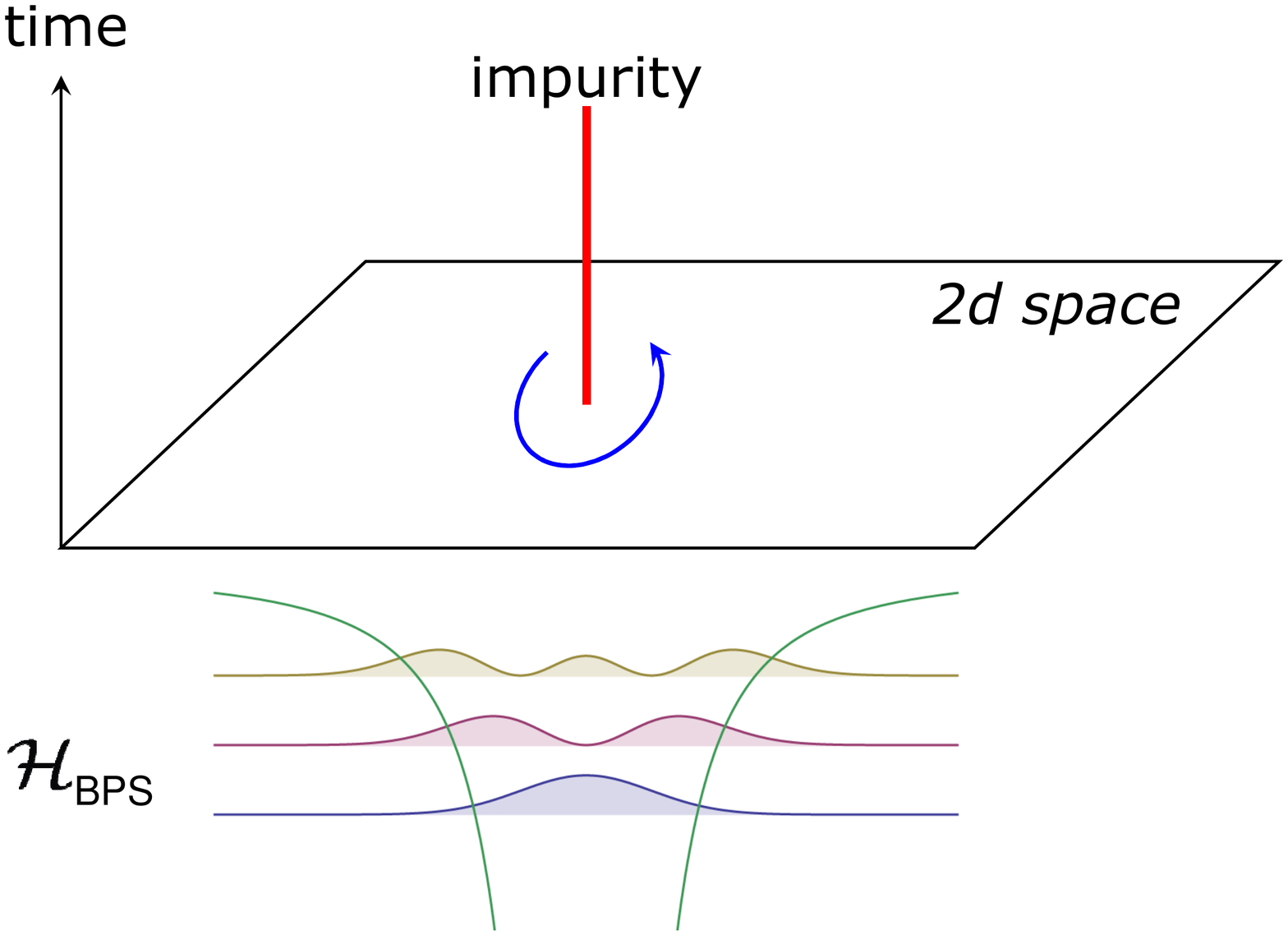}
\caption{An illustration (from \cite{Gukov:2017kmk}) of a supersymmetric spectrum in 3d $\CN=2$ theory with an impurity.}
\label{fig:impurity}
\end{figure}

Let us consider an abelian 3d $\CN=2$ gauge theory with gauge group $U(1)$, which is a basic
building block of many 3d $\CN=2$ theories $T[M_3]$.
A half-BPS electric impurity in this theory creates an equal source for the gauge field
as well as for the real scalar field $\sigma$ in the $U(1)$ vector multiplet:
\be
S_{\text{impurity}} \; = \; - \frac{i}{2\pi} \int d^3 x \, (A_0 - \sigma) \rho (x).
\label{Simpel}
\ee
Since $A_0$ and $\sigma$ are both sourced by the electric charge density $\rho (x)$
and obey the same equations of motion, the BPS (resp. anti-BPS) solutions have $A_0 = \sigma$ (resp. $A_0 = - \sigma$).
Indeed, the supersymmetry variation of the gaugino leads to the BPS equation:
\be
(\partial_i A_0 \gamma^0 - \partial_i \sigma) \epsilon = 0,
\ee
which is solved by $A_0 = \pm \sigma$ and the spinor $\epsilon$ that obeys $(1 \mp \gamma_0) \epsilon = 0$.
In the presence of the source term \eqref{Simpel}, the equation of motion for the temporal component of
the gauge field looks like:
\be
\frac{2 \pi}{g^2} \partial_i^2 A_0 \; = \; \rho(x),
\ee
and $\sigma$ obeys the same equation since $A_0 = \sigma$.
Later on, it will be convenient to replace the gauge coupling constant $g^2$
by the parameter $2\pi R = g^2$.

The simplest example of a half-BPS electric impurity is the familiar supersymmetric Wilson line.
If we assume that a Wilson line (of charge $q_0$) is extended along the ``time'' direction and
localized at the origin of the two-dimensional ``space'' $\Sigma$, it can be represented by the charge density:
\be
\rho (z, \bar z) \; = \; 2\pi q_0 \delta^2 (z, \bar z),
\ee
where $z$ is a complex coordinate on $\Sigma$.
(For the purposes of the present discussion, we can take $\Sigma$ to be simply a complex plane, $\Sigma \cong \C$.)
{}From the above equations, it is easy to see that such a delta-function source creates
an equal profile for the fields $A_0$ and $\sigma$:
\be
A_0 = \sigma = - \frac{1}{4\pi} g^2 q_0 \log |z|^2.
\label{elBPSimp}
\ee
We can write it in a more convenient form, that will be suggestive of generalizations in 2d $\CN=(0,2)$ theory $T[M_4]$,
by introducing the dual photon $X$,
\be
F_{\mu\nu} \; = \; \epsilon_{\mu \nu \rho} \partial^{\rho} X,
\ee
or, in holomorphic coordinates,
\be
\partial_z X = i \partial_z A_0
\qquad , \qquad
\partial_{\bar z} X = - \partial_{\bar z} A_0.
\ee
Then, the charge quantization implies that $X \sim X + g^2$ is periodic with period $g^2$,
and the solution \eqref{elBPSimp} can be written as:
\be
\sigma + i X \; = \; - \frac{g^2}{2\pi} q_0 \log (z).
\label{Bionsol}
\ee
In other words, in terms of $\sigma + i X$ the BPS condition is simply $\bar \partial (\sigma + i X) = 0$
away from the impurity.
This is precisely the form of the BPS equations \eqref{BPS2d} in 2d half-twisted $(0,2)$ model on $\Sigma$
that will be relevant to us in what follows; its solutions are holomorphic maps \eqref{SmaptoX}.

Moreover, by considering $z \mapsto e^{i \theta} z$, it is easy to see that going around
electric impurity of charge $q_0$, the dual photon has a monodromy $X \to X + q_0 g^2$.
In other words, a half-BPS electric impurity creates a ``winding'' of the dual photon and,
therefore, in the language of a two-dimensional theory on $\Sigma$ it is a winding-state operator.
In D-brane physics, a half-BPS solution \eqref{Bionsol} is known as the ``BIon'' spike \cite{Callan:1997kz,Gibbons:1997xz}
that describes charges carried by strings ending on D-branes and related configurations.

Magnetic impurities can be treated in a similar way and correspond to ``momentum'' operators for the dual photon.
For example, a half-BPS magnetic impurity (or, equivalently, an external vortex) in 3d $\CN=2$ SQED
is a source that equally couples to the magnetic field $B$ and to the auxiliary field $D$.
Thus, a delta-function source $B = 2\pi q_0 \delta^2 (z, \bar z)$ has $A = q_0 d \theta$
and defines a magnetic impurity of charge $q_0$. The ``mirror symmetry'' of 3d $\CN=2$ gauge theories
maps it to an electric impurity of the dual theory (see, {\it e.g.}, \cite{Hook:2013yda,Tong:2013iqa} for more details).

Dimensional reduction of electric and magnetic impurities discussed here gives local impurity
operators in 2d $\CN=(2,2)$ theories. For example, dimensional reduction of BPS vortex operators
gives the supersymmetric completions of the ``momentum'' operators for the dual photon:
\be
\exp \left( \pm 2\pi \frac{\sigma + i X}{g^2} \right),
\ee
that play an important role in mirror symmetry of 2d $\CN=(2,2)$ theories~\cite{Hori:2000kt}.

In fact, our next goal will be to consider two-dimensional versions of such electric and magnetic impurities
in theories with $\CN=(0,2)$ supersymmetry, first in general and then focusing more closely on applications to $T[M_4]$.

\subsection{Half-twisted model with target space $\C^* \simeq \R \times S^1$}

Let us consider a simple 2d $\CN=(0,2)$ theory
which consists of a single chiral superfield $\Phi$ with period $2 \pi i R$.
(A reader less familiar with 2d $\CN=(0,2)$ supersymmetry may find it convenient to think
of a chiral superfield either in 3d $\CN=2$ theory or, via dimensional reduction, in 2d $\CN=(2,2)$ theory;
this will also help to make contact with the above discussion of impurities in 3d.)
This simple theory is basically a model for $T[M_4]$ when $M_4 = S^2 \times S^2$,
{\it cf.} section~\ref{sec:example-S2xS2}.

Let $\sigma + i X$ be a complex scalar in 2d free chiral superfield $\Phi$, such that $X$ is periodic,
\be
X \; \sim \; X + 2\pi R,
\ee
and, as usual, we can introduce its left-moving and right-moving components on-shell:
\be
X (z, \bar z) \; = \; X_L (z) + X_R (\bar z).
\ee

\begin{figure}[ht]
\centering
\includegraphics[width=4.0in]{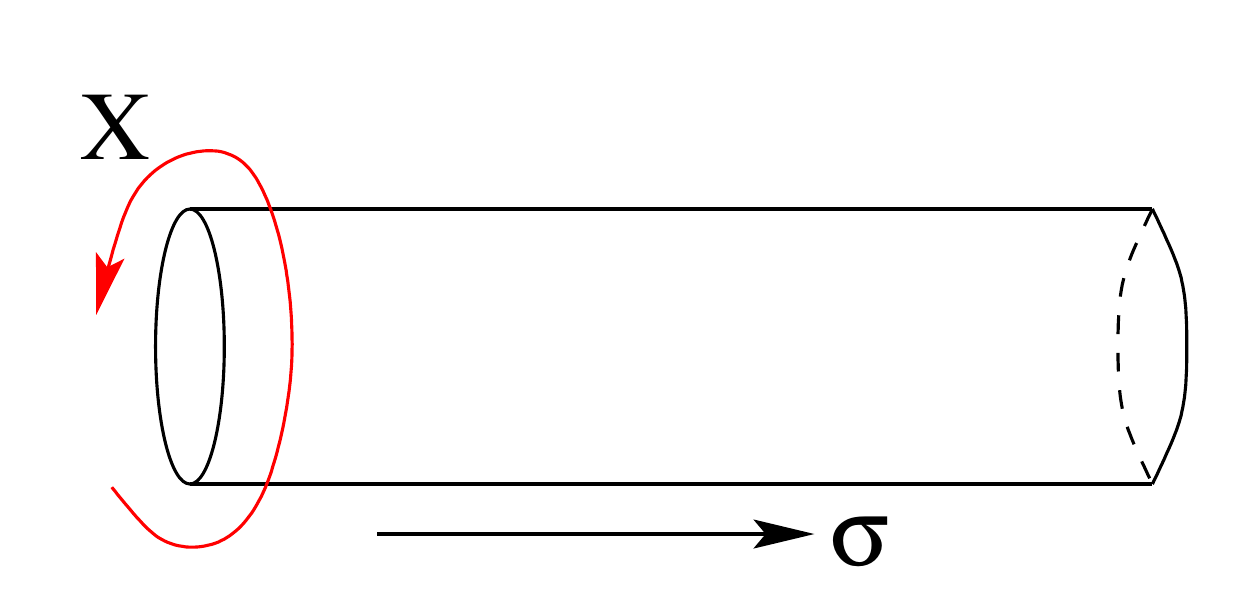}
\caption{Target space $\C^* \simeq S^1 \times \R$ parametrized by $\sigma + i X$.}
\label{fig:cylinder}
\end{figure}

To construct winding and momentum operators in this theory, we can consider
general vertex operators of the form:
\be
e^{i (k_L X_L + k_R X_R) + p \sigma},
\label{generalVkkp}
\ee
where for a moment we suppress fermions.
Note, since $X$ is periodic, the momenta $k_L$ and $k_R$ are quantized:
\be
\big( k_L, k_R \big) \; = \; \left( \frac{n}{R} + \frac{wR}{2} , \; \frac{n}{R} - \frac{wR}{2} \right),
\qquad n,w \in \Z.
\ee
Moreover, as in our previous discussion, we are interested in BPS operators.
In 2d $\CN=(0,2)$ theory, these are the operators in $\bar Q_+$-cohomology
of the right-moving $\CN=2$ supersymmetry algebra.
In particular, in our simple example this imposes a condition on the right-moving
components of the fields $X$ and $\sigma$ that must appear in a holomorphic combination
$\sigma + i X_R$ in order to be in $\bar Q_+$-cohomology.
Such operators have $p = k_R$, and we conclude that \eqref{generalVkkp}
can be a bosonic representative of a BPS vertex operator when it is right-holomorphic,
\be
e^{i k_L X_L + k_R (\sigma + i X_R)}.
\label{holVkk}
\ee
Another way to arrive at the same conclusion is to note that general operators \eqref{generalVkkp}
have scaling dimensions:
\be
h = \frac{1}{2} k_L^2 - \frac{1}{2} p^2
\qquad , \qquad
\bar h = \frac{1}{2} k_R^2 - \frac{1}{2} p^2,
\ee
and the BPS bound $\bar h = 0$ also leads to $p = k_R$. (Vertex operators with $p=-k_R$ are anti-BPS.)

In what follows, we denote the supersymmetric completion of the right-holomorphic vertex operators \eqref{holVkk}
by $V_{\lambda}$ and write:
\be
V_{\lambda} \; = \; e^{i k_L X_L + k_R \Phi },
\label{Vlambda}
\ee
where $\lambda := (k_L, k_R)$ and $\Phi$ is a peculiar version of a 2d $\CN=(0,2)$ chiral multiplet
discussed in section~\ref{sec:fluxes}; the lowest component of $\Phi$ is a complex scalar field $\sigma + i X_R$
whose real part is an ordinary, non-chiral scalar and the imaginary part is a chiral right-moving scalar.
Note, this form of winding and momentum operators has an obvious generalization to
a theory with multiple fields $\sigma$, $X_L$, and $X_R$, as in Table~\ref{tab:TMfields}.
The only modification is that $\lambda$ takes values in the winding/momentum lattice $\Gamma$,
which in theories $T[M_4]$ can be identified with the cohomology lattice of the 4-manifold $M_4$:
\be
\lambda := (k_L, k_R) ~\in~ \Gamma := H^2 (M_4, \Z).
\ee
By construction, the supersymmetric operators $V_{\lambda}$ have $\bar h = 0$
and $h = \frac{1}{2} k_L^2 - \frac{1}{2} k_R^2$, which we sometimes write as:
\be
h \; = \; \lambda_-^2 - \lambda_+^2,
\ee
using the fact \eqref{HviaXLR} that self-dual and anti-self-dual projections of $\lambda$
correspond to projecting on the right-moving and left-moving modes, respectively.
The supersymmetric completions of momentum and winding operators play an important role
in mirror symmetry~\cite{Hori:2000kt} and in various extensions of the sine-Liouville
theory.\footnote{Note, that in Liouville theory a correlation function of operators $V_{\lambda_i}$
is finite if and only if these operators satisfy the Seiberg bound \cite{Seiberg:1990eb}:
\be
\lambda_i < Q \qquad , \qquad \sum_i \lambda_i > 2Q.
\ee
In Liouville theory, normalizable states are called non-local and non-normalizable states
are called local; the latter correspond to curvature singularity on $\Sigma$.}

One of the main goals of this paper is to relate Seiberg-Witten invariants of $M_4$
(and their generalizations) to correlation functions of the operators $V_{\lambda}$
in the half-twisted theory $T[M_4]$. In the absence of a background charge,
all such correlators are subject to a ``neutrality condition'' which states
that correlation functions vanish unless the total charge $\lambda$ of the operators
is equal to zero.\footnote{Recall, that in a theory of a chiral boson $X$,
vertex operators $V_{\lambda} = e^{i \lambda X (z)}$ have conformal weight
$h = \frac{1}{2} \lambda^2$ and charge $Q(V_{\lambda}) = \lambda$
with respect to the $U(1)$ current $J = \partial X$,
\be
Q \; = \; \frac{1}{2\pi} \oint dw J(w).
\ee
In this case, the ``neutrality condition'' means that the total $U(1)$ charge must vanish.}
Sometimes, it will be convenient to consider the corresponding winding and momentum states:
\be
\vert \lambda \rangle \; \equiv \; \vert k_L , k_R \rangle \; := \;
V_{\lambda} \vert 0 \rangle \; \sim \;
e^{i k_L X_L(0) + k_R \sigma (0) + i k_R X_R (0)} \vert 0 \rangle.
\ee

So far, our discussion of BPS winding-momentum operators $V_{\lambda}$ was fairly general and applies
to a broad class of 2d $\CN=(0,2)$ theories. Our next task is to gain a more detailed understanding
of vertex operators in theories $T[M_4]$ by using the fivebrane setup~\eqref{Msetup}.

\subsection{Anomalies: M2-branes ending on M5-branes and embedded surfaces}
\label{sec:M2branes}

\begin{figure}[ht]
\centering
\includegraphics[width=1.8in]{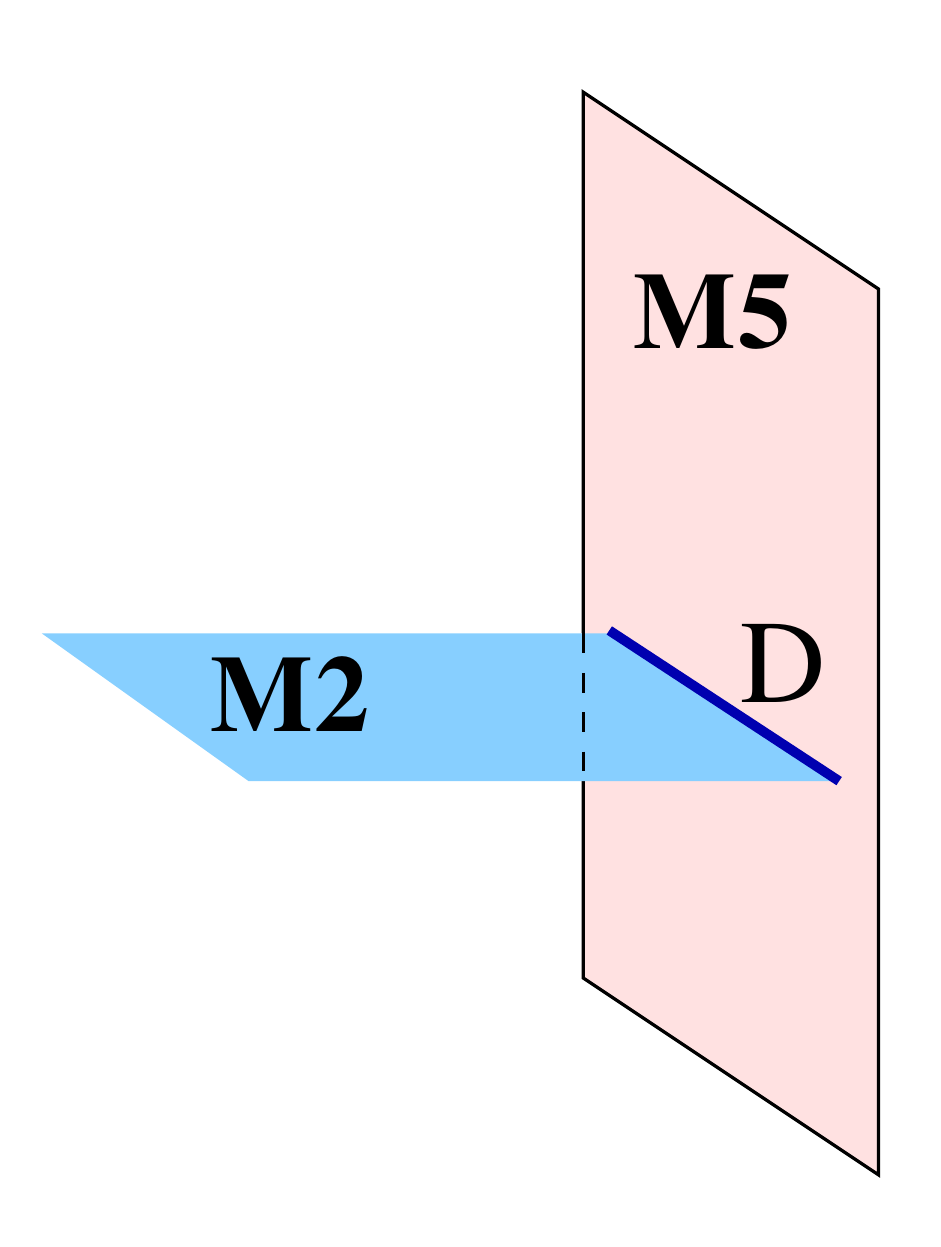}
\caption{M2-branes ending on M5-branes.}
\label{fig:M2M5}
\end{figure}

In this section, we propose to identify vertex operators $V_{\lambda}$ in 2d theory $T[M_4]$
with (non-dynamical) half-BPS self-dual strings in 6d $(2,0)$ fivebrane
theory --- or, equivalently, with M2-branes ending on M5-branes, as illustrated in Figure~\ref{fig:M2M5} ---
such that
\be
\lambda \; = \; [D]
\ee
is the Poincar\'e dual cohomology class of the string world-sheet, the embedded surface $D \subset M_4$.
Then, we discuss important consequences of this identification, including, {\it e.g.}, application to knot cobordisms.

First, since M2-brane ending on a fivebrane acts as a source for the 3-form field $H$,
\be
dH = \pi \delta^{(4)} (D \hookrightarrow M_4 \times \Sigma),
\ee
it follows from the Kaluza-Klein reduction \eqref{HviaXLR} that scalar fields
$X_L^j$ and $X_R^i$ are no longer single-valued, but rather have prescribed monodromies.
And, $V_{\lambda}$ are precisely the vertex operators that create such monodromies.

Secondly, both $V_{\lambda}$ described above as well as M2-branes
ending on $D \subset M_4$ preserve the same supersymmetry,
namely they are both BPS with respect to the same supercharge $\bar Q_+$ regardless of the choice of $D$.
We already demonstrated this for the vertex operators $V_{\lambda}$
and it is also true for the non-dynamical self-dual strings in 6d $(2,0)$ fivebrane theory.
The reason is that 6d theory is topologically twisted along the 4-manifold $M_4$,
so that any choice of metric on $M_4$ and any shape of $D \subset M_4$
preserve the topological supercharge.

{}From the vantage point of the brane system \eqref{Msetup}, the topological twist along the 4-manifold $M_4$
is implemented by embedding this part of the fivebrane world-volume in the $G_2$ manifold $\Lambda^{2,+} (M_4)$.
In this setup, non-dynamical self-dual strings supported on $D \subset M_4$ are realized by
M2-branes supported on associative 3-manifolds in $\Lambda^{2,+} (M_4)$ that meet $M_4$ along $D$.
In other words, given a surface $D \subset M_4$, we wish to associate to it an associative
submanifold inside $\Lambda^{2,+} (M_4)$ that meets $M_4$ along $D$.
The geometry of such associative submanifolds was briefly discussed in \cite{Gorsky:2013jxa}
and follows the ``conormal bundle'' construction of \cite{Ooguri:1999bv} when $M_4 = \R \times M_3$.

Recall that, in general, for a submanifold $K \subset M$, the conormal bundle $N^* K \subset T^* M$
is a subbundle defined as:
\be
N^* K \; := \; \{ (x,p) \in T^* M ~\vert~ x \in K \,, T_x K \subset \text{ker} (p) \}.
\ee
Similarly, given a 2-dimensional surface $D \subset M_4$, the span of $\text{vol}_D + *_4 \text{vol}_D$
defines a line bundle $L_D \subset \Lambda^{2,+} (M_4)$, where $\text{vol}_D$ is the induced volume form on $D$.
The total space of this line bundle defines an associative submanifold,
which meets the coassociative 4-manifold $M_4$ along $D$.
Indeed, by examining\footnote{Locally,
\be
\Phi \; = \;
e^{567}
- (e^{12} + e^{34}) \wedge e^5
- (e^{13} + e^{42}) \wedge e^6
- (e^{14} + e^{23}) \wedge e^7,
\ee
\be
* \Phi \; = \;
e^{1234}
+ (e^{12} + e^{34}) \wedge e^{67}
+ (e^{13} + e^{42}) \wedge e^{75}
+ (e^{14} + e^{23}) \wedge e^{56}.
\ee
Associative 3-manifolds are defined by the condition that $\Phi$ restricts to a volume form and
coassociative 4-manifolds are defined by a similar condition with respect to $* \Phi$.
Equivalently, an associative submanifold of $X$ can be characterized by vanishing of the restriction
of the $TX$-valued 3-form $\chi \in \Omega^3 (X,TX)$ defined by the identity \cite{MR666108}:
\be
\langle \chi (u,v,w) ,z \rangle  \; = \; * \Phi (u,v,w,z).
\ee}
the associative 3-form $\Phi$ and the coassociative 4-form $* \Phi$,
it is easy to see that associative and coassociative submanifolds can meet only over
subsets of dimension 0 or 2.

Now, it is easy to see that M2-branes supported on this associative submanifold
indeed preserve one supercharge $\bar Q_+$.
In the setup \eqref{Msetup}, the M-theory reduced on $\Lambda^{2,+} (M_4)$ preserves, in the remaining $\R^4$, the 4d $\CN=1$ supersymmetry parametrized by a spinor $\epsilon$. The fivebrane supported on a coassociative cycle
in a $G_2$ manifold $\Lambda^{2,+} (M_4)$ preserves half of it, namely:
\be
\gamma_{01} \epsilon = \epsilon,
\label{susym5}
\ee
where $x^{0,1}$ are the directions along the non-compact part of the M5-brane that we call $\Sigma$.
This is precisely 2d $\CN=(0,2)$ supersymmetry of the theory $T[M_4]$ on $\Sigma$.
Introducing an extra M2-brane supported on an associative 3-manifold breaks supersymmetry further.
The corresponding condition (where $\gamma_5$ is the 4d chirality gamma-matrix),
\be
\gamma_5 \epsilon = \epsilon,
\label{susym2}
\ee
leaves one Weyl spinor in 4d, (on which the full 4d Lorentz algebra is still represented,)
and then \eqref{susym5} leaves one of its components which has the required spin.\\

Now, once we covered the ``charge'' $\lambda \in \Gamma$ and supersymmetry of
the M2-branes ending on fivebranes, our next task is to discuss anomalies of
the 2d degrees of freedom localized on the membrane boundary $D \subset M_4$.
For applications to Seiberg-Witten invariants, we are mostly interested in
the setup \eqref{Msetup} with a single fivebrane ({\it i.e.}, $N=1$) and
a single M2-brane ending on it, as illustrated in Figure~\ref{fig:M2M5}.
Let $(x^0,x^1,x^2,x^3,x^4,x^5)$ be the coordinates along the M5-brane world-volume
and $(x^4,x^5,x^6)$ to be the coordinates along the M2-brane world-volume:
\medskip
\be
\begin{tabular}{l || c|c|c|c|c|c|c|c|c|c|c}
 & \multicolumn{4}{c}{$\overbrace{\rule{2.0cm}{0pt}}^{T}$} & \multicolumn{2}{c}{$\overbrace{\rule{1.0cm}{0pt}}^{D}$} & \multicolumn{5}{r}{$\overbrace{\rule{2.3cm}{0pt}}^{N}$} \\
Brane & 0 & 1 & 2 & 3 & 4 & 5 & 6 & 7 & 8 & 9 & 10 \\ \hline\hline
$M5$  & x & x & x & x & x & x &   &   &   &   &   \\
$M2$  &   &   &   &   & x & x & x &   &   &   &   \\
& \multicolumn{2}{c}{$\underbrace{\rule{1.0cm}{0pt}}_{\Sigma}$} & \multicolumn{4}{c}{$\underbrace{\rule{2.0cm}{0pt}}_{M_4}$} & \multicolumn{3}{c}{$\underbrace{\rule{1.5cm}{0pt}}_{\Lambda^{2,+}}$} & \multicolumn{2}{c}{$\underbrace{\rule{1.0cm}{0pt}}_{R}$}
\end{tabular}
\label{M2M5}
\ee
Before we implement partial topological twists along $\Sigma$ and $M_4$ by embedding
this brane system in a curved background, we can start with branes in flat space.
Such configuration has a symmetry group $SO(2)_D \times SO(4)_T \times SO(4)_N$ where,
following \cite{Berman:2004ew}, we denote:
\begin{eqnarray}
SO(2)_{D} & \; := \; & SO(2)_{45}, \nonumber \\
SO(4)_T & \; := \; & SO(4)_{0123}, \label{2dN44} \\
SO(4)_N & \; := \; & SO(4)_{789\, 10}. \nonumber
\end{eqnarray}
After the topological twist, $N$ and $T$ are replaced by non-trivial $SO(4)$ bundles.
For example, in our system $T = T\Sigma \oplus N_D$, where $N_D$ is the normal bundle to $D$ in $M_4$.

In general, the anomaly polynomial of the self-dual string with world-sheet $D \subset M_4$,
{\it i.e.}, M2-brane ending on a fivebrane, is given by \cite{Brax:1997ht,Berman:2004ew} (see also \cite{Haghighat:2013gba}):
\be
A_{M2} \; = \; e (N) - e (T),
\ee
where $e(T)$ is the Euler class of the $SO(4)_T$ bundle $T$, and similarly for the ``normal bundle'' $N$.
For simplicity, let us assume that $M_4$ is a complex surface.
Then, using the information about the topological twist along the $M_4$ in our system we can write:
\be
A_{M2} \; = \; 2 c_1 (TM_4) c_1 (R) - c_1 (N_D) c_1 (T\Sigma),
\ee
where $R$ is the $U(1)_R$ symmetry bundle and $T\Sigma$ is the $U(1)_q$ line bundle.
In identifying $c_1 (R)$ with the first Chern class of the $SO(2)_{9\, 10}$ bundle,
we introduced a factor of 2 in order to match the standard normalization of R-charges.
Integrating over $D$, we conclude that local BPS operators on $\Sigma$ constructed from M2-branes
ending on a fivebrane carry the R-charge:
\be
R (D) \; = \; 2 \int_D c_1 (TM_4),
\label{RforD}
\ee
and 2d spin:
\be
\deg_q (D) \; = \; - \int_D c_1 (N_D) = \chi (D) - \int_D c_1 (TM_4),
\label{qforD}
\ee
where we used $c_1 (TD) + c_1 (N_D) = c_1 (TM_4)$.
Note, since $\deg_q = h - \bar h$ and BPS operators in 2d $\CN=(0,2)$ theory have $\bar h = \frac{R}{2}$,
it follows that M2-branes ending on fivebranes along surface $D$ have $\bar h = \int_D c_1 (TM_4)$
and
\be
h \; = \; \chi (D).
\ee
Equivalently, this expression gives the value of $h - \bar h$ in the half-twisted model on $\Sigma$.

As a consistency check of this anomaly calculation,
we now consider ``surface operators'' (or ``foams'') in application of this formalism to knot homology. In particular,
using this approach, we make a concrete prediction for $q$-degree of foams colored by symmetric representations of $sl(2)$
which, to the best of our knowledge, has not appeared in math or physics literature so far.

\subsubsection{Foams and knot cobordisms}

Although somewhat outside our main line of development, it is instructive to consider
the configuration \eqref{Msetup} with multiple fivebranes ($N>1$).
When considered on a 4-manifold of the form $M_4 = \R \times M_3$, it provides a physical realization of
the Khovanov-Rozansky $sl(N)$ homology and its generalizations (see, {\it e.g.}, \cite{Gukov:2017kmk} and references therein).
For the purposes of the following discussion we can simply take $M_4 = \R^4$.

In application to knot homologies, the type of the surface operator or M2-brane boundary determines
the ``color'' of the foam $D$, which can be either a knot cobordism or a closed surface.
In part to establish the dictionary, let us start with a relatively well understood
{\it uncolored} version and then make predictions for {\it colored} foams and knot cobordisms.\footnote{The terminology is
such that ``uncolored'' refers to knots and their cobordisms colored by the fundamental representation, whereas invariants
associated with more general representations are called ``colored.''}
In the case of a fundamental $N$-dimensional representation of $sl(N)$, the 2d TQFT (Frobenius algebra)
that lives on $D$ is a B-model based on the Landau-Ginzburg potential $W = x^{N+1}$ \cite{KR1}.
In particular, a pair-of-pants cobordism $D = {\raisebox{-.4cm}{\includegraphics[width=1.0cm]{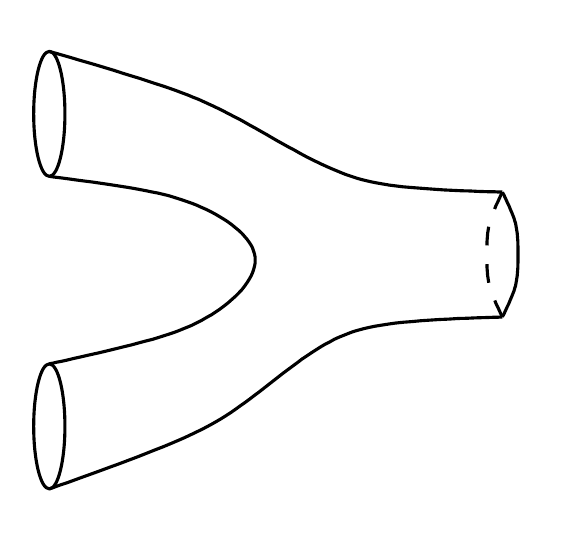}}\,}$
defines a product on the cohomology of the unknot,
\be
\CH ({\unknot}) \; = \; \C [x] / \langle x^N \rangle.
\ee
Note, this is the {\it classical} cohomology of $\cp^{N-1}$.
The chiral ring of the 2d TQFT on the surface operator $D$ is generated by the ``observable'' $x$.
In the convention $\deg_q (x) = 2$, a genus-$g$ surface $D$ has $\deg_q (D) = 2 (N-1) (g-1)$,
so that the closed surfaces which have non-zero evaluation are either of genus-0 with $N-1$ observables
or genus-1 with no extra insertions \cite{MR2113903,MR2174270,MR2213759,MR2171235,Rasmussen,MR3100886}:
\be
{\,\raisebox{-.8cm}{\includegraphics[width=1.8cm]{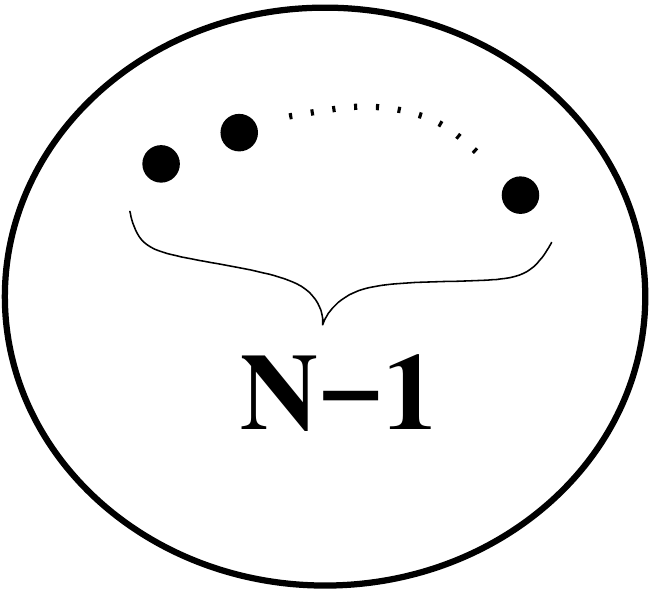}}\,} \; = \; 1
\qquad~~~,~~~\qquad
{\,\raisebox{-.4cm}{\includegraphics[width=1.8cm]{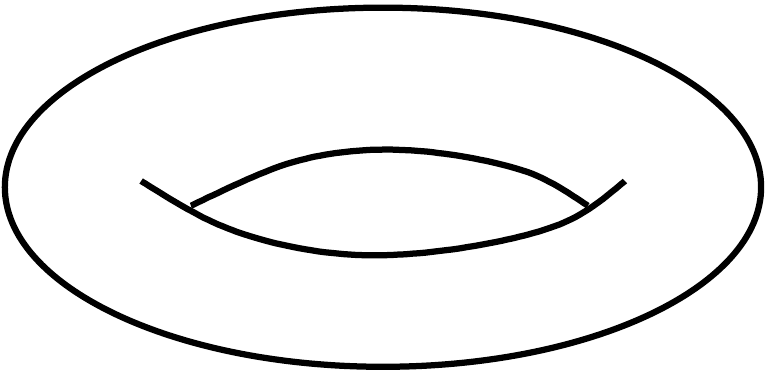}}\,} \; = \; N.
\ee
For a generalization to foams colored by antisymmetric representations of $sl(N)$ see, {\it e.g.}, \cite{MR2491657,MR3545951}.

This structure is in perfect agreement with anomalies of M2-branes ending on fivebranes.
Consider, for example, the case of $N=2$, just as in the ordinary Khovanov homology.
This means we have to study M2-brane boundaries / self-dual strings in $A_1$ fivebrane theory.
Using the anomaly polynomial of such strings (with multiplicity $r$) \cite{Shimizu:2016lbw}:
\be
A_{M2} \; = \; r^2 e (T) - r e(N),
\ee
we can repeat the steps that led to \eqref{RforD} and \eqref{qforD} and conclude that
multiplicity-$r$ closed surface in $SU(2)$ theory has zero R-charge (= homological grading) and
\be
\deg_q (D) \; = \; r^2 \chi (D).
\ee

\subsection{Anomalies: intersecting M5-branes and basic classes}
\label{sec:impurity}

Now we are ready to introduce the second key ingredient needed for realizing SW invariants in 2d theory $T[M_4]$,
namely a vertex operator localized at a point on $\Sigma$ which, if the order of compactification was reversed, would
produce a charged hypermultiplet in 4d $\CN=2$ theory on $M_4$.

Indeed, if we start with 6d $(2,0)$ tensor multiplet on a single fivebrane and first reduce it on $\Sigma$,
we obtain precisely the construction \cite{Witten:1997sc} of 4d $\CN=2$ low-energy effective physics with Seiberg-Witten curve $\Sigma$.
{}From the perspective of this low-energy 4d $\CN=2$ gauge theory, a further topological twist along the 4-manifold $M_4$
(that corresponds to embedding it as a coassociative 4-cycle in $G_2$ geometry) can be identified with
the standard topological twist which leads to Donaldson or Seiberg-Witten invariants (see, {\it e.g.}, \cite[sec.5]{Gukov:2016gkn}).
\be
\begin{array}{ccccc}
\; & \; & \text{6d $(2,0)$ theory} & \; & \; \\
\; & \; & \text{on $\Sigma \times M_4$} & \; & \; \\
\; & \swarrow & \; & \searrow & \; \\
\text{2d half-twisted} & \; & = & \; & \text{4d $\CN=2$ theory} \\
\text{model on $\Sigma$} & \; & \; & \; & \text{twisted on $M_4$}
\end{array}
\label{6d4d2d}
\ee

After reduction on $\Sigma$, the resulting 4d $\CN=2$ theory has R-symmetry $U(1)_R \times SU(2)_R$.
Let $R$ and $E$ denote the corresponding $U(1)_R$ and $SU(2)_R$ bundles.
Then, much like its simpler version \eqref{cRRR} that we encountered earlier, the anomaly polynomial of
a general 4d $\CN=2$ superconformal theory is a degree-6 polynomial in characteristic classes of $R$ and $E$,
whose coefficients are determined by the conformal central charges, $a$ and $c$, see, {\it e.g.}, \cite{Kuzenko:1999pi,Shapere:2008zf}:
\be
A_{4d} \; = \; (n_v - n_h) \left(  \frac{c_1 (R)}{12} p_1 (TM_4) - \frac{c_1 (R)^3}{3} \right)
- n_v c_1 (R) c_2 (E) + \frac{k}{2} c_1 (R) \text{ch}_2 (F).
\label{A4d}
\ee
Here, $F$ is the curvature of a $G$-bundle over $M_4$,
$n_v = 8a - 4c$ and $n_h = 20c - 16a$ denote the effective number of vector and hypermultiplets, respectively,\footnote{In terms of
the conformal anomaly coefficients, $c = \frac{1}{6} n_v + \frac{1}{12} n_h$ and $a = \frac{5}{24} n_v + \frac{1}{24} n_h$, the same
expression reads:
$$
A_{4d} \; = \; (a-c) \left( 2 c_1 (R) p_1 (TM_4) - 8 c_1 (R)^3 \right)
- (8a-4c) c_1 (R) c_2 (E) + \frac{k}{2} c_1 (R) \text{ch}_2 (F).
$$}
and the central charge $k$ of the symmetry group $G$ is defined via OPE of two $G$-symmetry currents:
\be
J_{\mu}^a (x) J_{\nu}^b (0) \; = \; \frac{3 k}{4 \pi^4} \delta^{ab} \frac{x^2 g_{\mu\nu} - 2 x_{\mu} x_{\nu}}{x^8} + \ldots
\ee

Since the R-symmetry $U(1)_R$ of the 4d $\CN=2$ theory on $M_4$ is precisely the R-symmetry (under the same name)
of 2d $\CN=(0,2)$ theory $T[M_4]$ on $\Sigma$, we can determine $U(1)_R$ anomaly of the latter
by integrating \eqref{A4d} over $M_4$, with $G=U(1)$ and $E = \Lambda^{2,+} (M_4)$.
In particular, if we choose $\Sigma$ such that 4d $\CN=2$ theory is a theory of a single vector multiplet, {\it i.e.}, $n_v = 1$ and $n_h = 0$,
we find precisely the $U(1)_R$ anomaly \eqref{RTM4b} of the theory $T[M_4]$, computed earlier from the 2d perspective:
\be
\Delta R_{T[M_4]} \; = \; - \frac{\chi + \sigma}{2},
\label{RTM4b}
\ee
where we used
\be
\frac{1}{12} \int_{M_4} p_1 (TM_4) \; = \; \frac{\sigma}{4}
\qquad , \qquad
\int_{M_4} c_2 (E) \; = \; \frac{2 \chi + 3 \sigma}{4},
\ee
when $E = \Lambda^{2,+} (M_4)$.

Similarly, an impurity on $\Sigma$ that adds a charged hypermultiplet to 4d $\CN=2$ theory on $M_4$
has $(n_v,n_h) = (0,1)$ and, according to \eqref{A4d}, changes the $U(1)_R$ anomaly by:
\be
\Delta R_{\text{impurity}} \; = \; \frac{\lambda^2 - \sigma}{4}.
\label{Rimpurity}
\ee
In what follows, we denote this kind of impurity by $\CS_+$ to distinguish it from the impurity $\CS$ defined by the intersection of two M5-branes.
Combining this with \eqref{RTM4b}, we learn that a Riemann surface $\Sigma$ that, via reduction \eqref{6d4d2d},
engineers $N_f$ topologically twisted hypermultiplets charged under the $U(1)$ vector multiplet on $M_4$, has $U(1)_R$ anomaly:
\be
\Delta R_{N_f} \; = \; \frac{N_f (\lambda^2 - \sigma) - 2 (\chi + \sigma)}{4}.
\label{RNf}
\ee
In particular, in the special case $N_f = 1$ we recover the well known expression for the virtual
dimension of the moduli space of solutions to the Seiberg-Witten equations \cite{Witten:1994cg,Labastida:2005zz}:
\be
{\rm VirDim}\, \CM_{\text{SW}} (\lambda) \; = \; \frac{\lambda^2 -2\chi - 3\sigma}{4} \,,
\label{SWdim}
\ee
whereas the general formula \eqref{RNf} gives a (less familiar) expression \cite{MR1392667} for the dimension
of the moduli space of generalized multi-monopole solutions that will be the subject of section~\ref{sec:NfSW}.
Note, in gauge theory on a 4-manifold $M_4$, the expression \eqref{SWdim} is (minus) the index of the deformation complex
for the Seiberg-Witten monopole equations:
\be
0 \longrightarrow \Omega^0 (M_4) \longrightarrow
\Omega^1 (M_4) \oplus \Gamma (M_4, S^+ \otimes L^{1/2}) \longrightarrow
\Omega^{2,+} (M_4) \oplus \Gamma (M_4, S^- \otimes L^{1/2}) \longrightarrow 0
\label{SWdefcomplex}
\ee
It splits into a complex for the Dirac operator, with the familiar index $\frac{\lambda^2 - \sigma}{4}$,
and a simple complex for the anti-self-duality equation \eqref{ASDeq}, with index $- b_0 + b_1 - b_2^+ = - \frac{1}{2} (\chi + \sigma)$.
Note, these two contributions are precisely the contributions to $U(1)_R$ anomaly from the impurity \eqref{Rimpurity}
and from 2d theory \eqref{RTM4b}, respectively. Adding them together we obtain \eqref{SWdim}.\\

As a byproduct of deriving \eqref{RNf} we also obtain the anomaly polynomial of two intersecting fivebranes:
\be
A_{\text{M5} \, \cap \, \text{M5}'} \; = \; \frac{1}{2} c_1(R) c_2(E) + \frac{1}{2} c_1(T\Sigma) c_2(E).
\label{AM5M5}
\ee
Indeed, the half-BPS impurity operator $\CS$ is realized by a codimension-2 defect in 6d $(2,0)$ theory or,
equivalently, by intersecting M5-branes in the M-theory setup \eqref{Msetup}, {\it cf.} \eqref{M2M5}:
\begin{equation}
\begin{tabular}{l || c|c|c|c|c|c|c|c|c|c|c}
 & \multicolumn{2}{c}{$\overbrace{\rule{1.0cm}{0pt}}^{\Sigma}$} & \multicolumn{4}{c}{$\overbrace{\rule{2.0cm}{0pt}}^{M_4}$} & \multicolumn{3}{c}{$\overbrace{\rule{1.5cm}{0pt}}^{E}$} & \multicolumn{2}{c}{$\overbrace{\rule{1.0cm}{0pt}}^{R}$} \\
Brane & 0 & 1 & 2 & 3 & 4 & 5 & 6 & 7 & 8 & 9 & 10 \\ \hline\hline
$M5$  & x & x & x & x & x & x &   &   &   &   &   \\
$M5'$ &  & & x & x & x & x &   &   &   & x & x
\end{tabular}
\label{M5impurity}
\end{equation}
By using \eqref{6d4d2d} and analyzing how the Kaluza-Klein spectrum of 4d $\CN=2$ theory on $M_4$
depends on the topology of $\Sigma$, it is easy to see that a puncture on $\Sigma$ effectively
carries $n_v = - \frac{1}{2}$ and $n_h = - \frac{1}{2}$. Then, from \eqref{A4d}
and the symmetry between two fivebranes in \eqref{M5impurity} we obtain the anomaly polynomial \eqref{AM5M5}.

Note that, for $E = \Lambda^{2,+} (M_4)$, integrating $c_2 (E)$ over $M_4$ in the first term of \eqref{AM5M5}
gives precisely the ``constant'' term $2\chi+3\sigma$ in \eqref{SWdim}.
Alternatively, from the second term in \eqref{AM5M5} we can find the 2d spin of the impurity operator $\CS$
defined as the intersection of two M5-branes. Comparing it with the spin of $V_{\lambda}$ we obtain
the condition $\lambda^2 = 2\chi+3\sigma$ on basic classes in Seiberg-Witten theory.

In this section, we described the R-charges of the impurity operators $\CS$ and $\CS_+$ that will play an important role in the rest of this paper. While it would be desirable to give a more detailed 2d characterization of these vertex operators (and their various generalizations, e.g. to Argyres-Douglas theories), what we have so far suffices to describe structural properties of their correlation functions. By matching the latter with a known answer in one simple case, say for $G = U(1)$ and $N_f = 1$, then determines more general correlators of $\CS$ and $\CS_+$, which in turn lead to new concrete predictions for 4-manifold invariants. This will be our strategy in the next subsection.


\subsection{Gauge theoretic invariants of 4-manifolds as 2d correlators}
\label{sec:structure}

We are ready to make a proposal for the structure of 4-manifold invariants, old and new,
by putting together all of the above ingredients:
half-twisted theory $T[M_4]$ summarized in section~\ref{sec:fluxes},
winding-momentum vertex operators (section~\ref{sec:M2branes}) and BPS impurities (section~\ref{sec:impurity}).

As in \cite{Gukov:2016gkn}, it will be useful to consider the following generating function of SW invariants:
\begin{equation}
 \hat{\text{SW}}(t) \; = \; \sum_{\lambda\in H^2(M_4,\Z)} \text{SW}(\lambda)\,t^{\lambda},
\end{equation}
where the expansion parameter $t$ takes values in  $\text{Hom}(H^2(M_4,\Z),\C^*)$.
One can use it to define the following local BPS operator in $T[M_4,U(1)]$:
\begin{equation}
 \CS(z) \; := \; \hat{\text{SW}} (e^{i X(z)})\equiv \sum_\lambda \text{SW}(\lambda) \,V_\lambda(z),
\label{SviaV}
\end{equation}
where
\begin{multline}
 X(z)\equiv (X_L(z),X_R(z)+i\sigma(z))\in H^{2,-}(M_4)\oplus (H^{2,+}(M_4)\otimes \C) \\
 \subset H^2(M_4,\C) \stackrel{Q}{\cong}  H_2(M_4,\C),
\label{Big-X-field}
\end{multline}
so that $e^{i X(z)}\in \exp i H_2(M_4,\C) \subset  \text{Hom}(H^2(M_4,\Z),\C^*)$.
Since we are interested in half-twisted correlators and the chiral ring of $T[M_4]$,
we always consider $\CS (z)$ and other BPS vertex operators modulo $\bar Q_+$-exact terms.
With this proviso, the insertion of the local operator $\CS(z)$ at a point in the 2d space-time $\Sigma$ of $T[M_4]$ corresponds
to the insertion of the fivebrane that also wraps $M_4$ but has two directions orthogonal to $\Sigma$,
as in the setup \eqref{M5impurity}:
\be
\CS (z) \quad \sim~~
{\,\raisebox{-0.6cm}{\includegraphics[width=1.4cm]{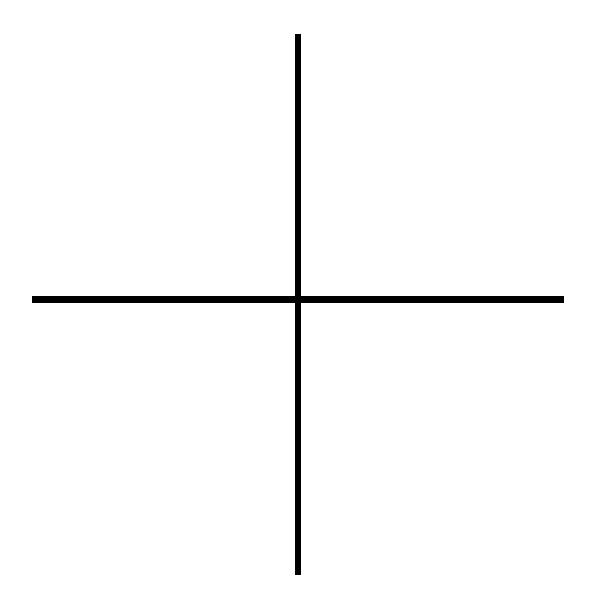}}\,},
\label{SX}
\ee
so that (3.58) can be interpreted as a ``decomposition'' of a codimension-2 defect in terms of codimension-4 defects in the fivebrane theory on $M_4$, again, up to $\bar Q_+$-exact terms.
This can be justified as follows.

\subsubsection{Seiberg-Witten invariants from half-twisted correlators}

\begin{figure}[ht]
\centering
\includegraphics[scale=1.1]{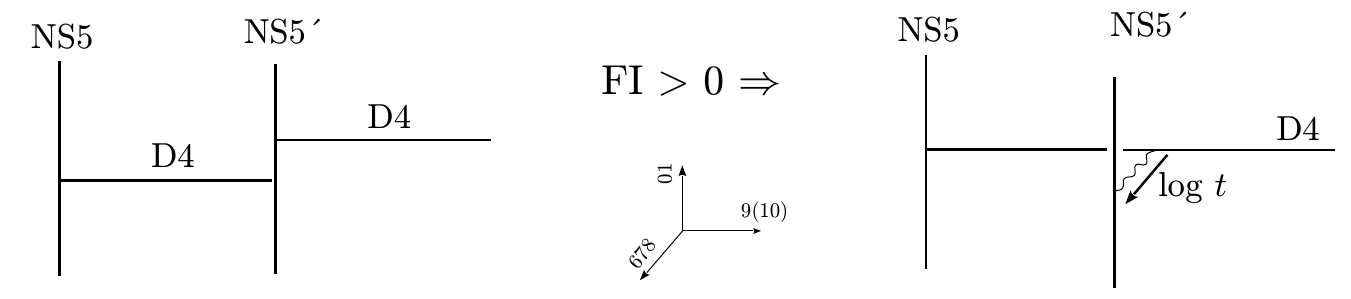}
\caption{Brane construction {\it a la} \cite{Witten:1997sc}
of an ordinary Seiberg-Witten theory (with $N_f = 1$) and its Higgs phase with non-zero FI parameter.}
\label{fig:branes-U1-1hyper}
\end{figure}

Consider the standard brane realization of the 4d $\CN=2$ gauge theory with gauge group $U(1)$ and
one charged hypermultiplet shown on Figure~\ref{fig:branes-U1-1hyper} (left panel).
In flat space, the $SU(2)_R$ triplet of the FI parameters corresponds
to the relative position of two NS5 branes in the directions 678.
After the topological twist along a 4-manifold $M_4$,
the directions 678 become the fiber directions of the $\Lambda^{2,+} (M_4)$ bundle.
And the FI parameter $\eta$ becomes an element of
$H^{2,+}(M_4) \subset H^2(M_4,\R) \stackrel{Q}{\cong} H_2(M_4,\R)$ (complexified). In the case of $M_4 = M_3 \times S^1$ this setup
was considered in \cite{Gukov:2016gkn,Mikhaylov:2015nsa}.
Setting $t=e^{i \eta}$, the partition function of the 4d SW theory then reads:
\begin{equation}
 \sum_{\lambda\in H^2(M_4,\Z)} \text{SW}(\lambda)\, e^{i \eta\cdot\lambda}\equiv\hat{\text{SW}}(t),
\label{FI-partition-function}
\end{equation}
where the sum represents the sum over all Spin$^c$ structures in the path integral.\footnote{For simplicity, we assume that $w_2=0$.}

Note, once we turn on the FI parameter, the relative position of the NS5 brane
in the direction $x^9$ becomes irrelevant and all supersymmetric configurations
that contribute to the partition function \eqref{FI-partition-function} are reprsented
by D2 branes stretched between NS5$'$ and D4 branes, as shown on the right side of Figure~\ref{fig:branes-U1-1hyper}.
In the Higgs phase of the 4d $\CN=2$ gauge theory, these D2 branes correspond to
supersymmetric vortex solutions \cite{Hanany:2003hp} supported on embedded surfaces inside $M_4$.
The mathematical counterpart of this statement is that Seiberg-Witten invariants of $M_4$
are equal to Gromov invariants that count (vortices localized on) embedded surfaces.
These two phases are related by a configuration with vanishing FI parameters,
in which supersymmetric contributions to the partition function \eqref{FI-partition-function}
are localized at the intersection of NS5$'$ and D4 branes.
In M-theory, it is lifted to the intersection of M5 and M5$'$ described in \eqref{M5impurity}.

The same partition function is given by a correlator of local operators in $T[M_4]$.
In the M-theory lift, the brane NS5$'$ is a fivebrane on $M_4 \times \Sigma$, where $\Sigma=\C$ spans directions 01.
Take $X(z)$ to be the collection of fields (\ref{Big-X-field}) of $T[M_4]$ living on $\Sigma$. Then, the components of $X(z)$ valued in $H^{2,+}(M_4)\otimes \C$ describe the local (complexified) position of the brane in the fiber directions of $\Lambda^{2,+} (M_4)$.
Namely, ``position'' refers to a choice of the harmonic section of $\Lambda^{2,+} (M_4)$.
This is same space where the FI parameter takes values.
Although we can have local fluctuations of the position of the NS5$'$ brane,
the brane is infinite and its position at infinity should be fixed (since the movement of the whole brane would require infinite energy).
As was discussed earlier, its position relative to other branes is given by the FI parameter $\eta$.
Similarly, we need to fix values of other fields at infinity of $\Sigma=\C$, in particular $X_L(\infty)=0$.
So, the partition function should be given by the following 2d correlator:
\begin{multline}
\langle \, \CS (z_1) \; \delta (X(\infty)-\eta) \, \rangle =\\
= \sum_\lambda \text{SW}(\lambda) \sum_\mu  \; \langle e^{i \lambda\cdot X(z_1)}  e^{- i \mu\cdot ( X(\infty)-\eta)}\rangle
= \sum_\lambda \text{SW}(\lambda)\, t^\lambda,
\end{multline}
where $z_1$ is the position of the brane intersection on $\Sigma$ and we used:
\begin{equation}
\langle \, e^{i \lambda\cdot X(z_1)} \, e^{- i \mu \cdot X(\infty)} \, \rangle \; = \; \delta(\mu-\lambda),
\end{equation}
up to normalization. This indeed reproduces (\ref{FI-partition-function}).
The insertion of $e^{- i \mu \cdot X(\infty)}$ in the correlation function can be interpreted
as the ``background charge'' at infinity.

Note, when the FI parameter is turned off (as on the left panel of Figure~\ref{fig:branes-U1-1hyper}),
there is a continuous Coulomb branch modulus that corresponds to the position of the D4 brane stretched between two NS5 brane.
When theory is put on a compact space, one has to integrate over it.
Turning on the FI parameter corresponds to localization of this integral
to the origin of the Coulomb branch, the point where the two D4 branes align.

\subsubsection{Multi-monopole invariants}

\begin{figure}[ht]
\centering
\includegraphics[scale=1.1]{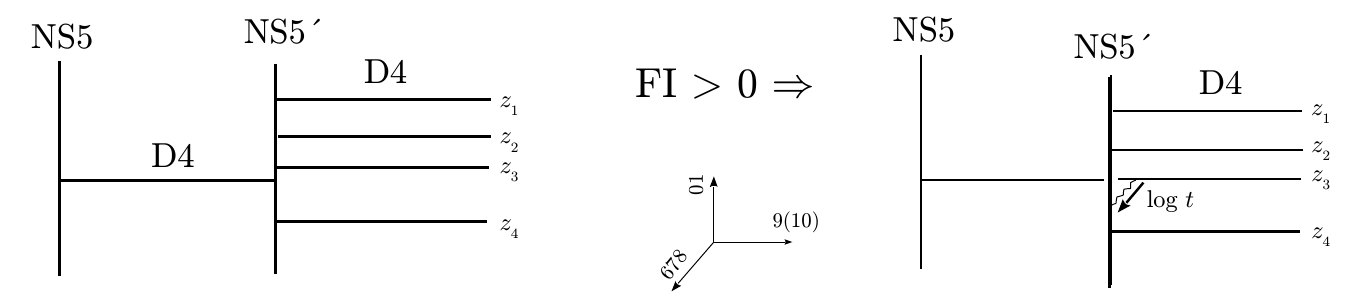}
\caption{Brane construction of $U(1)$ theory with $N_f$ hypermultiplets and its Higgs phase with non-zero FI parameter.}
\label{fig:branes-U1-Nhypers}
\end{figure}

Consider now $U(1)$ theory with $N_f$ hypermultiplets with non-zero masses $z_i$.
The corresponding brane construction is shown on Figure~\ref{fig:branes-U1-Nhypers}.
Turning on the FI parameter localizes the Coulomb branch ($u$-plane) integral
on one of $N_f$ possible brane configurations where the D4 brane stretched between the NS5 branes
aligns with one of the semi-infinite D4 branes with mass $z_i$.
In M-theory, the NS5$'$ brane is lifted again to an M5 brane on $M_4\times \Sigma$,
where $\Sigma=\C$ spans directions $(x^0,x^1)$.
The masses of hypermultiplets correspond to positions $z_j$ of semi-infinite D4 branes on $\Sigma$.
The  partition function of the fivebrane then has the following realization in terms of $T[M_4]$ on $\Sigma$:
\begin{equation}
\langle \, \CS (z_i) \, \prod_{j\neq i} \CS_+(z_j) \, \delta (X(\infty)-\eta) \rangle
\; = \; \sum_{\mu\in H^2(M_4,\Z)} \langle \CS (z_i) \prod_{j\neq i} \CS_+(z_j) \; e^{- i \mu\cdot X(\infty)} \rangle \, t^{\mu},
\label{multi-corr}
\end{equation}
where $\CS_+(z)$ is the operator that corresponds to adding a semi-infinite D4 brane located at $z$.
Schematically,
\be
\CS_+ (z) \quad \sim~
{\,\raisebox{-0.5cm}{\includegraphics[width=1.0cm]{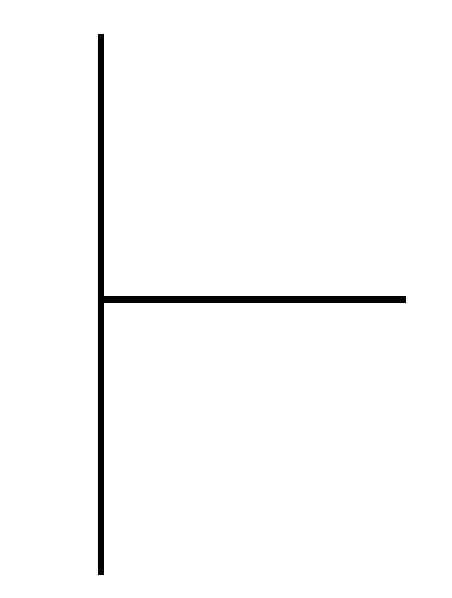}}\,}.
\label{ST}
\ee
In M-theroy lift, it corresponds to creating a semi-infinite neck (cusp) on $\Sigma$.
Inserting such operators at points $z_j,\,j\neq i$  is equivalent to deforming the geometry of $\Sigma$ from
$\C\cong \{y=\text{const}\} \subset \C \times \C_*$ to:
\be
\{ y=\text{const}\prod_{j\neq i}(z-z_j) \} \; \subset \; \C\times\C^*,
\ee
where $z=x_0+ix_1$ and $y=e^{x_9+ix_{10}}$ are the coordinates parametrizing $\C$ and $\C^*$, respectively.

The total partition function of the 4d theory on $M_4$ should be given by the sum over $N_f$ such configurations:
\begin{equation}
\sum_{i=1}^{N_f} \, \langle \, \CS (z_i) \, \prod_{j\neq i} \CS_+(z_j) \, \delta (X(\infty)-\eta) \, \rangle.
\end{equation}
In particular, using \eqref{multi-corr} we obtain the structure of the multi-monopole
invariant of $M_4$ with a given Spin$^c$ structure $\lambda$:
\be
\sum_{i=1}^{N_f} \, \langle \CS (z_i) \prod_{j\neq i} \CS_+ (z_j) \; e^{- i \lambda \cdot X(\infty)} \rangle.
\label{multi-prediction}
\ee
{}From the vantage point of the topological theory on $M_4$, the parameters $z_i$ can be understood
as the equivariant parameters for the maximal torus of $SU(N_f)$ symmetry.
This will be the subject of section~\ref{sec:NfSW}, where we also present a verification
of the prediction \eqref{multi-prediction} by direct calculation in gauge theory.

\begin{figure}[ht]
\centering
\includegraphics[width=3.0in]{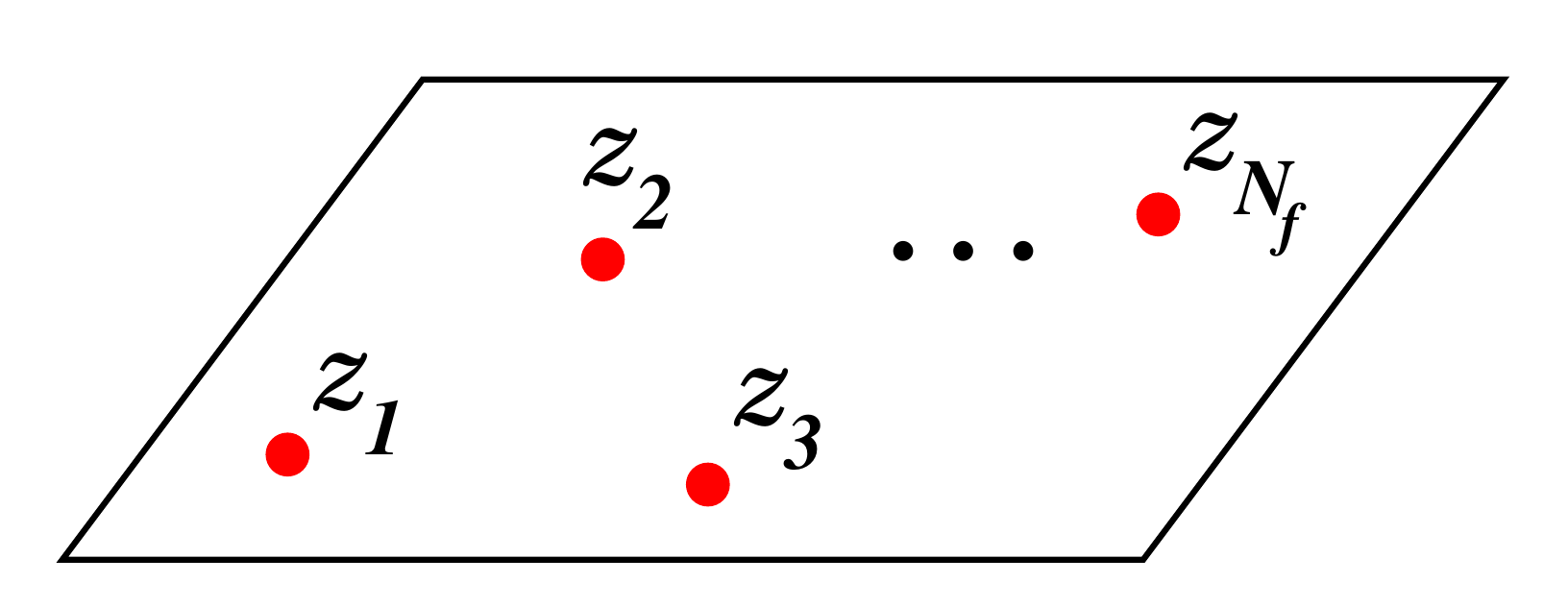}
\caption{Impurity vertex operators at points $z_i$, $i=1, \ldots, N_f$.}
\label{fig:4mfldplane}
\end{figure}

Note, the 2d correlator \eqref{multi-prediction} that computes the desired topological invariant of $M_4$
has another representation, which is completely symmetric in all $N_f$ impurity operators and corresponds
to the Coulomb phase / brane configuration on the left panel of Figure~\ref{fig:branes-U1-Nhypers}:
\be
\langle \, \oint \frac{dz}{2\pi i} J(z) \, \CS_+ (z_1) \ldots \CS_+ (z_{N_f}) \; e^{- i \lambda \cdot X(\infty)} \rangle.
\ee
Here, the $z$-integral is performed along a large contour that encloses all of the points $z_i$, $i=1, \ldots, N_f$
and
\be
J(z) \CS_+ (w) \; \sim \; \frac{1}{z-w} \CS (w) + \text{regular}.
\label{JSSrel}
\ee
In other words, $\CS (w) = [\CQ , \CS_+ (w)]$, where the charge $\CQ = \frac{1}{2\pi i} \oint J(z) dz$.
Deforming the integration contour into $N_f$ smaller contours, each of which encircles only one of
the vertex operators $\CS_+ (z_i)$, $i=1, \ldots, N_f$, we obtain \eqref{multi-prediction}:
\be
\langle \, \oint \frac{dz}{2\pi i} J(z) \, \CS_+ (z_1) \ldots \CS_+ (z_{N_f}) \; e^{- i \lambda \cdot X(\infty)} \rangle \;
= \; \sum_{i=1}^{N_f} \, \langle \CS (z_i) \prod_{j\neq i} \CS_+ (z_j) \; e^{- i \lambda \cdot X(\infty)} \rangle.
\label{mm-prediction}
\ee
It would be interesting to give a more intrinsic 2d characterization of the impurity vertex operators $\CS$ and $\CS_+$,
along with the explanation of the relation \eqref{JSSrel} between them. We hope to return to this problem in the future work.

\subsubsection{Multiple $U(1)$ groups and multiple monopoles}

\begin{figure}[ht]
\centering
\includegraphics[scale=1.1]{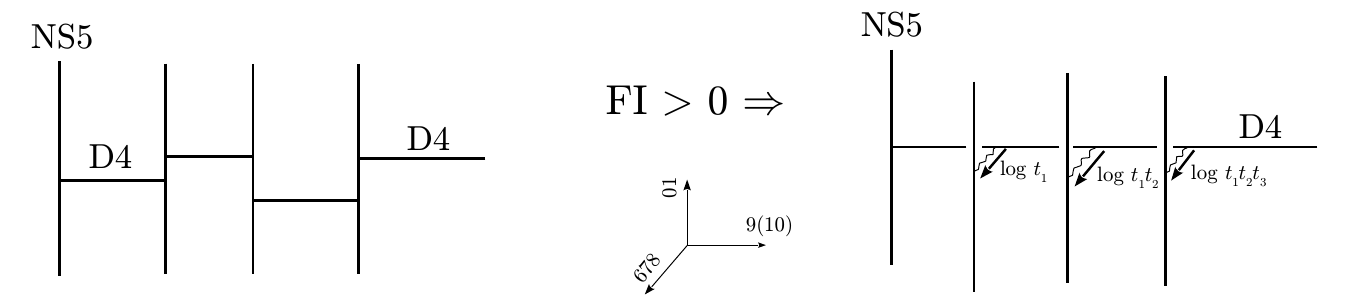}
\caption{Brane construction of a $U(1)^n$ quiver gauge theory with $n=3$ and its Higgs phase with non-zero FI parameters.}
\label{fig:branes-U1-quiver}
\end{figure}

Consider another example, a $U(1)^n$ quiver gauge theory, which is realized by a brane configuration
with $n+1$ parallel NS5 branes shown in Figure~\ref{fig:branes-U1-quiver}.
The matter consists of $n-1$ bifundamental hypermultiplets and one fundamental hyper.
Denote the FI parameters for $U(1)$ gauge factors by $\eta_j$ and their exponentiated versions by $t_j=e^{i \eta_j}$.
They correspond to relative positions of adjacent NS5 branes in the directions 678.

As in our previous examples, turning on the FI parametes for all of the $U(1)$ gauge factors
gives us $T[M_4]$ on $n$ disconnected copies of $\C$, each with one insertion of the impurity operator $\CS (z)$.
Therefore, the partition function should be given by the product of $n$ copies of
the partition function (\ref{FI-partition-function}), with FI parameters
given by the positions of NS5 branes relative to the semi-infinite D4 brane:
\begin{equation}
  \hat{\text{SW}}(t_1)\,\hat{\text{SW}}(t_1t_2)\ldots \hat{\text{SW}}(t_1t_2\ldots t_n).
\end{equation}
This prediction is consistent with the fact that the quiver theory described here
is dual to the tensor product of $n$ copies of $U(1)$ gauge theory with $N_f = 1$.
The equivalence between theories and the corresponding identification of the FI parameters
can be seen by integrating out $U(1)$ gauge multiplets one-by-one starting from the left end of the quiver.

Hopefully, by now it is clear how to use the basic rules \eqref{SX} and \eqref{ST}
to convert gauge theoretic invariants of 4-manifolds to correlators of impurity operators $\CS (z)$ and $\CS_+ (z)$.
We present one more illustration in section~\ref{sec:non-abelian}, with a proposal for the structural properties
of non-abelian gauge theoretic invariants, and now focus on verification of the abelian ones.


\section{Equivariant multi-monopole invariants}
\label{sec:NfSW}

In this section we explore 4-manifold invariants which, on the one hand,
share some tractability of the Seiberg-Witten invariants and, on the other hand,
provide the simplest context in which one can concretely see the connection with 2d correlators.
We then compute the new equivariant invariants for many 4-manifolds and confirm that
they indeed have the expected structure \eqref{mm-prediction}.

\subsection{A cure for non-compactness}

Very much like single-monopole Seiberg-Witten equations, the multi-monopole equations on $M_4$ are formulated in terms of a $U(1)$ connection $A$ and $N_f$ commuting left-handed ({\it i.e.}, positive chirality) spinors $\Psi_i,\, i=1,\dots,N_f$. As usual, one can combine the ${\rm Spin}$-connection and ($\frac12$ times) the $U(1)$ connection into a single ${\rm Spin}^c$ connection and thus see that only a ${\rm Spin}^c$ structure is required, which exists on an arbitrary 4-manifold $M_4$. Of course, in all local expressions we are free to separate ${\rm Spin}^c$ connection into a ${\rm Spin}$ and a $U(1)$ part, remembering that only their sum makes sense globally, unless $M_4$ is a ${\rm Spin}$ manifold.

The ${\rm Spin}^c$ structure on $M_4$ is given by a pair of $U(2)$ bundles $W^\pm$ with a Clifford multiplication map $c: \Omega^1(M_4)\otimes \Gamma(M_4,W^\pm) \to \Gamma(M_4, W^\mp)$. The determinant bundle of $W^+$ is denoted $L$ and $A$ is a connection on $L$. Also, we denote $c_1(L)=\lambda \in H^2(M_4)$ and abbreviate $Q^{-1}(\lambda,\lambda)\equiv \lambda^2$. (This is the same $\lambda$ as in the previous two sections.) With $\Psi_i, \dots, \Psi_{N_f}\in \Gamma(M_4, W^+)$, the $N_f$-monopole Seiberg-Witten equations are written as:
\begin{eqnarray}
F_A^+ & = & i\sum_{i=1}^{N_f} (\Psi_i \bar \Psi_i)_0, \label{NfSWeq} \\
D \!\!\!\! \slash \, \Psi_i & = & 0, \qquad i = 1, \ldots, N_f, \nonumber
\end{eqnarray}
where the notation $(\Psi \bar \Psi)_0$ means the traceless part of $\Psi\otimes \bar\Psi$, that is $\Psi\otimes \bar\Psi - \frac12 (\overline{\Psi}\Psi){\rm id}$, and $(\overline{\Psi}\Psi)$ is an inner product on spinors. With the spinor indices $\alpha,\beta=1,2$ made explicit, these objects are simply given by $(\overline{\Psi}\Psi)=\varepsilon^{\alpha\beta}\overline{\Psi}_\alpha\Psi_\beta$ and $[(\Psi\overline{\Psi})_0]_{\alpha\beta}=\Psi_{(\alpha} \bar \Psi_{\beta)}$.

We denote by $\CM_{N_f} (M_4; \lambda)$, or simply by $\CM_{N_f}$, the moduli space of solutions
to these equations modulo gauge transformations. By index theorem, this moduli space has virtual dimension equal to \eqref{RNf},
which we have already identified as the ghost number anomaly of $T[M_4]$ with $N_f$ copies of the impurity operator. We will also study the perturbed version of these equations, in which the first equation is replaced by:
\be
F^+_A + \eta = i\sum_{i=1}^{N_f} (\Psi_i \bar \Psi_i)_0,
\ee
where $\eta\in\CH^{2,+}(M_4)$ is a generic self-dual harmonic perturbation.
In such situations we will call the moduli space $\CM_{N_f}(\lambda, \eta)$, or again simply $\CM_{N_f}$ whenever it creates no confusion.

Although the system of equations \eqref{NfSWeq} is well motivated
in physics (see, {\it e.g.}, \cite{Witten:1994cg,Losev:1997tp,Labastida:2005zz} where closely related questions were studied),
its mathematical study presents a largely uncharted territory,
with only a few brave ventures in this direction \cite{MR1392667,haydysPhD,MR3432158}.
Part of the reason is that $\CM_{N_f}$ can be non-compact when $N_f > 1$,
and naively following the same steps as in the $N_f = 1$ case can lead to an ill-defined integral:
\be
\int_{\CM_{N_f}} c_1 (\CL)^{d/2},
\label{naiveNfSW}
\ee
where $\CL \to \CM_{N_f}$ is a universal line bundle over the moduli space (to be defined shortly).

\subsection*{Moduli space}

In this subsection we briefly review the key ingredients in the construction of moduli space. The $N_f$-monopole case parallels the 1-monopole theory in most aspects, except that now the moduli space can become non-compact. Non-compactness surely makes the definition of invariants more subtle. Also, in this non-compact space there are sequences of irreducible solutions that would converge to reducible solutions, had we included the latter in the moduli space, which is another potential subtlety.

Denote the space of $U(1)$ connections on $L$ by $\CA$. The configuration space of the multi-monopole problem is defined as:
\be
\CC = \CA \times \Gamma(M_4, W^+)^{N_f}.
\ee
The group of gauge transformations $\CG={\rm Hom}(M_4, U(1))$ acts on $\CC$, but this action is not free. So the most general moduli space of connections and spinors, the one given by:
\begin{equation}
\CM=\CC/\CG,
\end{equation}
is singular, with singularities at reducible solutions. However, since we work under the assumption that $b_2^+>1$, generic metrics do not admit abelian instantons ({\it i.e.}, solutions to $F^+_A=0$), except for trivial solutions. (Trivial solutions include $A=0$, $\Psi_i=0$ and, if $H_1(M_4)\ne 0$, flat connections.) Therefore, equations \eqref{NfSWeq} only have solutions with some $\Psi_i$ nonzero, and an additional locus of trivial solutions. Hence, it makes sense to consider a subspace:
\be
\CC^* = \{(A,\Psi_1,\dots,\Psi_{N_f})\in\CC\, |\, \exists i: \Psi_i\ne 0\},
\ee
on which $\CG$ acts freely. It gives rise to a smooth moduli space of spinors and connections:
\be
\CM^* = \CC^*/\CG.
\ee
The multi-monopole moduli space, with or without perturbation, is a subspace $\CM_{N_f}\subset \CM^*$ of this moduli space determined by the equations \eqref{NfSWeq} (or their perturbed version). For the perturbed problem, this $\CM_{N_f}$ is really all we have, because in the case of generic perturbation, there are no reducible solutions. For the unperturbed problem \eqref{NfSWeq}, in additions to $\CM_{N_f}$, there is a locus of reducible solutions that have all $\Psi_i=0$ and $A$ flat.

Under the standard technical assumptions (that various infinite-dimensional spaces we are working with can be completed in appropriate Sobolev norms), which are known to hold for the problem at hand \cite{MR1392667}, we can study the moduli spaces locally.  The linearization of the multi-monopole equations, with or without perturbation, is then described by the familiar deformation complex:
\begin{align}
0 \rightarrow \Omega^0(M_4)\, \stackrel{C}{\longrightarrow}\, \Omega^1(M_4)\oplus \Gamma(M_4,W^+)^{N_f} \,\stackrel{ds}{\longrightarrow}\, \Omega^{2,+}(M_4)\oplus \Gamma(M_4, W^-)^{N_f} \rightarrow 0,
\end{align}
where the first map $C(\chi) = (-d\chi, i\chi \Psi_i)$ derscribes gauge transformations, and the second one $ds(\psi,\mu)=((d\psi)^+_{\alpha\beta} -i (\overline{\Psi}^i_{(\alpha}\mu_{\beta)i} + \overline{\mu}^i_{(\alpha}\Psi_{\beta)i}), D_{\alpha\dot{\alpha}}\mu^\alpha_i + i\psi_{\alpha\dot{\alpha}}\Psi^\alpha_i )$ is the linearization of the multi-monopole equations. An equivalent complex is:
\begin{align}
0 \rightarrow \Omega^1(M_4)\oplus \Gamma(M_4,W^+)^{N_f} \,\stackrel{C^\dagger \oplus ds}{\longrightarrow}\, \Omega^0(M_4)\oplus \Omega^{2,+}(M_4)\oplus \Gamma(M_4, W^-)^{N_f} \rightarrow 0,
\end{align}
where $C^\dagger(\psi,\mu)=d^*\psi +\frac{i}2(\overline{\Psi}^i\mu_i - \overline{\mu}^i\Psi_i)$. The index of this complex gives the virtual dimension of $\CM_{N_f}$:
\be
{\rm VirDim}\, \CM_{N_f}(M_4; \lambda) = \frac{N_f(\lambda^2 - \sigma) - 2(\chi+\sigma)}{4}.
\ee
We will work under the usual assumption that for generic metric (or, possibly, generic perturbation, when needed), the actual dimension of $\CM_{N_f}$ will coincide with its virtual dimension. Any possible topological invariants of $M_4$ should be formulated in terms of such generic metrics, and it should be possible to connect two generic metrics by a family of generic metrics, which works well for $b_2^+>1$.

There is one new effect specific to the $N_f>1$ case: $\CM_{N_f}$ can become non-compact. Such non-compactness is controlled by non-zero solutions to the following system of equations:
\begin{align}
\label{Noncomp}
\sum_{i=1}^{N_f}(\Psi^{nc}_i\overline\Psi^{nc}_i)_0&=0,\cr
D \!\!\!\! \slash \Psi^{nc}_i&=0,\qquad i=1,\dots,N_f.
\end{align}
This point has been studied in the 3d case in \cite{MR3432158, 2016arXiv160701763H}.\footnote{We thank A.~Haydys for illuminating explanations.} The way it works is as follows. Suppose we have a sequence of solutions $(A,\Psi_i)$. Non-compactness means there exists such a sequence with no convergent subsequence, so the norms $||\Psi_i||_{L^2}\to\infty$ (at least for some subsequence). After renormalizing $\Psi_i \to \widetilde{\Psi}_i = \frac{\Psi_i}{\sqrt{\sum_j||\Psi_j||^2_{L^2}}}$, we get a sequence with bounded (unit) norm of $\widetilde{\Psi}_i$'s. Its subsequence will converge to some $\Psi_i^{nc}$, which has to satisfy equations \eqref{Noncomp}. The reason is that $\widetilde{\Psi}_i$ satisfy SW equations, with the first one replaced by $\frac{1}{\sum_j||\Psi_j||_{L_2}^2}(F^+ + \eta)=i\sum_{i=1}^{N_f}\left( \widetilde\Psi_i\overline{\widetilde\Psi}_i \right)_0$. In the limit, since norms of $\Psi_i$ diverge, the left-hand side tends to zero, and we get the first equation of \eqref{Noncomp}.

More precise treatment has to take into account that convergence to $\Psi^{nc}$ takes place in a complement of some codimension-two locus $\CZ\subset\CM_{N_f}$ of the moduli space. We are not going to investigate this question in any details, postponing 4d case to the other studies, as well as referring to the existing literature on the 3d case \cite{MR3432158, 2016arXiv160701763H}. For now, we will only describe a qualitative (and conjectural) picture of how $\CM_{N_f}$ becomes non-compact.

Suppose that equations \eqref{Noncomp} have some non-trivial solutions $\Psi_i,\, i=1,\dots,N_f$ with flat connections, in particular $F_A^+=0$. Such solutions will also satisfy SW equations without perturbation \eqref{NfSWeq}. Moreover, if $\Psi$ is a solution, then $t\Psi$ is also a solution. By taking $t\to 0$ we get a sequence of solutions which converges to the reducible solution with $\Psi=0$ and $A$ a flat connection. Such solutions, when exist, form a cone inside of $\CM_{N_f}(\lambda=0, \eta=0)$:
\begin{figure}[ht]
	\centering
	\includegraphics[width=4.0in]{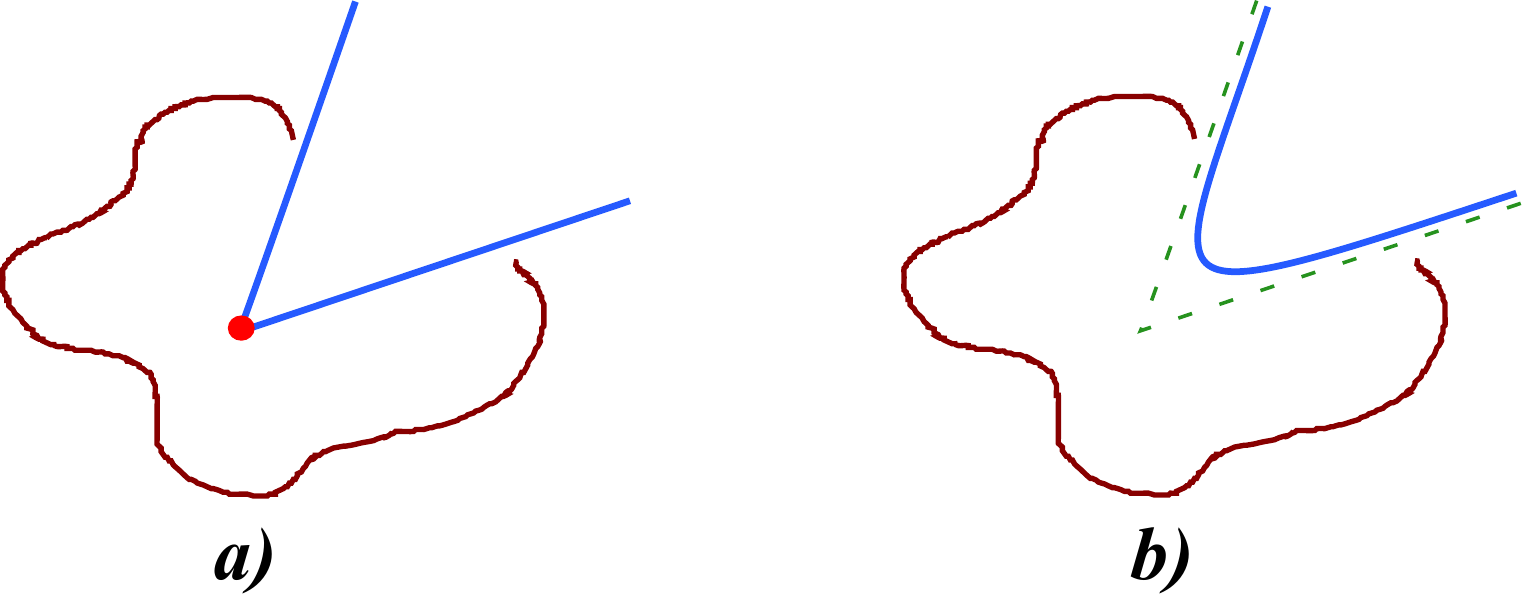}
	\caption{$a)$ The cone (shown in blue) is a non-compact subspace in $\CM_{N_f}(\lambda=0, \eta=0)$; its origin (represented by a red dot) is a singular point corresponding to the reducible solution and therefore not included in a smoothly defined $\CM_{N_f}(\lambda=0, \eta=0)$. $b)$ For non-trivial flux $\lambda\ne 0$ or in the presence of perturbation $\eta\ne0$, the cone looks the same at infinity, but smoothes out near its origin because reducibles do not satisfy SW equations any more.}
	\label{fig:Cone}
\end{figure}

When we consider a non-zero flux sector $\lambda\ne 0$ or turn on a generic perturbation $\eta$, the multi-monopole SW equations do not have reducible solutions any more. However, the asymptotic structure of the moduli space is still controlled by solutions to \eqref{Noncomp}, albeit they cannot satisfy SW equations now. In particular, if such solutions exist, the moduli space is still non-compact, and we expect that it is due to asymptotic cones as in Figure \ref{fig:Cone}$b$. Since reducible point is not a solution any more, we expect that the origin of this cone is smoothed out, so that the moduli space remains regular.

For $M_4$ admitting a Spin structure, in the $\lambda=0$ sector with the vanishing perturbation $\eta=0$, we can build solutions forming a cone as in Figure \ref{fig:Cone}$a$ explicitly. (We thank A.Haydys for the following example). Put $A=0$. Suppose $\Xi_1,\dots, \Xi_\ell$ are harmonic spinors on $M_4$, {\it i.e.}, they satisfy:
\begin{equation}
D \!\!\!\! \slash\, \Xi_j =0.
\end{equation}
There is a quaternionic structure $J$ acting on spinors by an anti-linear endomorphism\footnote{In physics language, it is given by charge conjugation, {\it i.e.}, complex conjugation composed with the multiplication by the matrix $\left(\begin{matrix}\,\,\,\,0 & 1\cr -1 & 0\end{matrix}\right)$.} which commutes with $D \!\!\!\! \slash$. Thanks to that, any $\Xi_j$ with $J\Xi_j$ form an orthogonal basis for positive-chirality spinors at any point of $M_4$ where $\Xi_j$ is non-zero. Using this, we can construct $N_f=2$ solutions by taking $\Psi_1=\Xi_j$ and $\Psi_2=J\Xi_j$. They will satisfy $(\Psi_1\overline{\Psi}_1)_0 + (\Psi_2\overline{\Psi}_2)_0=0$ and indeed solve the multi-monopole equations with zero gauge field $A=0$. It is trivial to build such solutions for any $N_f>1$, {\it e.g.}, for $N_f=3$ we could simply complete the $N_f=2$ solution by $\Psi_3=0$, and for $N_f=4$ we could do the same or, alternatively, pick $\Psi_3=\Xi_k$, $\Psi_4=J\Xi_k$ for some $k$. All these solutions can be rescaled $\Psi_i \to t\Psi_i$, with $t\in \C^*$.
Therefore, their space forms a cone, illistrated in Figure \ref{fig:Cone}$a$.

Existence of such an example depends on the existence of harmonic spinors, which, for generic metric, is determined by the index ${\rm Ind}(D \!\!\!\! \slash)$. For $\lambda=0$, ${\rm Ind}_\C(D \!\!\!\! \slash)=-\frac18 \sigma$. If it is positive, harmonic spinors exist for any metric. If it is non-positive, harmonic spinors appear in the codimension $1-{\rm Ind}_\C(D \!\!\!\! \slash)$ subspaces of the space of metrics. In particular, for vanishing index, they appear in codimension-1, and the space of metrics is divided into chambers of compactness by the walls at which $\CM_{N_f}$ might become non-compact. This kind of ``wall-crossing'' might take place in 3d case \cite{MR3432158, 2016arXiv160701763H} (though its consequences are not clear yet). Its existence and possible implications in 4d have to be investigated elsewhere.

Fortunately, this non-compactness does not affect equivariant quantities that we are going to define later. The reason is that they will be related to fixed points of the maximal torus $U(1)^{N_f-1}$ of the flavor symmetry $SU(N_f)$ acting on $\CM_{N_f}$. As we will see, fixed points are given by solutions that have $\Psi_i$ non-zero only for one value of $i$. Such solutions cannot satisfy \eqref{Noncomp}: indeed, $(\Psi_i \overline{\Psi}_i)_0=0$ (without summation over $i$) implies $\Psi_i=0$, because $\left((\Psi_i \overline{\Psi}_i)_0 \right)^2\propto \left(|\Psi_i|^2\right)^2$. Hence fixed points do not satisfy the ``non-compactness'' equation \eqref{Noncomp}, and so they cannot ``run away'' to infinity along such asymptotic cones.

\subsection*{A vanishing theorem}

Arguably, the most popular vanishing theorem in the ordinary Seiberg-Witten theory is the statement that
on a 4-manifold with $b_2^+ > 0$ that admits a metric of positive scalar curvature all Seiberg-Witten
invariants must vanish \cite{Witten:1994cg}.
The same vanishing theorem holds for multi-monopole invariants ($N_f > 1$) and also follows from
the Weitzenb\"ock formula:
\begin{multline}
\int_{M_4} \frac{1}{4} \vert F_A^+ - i\sum_{i=1}^{N_f} (\Psi_i \bar \Psi_i)_0 \vert^2
+ \sum_{i=1}^{N_f} \vert D \!\!\!\! \slash \, \Psi_i \vert^2
= \\
= \int_{M_4} \frac{1}{4} \vert F_A^+ \vert^2
+ \sum_{i=1}^{N_f} \left( \vert \nabla_A \, \Psi_i \vert^2 + \frac{s}{4} \vert \Psi_i \vert^2 \right)
+ \frac{1}{4} \left| \sum_{i=1}^{N_f} (\Psi_i \bar \Psi_i)_0 \right|^2,
\label{Weitzenbock}
\end{multline}
which, for simplicity, we write without perturbation (see \cite{MR1392667} for a version with perturbation).
When the scalar curvature $s$ is strictly positive, all terms on the right-hand side are non-negative and,
as a result, there are no non-zero solutions to \eqref{NfSWeq}.
An example of a 4-manifold that admits a positive curvature metric and, therefore,
vanishing multi-monopole invariants is:
\be
M_4 \; = \; \ell \cp^2 \; \# \; m \bar{\cp}^2,
\ee
which is homeomorphic but not diffeomorphic to a degree $d>4$ hypersurface in $\cp^3$ with
\be
\ell \; = \; \frac{d^3 - 6 d^2 + 11 d - 3}{3}
\qquad,\qquad
m \; = \; \frac{2 d^3 - 6 d^2 + 7 d - 3}{3},
\ee
and non-trivial Seiberg-Witten invariants.

Note, since $\sum_{i=1}^{N_f} (\Psi_i \bar \Psi_i)_0 \ne \sum_{i=1}^{N_f} \vert \Psi_i \vert^2$ when $N_f > 1$,
we can not immediately establish separate bounds on $\lambda_+^2$ and $\lambda_-^2$ which in the case of
the ordinary Seiberg-Witten theory also follow from \eqref{Weitzenbock} and the virtual dimension formula \eqref{RNf}.
However, we can still use a version of this argument when $M_4$ admits a metric of zero scalar curvature.
Then, from \eqref{Weitzenbock} we learn that $\lambda_+ = 0$, which, as discussed in section~\ref{sec:fluxes},
also implies that $\lambda = 0$ when $b_2^+ > 1$ and when the metric on $M_4$ is generic.
Zero scalar curvature metrics are not generic, however. A good example is a a $K3$ surface
with its Ricci-flat metric, which has the geometric genus $p_g = h^{2,0} = 1$ and:
\be
\Gamma \; = \; U \oplus U \oplus U \oplus (- E_8) \oplus (- E_8),
\ee
where $E_8$ is the unique even unimodular positive definite lattice of rank 8,
and the {\it hyperbolic plane} $U$ is a rank-2 lattice with a bilinear form:
\be
U \; = \;
\begin{pmatrix}
	0 & 1 \\
	1 & 0
\end{pmatrix}.
\ee
By the Lefschetz theorem on $(1,1)$-classes,
the N\'eron-Severi lattice can be realized as $\text{NS} (M_4) = H^2 (M_4, \Z) \cap H^{1,1} (M_4)$
and its rank $\rho (M_4) = \text{rank} \, \text{NS} (M_4)$ is called the Picard number of $M_4$.
All values of $0 \le \rho \le 20$ can be realized by complex K3 surfaces.
Moreover, when K3 is algebraic, the N\'eron-Severi lattice has signature $(1, \rho - 1)$.
Note, when combined with \eqref{RNf}, the fact that K3 surface admits a metric of zero scalar curvature
implies $\lambda_+ = 0$ and $\lambda_-^2 \le 16 \big(1 - \frac{1}{N_f} \big)$.

\subsection*{The universal bundle and its Chern class}

The space $\CM^*\times M_4$ is canonically equipped with a line bundle $\CL$ called the universal line bundle. It is constructed in the following way. Since we have a line bundle $L$ on $M_4$, we also have the line bundle:
\begin{align}
\label{CtimesL}
\CC^*&\times L\cr
&\downarrow\cr
\CC^*&\times M_4.
\end{align}
The group of gauge transformations $\CG$ acts on $\CC^*$ and (fiber-wise) on $L$, so we can take a quotient:
\begin{align}
\CL \cong \CC^* &\times_\CG L\cr
&\downarrow\cr
\CM^*&\times M_4.
\end{align}
This $\CL$ --- the universal line bundle\footnote{In the literature, $\CL$ is sometimes defined in an equivalent way: for fixed $x\in M_4$, define the total space of $\CL$ as the framed moduli space trivialized over $x$. (That is, one factors over gauge transformations that are trivial at $x$.)} --- allows in the theory of Seiberg-Witten invariants, just like in the Donaldson theory, to define a map $\mu: H_i(M_4) \to H^{2-i}(\CM^*)$ by integrating the first Chern class $c_1(\CL)$ over an $i$-cycle of $M_4$. The result is a form on $\CM_{N_f}$ of degree $2-i$ that can be integrated over $\CM_{N_f}$ to yield topological invariants (at least in the 1-monopole case, where $\CM_1$ is compact).

Proceeding in this direction, we need to construct a differential form on $\CM^*\times M_4$ that represents $c_1(\CL)$. This can be done using the Chern-Weil theory, because the canonical line bundle $\CL$ can be naturally equipped with the connection. Since $\CL$ is defined as a quotient of the bundle \eqref{CtimesL}, the connection on $\CL$ can be constructed as a quotient connection. It requires two pieces of data: one is the connection on \eqref{CtimesL}, another one is the connection on $\CG$-bundle:
\begin{align}
\label{CGbundle}
\CC^*&\times M_4\cr
&\downarrow\pi\cr
\CM^*&\times M_4.
\end{align}
The first connection is canonically defined: at a point $(A, \Psi_1,\dots, \Psi_{N_f},x)\in\CC^*\times M_4$, the connection one-form is simply given by $A$. One can write a covariant derivative $\widehat{\nabla}$ acting on the section $\widehat{s}:
\CC^*\times M_4 \to \CC^*\times L$ by the formula:
\begin{equation}
\widehat{\nabla}\widehat{s} = \int_{M_4}d^4x\, \left[ \delta A_\mu(x) \frac{\delta \widehat{s}}{\delta A_\mu(x)} + \delta \Psi_i(x) \frac{\delta \widehat{s}}{\delta \Psi_i(x)} \right] + \nabla^A \widehat{s},
\end{equation}
where $\delta$ represents the de Rham differential on the infinite-dimensional functional space $\CC^*$, and $\nabla^A$ is the connection on $L$.

To define the second ingredient --- the connection on \eqref{CGbundle}, which is a principal $\CG$-bundle --- we have to pick a horizontal subspace in every tangent space to $\CC^*\times M_4$, that is a direction transversal to the fiber ({\it i.e.}, transversal to the gauge orbit). Such connection will be represented by a one-form $\theta \in \Omega^1(\CC^*\times M_4, \mathfrak{g})$, where $\mathfrak{g}=Lie(\CG)= i C^\infty(M_4)$. The kernel of $\theta$ at each point of $\CC^*\times M_4$ is the horizontal subspace, and for the vertical vector field $\xi_X$ corresponding to the Lie algebra element $X\in \mathfrak{g}$, one has $\theta(\xi_X)=X$.

In physics, the choice of such a connection $\theta$ goes under the name of gauge fixing. A convenient choice in the multi-monopole case is as follows. For infinitesimal variations $\delta A$, $\delta\Psi_i$ representing a tangent vector to $\CC^*$, we require:
\begin{equation}
d^* \delta A + \frac{i}{2}\left( \overline{\Psi}_i \delta \Psi_i - \delta\overline{\Psi}_i \Psi_i \right)=0.
\end{equation}
Since the group $\CG$ acts on fields according to $A \to A - d\chi(x)$, $\Psi_j \to e^{i\chi(x)}\Psi_j$, where $\chi(x)\in C^\infty(M_4)$, for the vertical direction in tangent space we have $\delta A = -d\chi$ and $\delta\Psi_j = i\chi\Psi_j$. The left-hand side of the above equation becomes simply $-d^*d \chi - \sum_i \overline{\Psi}_i \Psi_i \chi$. To build $\theta$ that satisfy condition $\theta(\xi_X)=X$, we need to invert the operator $d^* d + \sum_i \overline{\Psi}_i\Psi_i$ (which is positive and hence invertible, whenever at least one $\Psi_i\ne0$), that is introduce the Green's function:
\begin{equation}
\label{GreenF}
\left(d_x^*d_x + \sum_i\overline{\Psi}_i(x)\Psi_i(x)\right)G(x,y)=\delta^{(4)}(x-y).
\end{equation}
In terms of it, the connection is defined as:
\begin{equation}
\theta = -i\int_{M_4}d^4y\, G(x,y)\left( d^* \delta A + \frac{i}{2}\left( \overline{\Psi}_i \delta \Psi_i - \delta\overline{\Psi}_i \Psi_i \right) \right).
\end{equation}
It is a one-form on $\CC^*\times M_4$, so the right-hand side depends on $x\in M_4$ as well as $(A, \Psi_1,\dots, \Psi_{N_f})\in \CC^*$ (even though it actually is $A$-independent). The connection $\theta$ is a $\CG$-invariant connection on the principal bundle $\CC^*\times M_4 \to \CM^*\times M_4$, so it allows to lift vector fields $X\in \Gamma\left(T(\CM^*\times M_4)\right)$ to horizontal vector fields on $\CC^*\times M_4$, which we denote as $X^h$.

Now we can define the quotient connection \cite{DonKron} on the universal line bundle $\CL$. A section $s: \CM^* \times M_4 \to \CL$ can be lifted to a $\CG$-equivariant section $\widehat{s}:\CC^*\times M_4 \to \CC^*\times L$, and a vector field $X\in \Gamma\left(T(\CM^*\times M_4)\right)$ -- to the horizontal vector field $X^h\in \Gamma\left(T(\CC^*\times M_4)\right)$. Since $\widehat{\nabla}$ is a $\CG$-invariant connection, $\widehat{\nabla}_{X^h}\widehat{s}$ is also a $\CG$-equivariant section of $\CC^*\times L$. It then corresponds to some section of $\CL$, which we define to be $\nabla_X s$. So,
\begin{equation}
\label{QuotNablaDef}
\widehat{\nabla_X s} = \widehat{\nabla}_{X^h}\widehat{s}.
\end{equation}
A simple calculation (see Appendix~\ref{sec:CurvApp}) shows that the curvature is:
\begin{equation}
F^\nabla = dA + \delta A - 2\Omega,
\end{equation}
where $dA\in \Omega^2(M_4)$ is the usual field strength, $\delta A = \delta A_\mu(x)\wedge dx^\mu$ and $\Omega$ is a two-form on $\CM^*$ depending on the point $x\in M_4$ and defined in the following way. For any two vectors $U,V \in T(\CM^*)$, we take their horizontal lifts $U^h, V^h$ and define $\Omega(U,V)=\Theta(U^h,V^h)$, where $\Theta$ is a two-form on $\CC^*$ given by:
\begin{equation}
\label{ChernClassTh}
\Theta=\int_{M_4}d^4y\, G(x,y)\,\sum_i\left(\delta \overline{\Psi}_i(y)\wedge \delta\Psi_i(y)\right),
\end{equation}
where parenthesis on the right denote the inner product on spinors, as before. The form $\Omega$ is a closed two-form on $\CM^*$, even though it is not manifest from this expression. (See Appendix~\ref{sec:CurvApp} for details.)

\subsection*{Equivariant form and the integral}

In the one-monopole problem, which has been studied a lot in the literature, one simply takes this $\Omega$, raises it to the power $\frac12 \dim \CM_1$, and integrates over $\CM_1\subset \CM^*$. The result is usually taken as a definition of the Seiberg-Witten invariant. Because for $N_f>1$ the space $\CM_{N_f}$ is non-compact, we cannot really follow this route any more.

A way out is related to the group $SU(N_f)$ that acts naturally (through its defining representation) on $\CM_{N_f}$. What we can do is extend all the cohomology classes we are dealing with to $SU(N_f)$-equivariant cohomology, in particular construct an equivariant version of $\Omega$, that we denote $\Omega(\xi)$, representing the equivariant Chern class of $\CL$. Then, we can perform an equivariant integration over $\CM_{N_f}$, which is possible even if $\CM_{N_f}$ is non-compact. All we need is that the relevant fixed point sets of the maximal torus action of $SU(N_f)$ on $\CM_{N_f}$ are compact. Then the equivariant integration produces a Laurent polynomial in the equivariant parameters as the answer. This will be our definition of the multi-monopole invariants.

The group $SU(N_f)$ starts its life through the action on $\CC^*$ by the vector field:
\begin{equation}
\label{SUNvectorField}
v_a = \sum_{i,j}(T_a)^{ij} \int_{M_4}d^4x\, \left( \Psi_{j\alpha}\frac{\delta}{\delta\Psi_{i\alpha}(x)} - \overline{\Psi}^\alpha_i(x) \frac{\delta}{\delta\overline{\Psi}^\alpha_j(x)} \right),
\end{equation}
where $T_a$ is a generator of $SU(N_f)$ in the fundamental representation. Using the projection $\pi: \CC^*\to \CM^*$, we get a vector field $d\pi(v_a)$ on $\CM^*$. Denoting the equivariant parameters of $SU(N_f)$ by $\xi^a$, we introduce the equivariant differential of the Cartan model on $\CM^*$:
\begin{equation}
D = d + \xi^a \iota_{d\pi(v_a)}.
\end{equation}
If $\Omega$ is $SU(N_f)$-invariant, one can construct the $D$-closed extension of $\Omega$ as:
\begin{equation}
\Omega(\xi) = \Omega + \xi^a H_a,
\end{equation}
where the Hamiltonian,
\begin{equation}
H_a = \int_{M_4} d^4y\, G(x,y)\sum_{i,j}\left(\overline{\Psi}_i(y) T_{a}^{ij}\Psi_j(y)\right),
\end{equation}
is a function on $\CM^*\times M_4$.
The proof of this expression for $H_a$, as well as $SU(N_f)$-invariance of $\Omega$ can be found in the Appendix \ref{sec:CurvApp}.

Our goal is to define equivariant multi-monopole invariants as:
\begin{equation}
\text{equivariant} \int_{\CM_{N_f}} f\left[ \Omega(\xi) \right],
\end{equation}
where $f$ is some function of $\Omega(\xi)$, {\it e.g.}, $\left(\Omega(\xi)\right)^{d/2}$ or $\exp\left[\Omega(\xi) \right]$.
In general, the result is an $SU(N_f)$-invariant rational function of the equivariant parameters, {\it i.e.}, the element of $\C(\mathfrak{g}^*)^{SU(N_f)} =\C(\mathfrak{t}^*)^W\subset \C(\mathfrak{t}^*)$,
where $\mathfrak{g}$ is the Lie algebra of $SU(N_f)$, $\mathfrak{t}$ is the Cartan subalgebra of $\mathfrak{g}$, and $W$ is the Weil group.
Moreover, following \cite{Moore:1997dj,Nekrasov:2002qd}, we can identify the coordinates on the Cartan subalgebra
of $\mathfrak{g}$ with the hypermultiplet masses in the topologically twisted 4d $\CN=2$ gauge theory on $M_4$.
It is enough to think of the integral as equivariant with respect to the maximal torus $U(1)^{N_f-1}$,
the answer will automatically lie in the $W$-invariant subspace $\C(\mathfrak{t}^*)^W\subset \C(\mathfrak{t}^*)$.

The maximal torus acts on $\Psi_j$ as $\Psi_j \mapsto e^{i\varphi_j}\Psi_j$, where $\sum_i \varphi_i=0$. At the fixed point set of the $U(1)^{N_f-1}$ action on the moduli space, this should be equivalent to the gauge transformation $\Psi_j \mapsto e^{i\varphi}\Psi_j,\, j=1,\dots,N_f$. This is possible only if $\Psi_j$ vanish for all $j$ except $j=i$ with some $1\leq i\leq N_f$. In other words, only one of the $N_f$ monopoles $\Psi_i$ is non-zero. We call the corresponding component of the fixed point set $F_i$. There are $N_f$ such disjoint components, and each one is isomorphic to the 1-monopole moduli space, $F_i\cong \CM_1$. Denote the inclusion of the $i$-th component as:
\begin{equation}
s_i: \CM_1 \xhookrightarrow{} \CM_{N_f}.
\end{equation}

Recall that the equivariant integral for non-compact spaces is \emph{defined} by the Atiyah-Bott localization formula. So it is given by:
\begin{equation}
\sum_{i=1}^{N_f} \int_{\CM_1} \frac{(s_i)^* f\left[ \Omega(\xi) \right]}{Eul(N_i)(\xi)},
\end{equation}
where $Eul(N_i)(\xi)$ is the equivariant Euler class of the normal bundle to $s_i(\CM_1)\subset \CM_{N_f}$.

\subsection{Computation for $M_4$ of simple type}
\label{sec:simpletype}

To move further, recall that there exist notions of Kronheimer-Mrowka (KM) and Seiberg-Witten (SW) simple type. We need the latter notion, the SW simple type, which requires that the manifold has only zero dimensional one-monopole moduli spaces $\CM_1(M_4; \lambda)$ for all $\lambda$. The Simple Type Conjecture states that every closed simply-connected oriented Riemannian 4-manifold with $b_2^+>1$ is of simple type -- both KM and SW.

No matter whether the Simple Type Conjecture holds or not, if we have a 4-manifold of SW simple type, we can compute the above equivariant integral explicitly. In this case, the space $\CM_1$ consists of isolated points with signs. The normal bundle $N_i$ to each of these points is trivial. Denote equivariant parameters for $U(1)^{N_f}$ acting naturally on $\CM_{N_f}$ by $z_1, z_2,\dots, z_{N_f}$,
and require that\footnote{In physics literature, mass parameter are traditionally denoted $m_i$. Nevertheless,
we remain faithful to the two-dimensional perspective, where a standard notation for these
parameters would be $z_i$, as they denote positions of impurity vertex operators on $\Sigma$.
A physicist more familiar with gauge theory may find it comforting to read ``$z_i$'' as ``$m_i$''.}
\begin{equation}
\sum_{i=1}^{N_f} z_i =0,
\end{equation}
so these are really equivariant parameters for the maximal torus $U(1)^{N_f-1}\subset SU(N_f)$ acting on $\CM_{N_f}$.  Then the equivariant parameters for the maximal torus $U(1)^{N_f-1}$ acting on $N_i$ are given by $z_j - z_i$ for $j\ne i$.

Since $M_4$ is of SW simple type (and we use generic metric), $\dim \CM_1=0=\frac14 (\lambda^2-\sigma)-\frac12 (\chi + \sigma)$. In this case:
\begin{equation}
\dim_\C \CM_{N_f} = \frac{N_f-1}{8}(\lambda^2-\sigma) = (N_f -1)\, {\rm Ind}_\C( D \!\!\!\! \slash).
\end{equation}
So, every $N_i$ is a direct sum of $N_f-1$ copies of a trivial complex bundle whose rank\footnote{We use a subscript $\C$ to emphasize that it is the complex dimension that is relevant here, while when we omit $\C$, we always mean real dimension.} is ${\rm Ind}_\C( D \!\!\!\! \slash)$. Each $U(1)$ factor in the maximal torus $U(1)^{N_f-1}$ acts on the corresponding copy of this trivial bundle, with the equivariant parameter $z_j-z_i$. Thus, the equivariant Euler class is:
\begin{equation}
Eul(N_i)(\xi) = \prod_{j\ne i}(z_j - z_i)^{{\rm Ind}_\C(D \!\!\!\! \slash)}= \prod_{j\ne i}(z_j - z_i)^{\frac18 (\lambda^2-\sigma)}.
\end{equation}
As for the pull-back of $\Omega(\xi)$, since $\dim F_i=0$, we have: $(s_i)^* (\Omega + \xi^a H_a)= (s_i)^* (\xi ^a H_a)$. Because at $F_i$ only $\Psi_i\ne0$, we have:
\begin{equation}
(s_i)^* H_a = T_a^{ii}\int_{M_4} d^4y\, G(x,y) \overline{\Psi}_i(y) \Psi_i(y),\qquad \text{no sum over $i$},
\end{equation}
and from:
\begin{equation}
\int_{M_4} d^4y\, G(x,y) \overline{\Psi}_i(y) \Psi_i(y) =  \int_{M_4} d^4y\, G(x,y) \left(d_y^* d_y +  \overline{\Psi}_i(y) \Psi_i(y)\right)1=1,
\end{equation}
we conclude:
\begin{equation}
(s_i)^* H_a = T_a^{ii},\qquad \text{no sum over $i$},
\end{equation}
and this gives simply
\begin{equation}
\sum_a \xi^a  s_i^* H_a = \sum_a \xi^a T_a^{ii} = \sum_{j=1}^{N_f}(z_j-z_i)\frac{-1}{N_f}=z_i.
\end{equation}
So the pull-back is simply $(s_i)^* \Omega(\xi) = z_i$. We are almost done, all we have to compute is the 0-dimensional integral:
\begin{equation}
\sum_{i=1}^{N_f} \frac{f[z_i]}{\prod_{j\ne i}(z_j-z_i)^{\frac18 (\lambda^2-\sigma)}}\int_{\CM_1} 1.
\end{equation}
The only non-trivial step left here is to recall that isolated points in $\CM_1$ come with signs. These signs are induced from the orientation of the moduli space $\CM_{N_f}$, which in turn is induced by the orientation of $\CM^*$ in which $\CM_{N_f}$ is embedded. The space $\CM_{N_f}$ might have several connected components, each with its own orientation. This is why points in the fixed set $F_i\cong \CM_1$ might contribute with different signs.

Of course, points in $F_i$ and $F_j$ that are identified by the isomorphism $F_i \cong \CM_1 \cong F_j$ belong to the same connected component and thus come with the same orientation. This is because they correspond to the same solution $\Psi^{(1)}$ of the 1-monopole problem and can be connected by the path:
\begin{align}
\Psi_i &= \Psi^{(1)}\cos\alpha,\quad \Psi_j = \Psi^{(1)}\sin\alpha,\quad \alpha \in [0,\pi/2],\cr
\Psi_k&=0,\quad k\ne i \text{ and } k\ne j.
\end{align}
However, different points inside $F_i$ might belong to different connected components of $\CM_{N_f}$ and come with different signs. The way these signs should be determined is precisely as in the 1-monopole problem \cite{Witten:1994cg}. In fact, thanks to the equivariant localization, we actually have completely eliminated the extra $N_f-1$ spinors and have reduced the $N_f$-monopole equations to the 1-monopole problem. Therefore, the numbers:
\begin{equation}
\text{SW} (\lambda) = \int_{\CM_1} 1
\end{equation}
are nothing else but the 1-monopole Seiberg-Witten invariants.
Even if hard-boiled skeptics may find this claim not completely convincing,
in Appendix~\ref{sec:twist} we present another derivation of it that does not assume $M_4$ to be of simple type.
There, it will be very clear that after applying equivariant localization, we are left precisely with the 1-monopole version of the problem.

Note also, that for $M_4$ of SW simple type, there is no real value in keeping function $f$ in the equivariant integral arbitrary.
We can just choose $f=1$ and define:
\begin{equation}
\text{ESW}_{M_4} (\lambda, z_i) \; := \; \text{equivariant} \int_{\CM_{N_f}} 1.
\end{equation}
The above computation shows that the answer is:
\begin{equation}
\text{ESW}_{M_4} (\lambda, z_i) \; = \; \text{SW} (\lambda) \, \sum_{i=1}^{N_f} \frac{1}{\prod_{j\ne i}(z_j-z_i)^{\frac18 (\lambda^2-\sigma)}}.
\end{equation}
This result has precisely the expected structure \eqref{multi-prediction}, where two vertex operators $\CS_+ (z)$ and $\CS_+ (w)$ have
non-singular OPE and:
\be
\CS_+ (z) \CS (w) \; \sim \; \frac{1}{(z-w)^{\frac18 (\lambda^2-\sigma)}} \, \CS (w),
\ee
with the background charge $\lambda$.

\subsection{Simply-connected K\"ahler surfaces}

In this subsection we review the structure of multi-monopole moduli spaces for $M_4$ that is K\"ahler and simply-connected, following \cite{MR1392667}, as well as discuss the perturbed problem and some further properties of $\CM_{N_f}$.

Any K\"ahler manifold has the canonical ${\rm Spin}^c$ structure:
\begin{align}
W^+_0&=\Lambda^0 \oplus \Lambda^{0,2},\cr
W^-_0&=\Lambda^{0,1},
\end{align}
with Levi-Civita connection $\nabla_0$ playing the role of ${\rm Spin}^c$ connection, and the Dirac operator $\partial \!\!\! \slash{}_0=\sqrt{2}(\overline{\partial} + \overline{\partial}^*)$. Every other ${\rm Spin}^c$ structure with connection is determined by a unique line bundle with connection $(E, A)$:
\begin{align}
W^+_E &= W^+_0\otimes E,\cr
W^-_E &= W^-_0 \otimes E,\cr
\nabla_E &= \nabla_0\otimes A.
\end{align}
The multi-monopole equations are written in terms of pairs $(\alpha_i, \beta_i), i=1\dots N_f$, where $\alpha_i\in\Omega^0(E)$ and $\beta_i\in\Omega^{0,2}(E)$:
\begin{align}
\overline\partial_A\alpha_i + \overline\partial^*_A \beta_i &=0,\cr
2 F^{0,2}_A + \overline\eta &= \sum_{i=1}^{N_f} \alpha_i^* \beta_i,\cr
-2i\Lambda_\omega F_A &= \frac12 \sum_{j=1}^N (|\alpha_j|^2 -|\beta_j|^2) - i\Lambda_\omega F_{\nabla_0}-r,
\end{align}
where $\Lambda_\omega$ is a dual of the Hodge operator $\alpha\mapsto\omega\wedge\alpha$, and $\omega$ is a Kahler form. We slightly abuse notations for the perturbation here, and $\eta$ is (twice) a $(2,0)$-part of what was a general self-dual harmonic perturbation $\eta$ before, while the $(1,1)$-part of that perturbation is now chosen to be $r\omega$ with $r\in\R$.

Following \cite{MR1392667}, the moduli space at $\eta=0$ has the following description. Define $V_1=\oplus_{N_f} H^0(E)\cong \C^{N_f h^0(\lambda)}$ and $V_2=\oplus_{N_f} H^0(K\otimes E^*)\cong \C^{N_f h^2(\lambda)}$, where we denote:
\be
h^0 (\lambda) \; = \; \dim_\C H^0 (E)
\qquad , \qquad
h^2 (\lambda) \; = \; \dim_\C H^0 (K \otimes E^*).
\ee
Define a map $S: V_1\oplus V_2 \to H^0(K)$ by the following equation:
\begin{equation}
S(\alpha_1,\dots \alpha_{N_f}, \beta_1^*,\dots, \beta_{N_f}^*)=\sum_{i=1}^N \alpha_i\beta_i^*,
\end{equation}
and define $Z\subset V_1\oplus V_2$ as a zero set of $S$. Next, for $2\pi c_1(L)\cdot [\omega] + r[\omega]\cdot [\omega]>0$, define $Z^s\subset Z$ as a subset with $(\alpha_1,\dots,\alpha_N)\ne (0,\dots,0)$, {\it i.e.}, $Z^s \cong Z\cap \left[(V_1\setminus \{0\})\oplus V_2\right]$, while for $2\pi c_1(L)\cdot [\omega] + r[\omega]\cdot [\omega]<0$, the role of $\alpha_i$ and $\beta_i^*$ is interchanged: $Z^s \cong Z\cap \left[V_1\oplus(V_2\setminus \{0\})\right]$. Finally, there is a $\C^*$ action on $V_1$ an $V_2$ via multiplication by $\lambda$ and $\lambda^{-1}$ respectively. Then the moduli space at $\eta=0$ is given by:
\begin{equation}
\CM_{N_f}(\lambda,\eta=0)= Z^s/\C^*.
\end{equation}

It is straightforward to generalize this to non-zero $\eta\in H^0(K)$. At $\eta=0$, as was shown in \cite{MR1392667}, the monopole equations imply $F_A^{0,2}=0$. In other words, for K\"ahler $M_4$, only bundles with $c_1(L)=\lambda$ of type $(1,1)$ contribute to the moduli space $\CM_{N_f}$. This property must hold after turning on the perturbation as well, even if it is not entirely obvious from the equations: for $\lambda$ not of type $(1,1)$, moduli space must be empty. Therefore, we might restrict to $\lambda$ of type $(1,1)$ from the very beginning. Then:
\begin{equation}
\int_{M_4} \eta\wedge F^{0,2} = \int_{M_4} \overline{\eta}\wedge F^{2,0}=0.
\end{equation}
Using this and repeating manipulations from \cite{MR1392667}, one arrives at the following description:
\begin{equation}
\CM_{N_f}(\lambda, \eta\ne 0) = S^{-1}(\eta)/\C^*.
\end{equation}
Now we wish to discuss various properties of $\CM_{N_f}(\lambda,\eta=0)$ and $\CM_{N_f}(\lambda,\eta\ne0)$.

\subsubsection*{Dimensionality}

As we change $\eta$ from a generic non-zero value (which is possible only if $b_2^+>1$) to zero,
the dimensionality of the space $\CM_{N_f}$ might change.
Since K\"ahler metrics are non-generic, we expect that the actual dimension of $\CM_{N_f}$ might be higher than the virtual dimension,
as well as the dimension at $\eta=0$ might be higher than that at generic $\eta$.

The space $V_1 \oplus V_2$ has dimension $N_f (h^0(\lambda) + h^2(\lambda))$, and we denote the dimension of $H^0(K)$ by $h^{2,0}$. The condition $S(\alpha_1,\dots \alpha_{N_f}, \beta_1^*,\dots, \beta_{N_f}^*)=\eta$ consists of $h^{2,0}$ equations. So, whenever $\eta$ is a regular value of the map $S$, the complex dimension of $\CM_{N_f}$ is given by what we call regular dimension:
\begin{equation}
\label{RegDim}
{\rm RegDim}_\C\, \CM_{N_f}=N_f (h^0(\lambda) + h^2(\lambda)) - h^{2,0} -1,
\end{equation}
where ``$-1$'' comes from taking a quotient by $\C^*$.
For example, for a $K3$ surface with a K\"ahler metric and $\lambda = 0$, we have $h^0 (\lambda) = h^2 (\lambda) = 1$ as well as $h^{2,0}=1$.
Therefore, the (complex) regular dimension is $4 (N_f - 1)$, {\it cf.} \eqref{RNf}.

On the other hand, consider $\eta=0$ and $2\pi c_1(L)\cdot [\omega] + r[\omega]\cdot [\omega]>0$. In this case $\CM_{N_f}$ was constructed as a subvariety in $\left[(V_1\setminus\{0\})\oplus V_2\right]/\C^*$. Let us look at the geometry of $\CM_{N_f}$ near the fixed point locus of the maximal torus $U(1)^{N_f-1}\subset SU(N_f)$. One component of such fixed point locus is given, as we know, by solutions with only $\alpha_i$ and $\beta_i^*$ possibly non-zero, while for $j\ne i$, $\alpha_j=\beta^*_j=0$. The equation $S=0$ becomes simply $\alpha_i \beta_i^*=0$. Since $\alpha_i\ne 0$ for  $2\pi c_1(L)\cdot [\omega] + r[\omega]\cdot [\omega]>0$, we conclude $\beta_i=0$ and $\alpha_i \in \C^{h^0(\lambda)}$ has an arbitrary non-zero value. Factoring by $\C^*$, the fixed point locus is isomorphic to $\C P^{h^0(\lambda)-1}$.

Let us determine the normal bundle to this fixed point set inside $\CM_{N_f}(\lambda,\eta=0)$, just for fun and to have a better understanding of the geometry of $\CM_{N_f}$. Points of this $\C P^{h^0(\lambda)-1}$ are given by homogeneous coordinates $(0:\dots:0:\alpha_i:0:\dots :0)$. Moving away from this point by $\delta\alpha_j$, $\delta\beta_j^*$, the equation $S=0$ only implies that $\alpha_i \delta \beta_i^*=0$, hence $\delta\beta_i^*=0$ and $\delta\alpha_j$, $\delta\beta_j^*$ with $j\ne i$ are not constrained. We also put $\delta\alpha_i=0$ since we do not want to move in the tangent direction. Recalling that $\C^*$ acts on $\delta\alpha_j$ in the same way as on $\alpha_i$, and in the opposite way on $\delta\beta_j^*$, we can find that the normal bundle looks as follows:
\begin{align}
\CO(1)^{\oplus (N_f-1)h^0(\lambda)} &\oplus \CO(-1)^{\oplus (N_f-1)h^2(\lambda)}\cr
&\,\downarrow\cr
&\C P^{h^0(\lambda)-1}.
\end{align}
This describes the geometry of $\CM_{N_f}$ near the fixed point set, and one can read off $\dim_\C \CM_{N_f}(\lambda, \eta=0)$ from this, at least near the fixed point set:
\begin{multline}
\dim_\C \CM_{N_f}(\lambda, \eta=0) = (N_f-1)\left(h^0(\lambda)+h^2(\lambda)\right) + h^0(\lambda)-1= \\
= N_f\left(h^0(\lambda)+h^2(\lambda)\right)-h^2(\lambda)-1 \,.
\end{multline}
This differs from \eqref{RegDim} by $h^{2,0}- h^2(\lambda)$. It is not hard to determine the sign of this difference. Since $h^0(\lambda)>0$, there exists a section $\alpha\in H^0(E)$. Multiplication by $\alpha$ defines a map of sheaves $E^* \to \CO$, (where we identify the bundle $E$ with its sheaf of holomorphic sections) which can be further completed into the short exact sequence:
\begin{equation}
0 \longrightarrow E^* \longrightarrow \CO \longrightarrow \CO/E^* \longrightarrow 0.
\end{equation}
Taking a tensor product with $K$ (which is locally free) and taking long exact sequence in the cohomology implies an injective map $H^0(K\otimes E^*) \to H^0(K)$. Therefore it must be that $h^{2,0} = \dim_\C H^0(K) \ge \dim_\C H^0(K\otimes E^*)= h^2(\lambda)$. So we conclude $h^{2,0}- h^2(\lambda)\ge 0$, {\it i.e.}, the actual dimension $\dim_\C \CM_{N_f}(\lambda, \eta=0)$ can only jump upward compared to the expected dimension \eqref{RegDim} which was derived for $\eta$ a regular value of $S$.

Let us also compare this to the virtual dimension. From the Riemann-Roch theorem:
\begin{align}
h^0(\lambda) + h^2(\lambda) - h^1(\lambda) &=  1 + h^{2,0} + \frac12 \left[c_1(E)^2 - c_1(E)\cdot c_1(K)\right]\cr
&=1 + h^{2,0} + \frac18 \left[ c_1(L)^2 - c_1(K)^2\right]\cr
&= 1 + h^{2,0} + {\rm VirDim}_\C\, \CM_1,
\end{align}
where $h^1(\lambda)=\dim_\C H^1(E)$. From this and $\frac14 (\chi+\sigma)=1+h^{2,0}$, one can write the complex virtual dimension of $\CM_{N_f}$ as follows:
\begin{multline}
{\rm VirDim}_\C\, \CM_{N_f}=N_f {\rm VirDim}_\C\, \CM_1 + (N_f-1)\frac{\chi+\sigma}{4}= \\
= N_f(h^0(\lambda)+h^2(\lambda)-h^1(\lambda))-h^{2,0}-1 \,.
\end{multline}
Because $h^1(\lambda)\ge 0$, this cannot be larger than ${\rm RegDim}_\C\, \CM_{N_f}$ from \eqref{RegDim}. So far we have found the following inequalities:
\begin{equation}
{\rm VirDim}_\C\, \CM_{N_f}\leq {\rm RegDim}_\C\, \CM_{N_f}\leq \dim_\C \CM_{N_f},
\end{equation}
where the first one follows from the definition, and the second one holds because ${\rm RegDim}$ counts dimension in the assumption that $h^{2,0}+1$ equations in $S$ are independent. If they happen to be dependent, the actual dimension of $\CM_{N_f}$ can only be larger than ${\rm RegDim}$. \\
To make one more estimate, we recall the following result:

\textbf{Proposition 1 \cite{Witten:1994cg}:} Pick a non-zero $\eta\in H^0(K)$ and consider equation on $\alpha\in H^0(E)$, $\beta^*\in H^0(K\otimes E^*)$:
\begin{equation}
\alpha\beta^*=\eta.
\end{equation}
Its space of solution modulo the $\C^*$ action $(\alpha,\beta^*)\to (t\alpha, t^{-1}\beta^*)$ is either empty or consists of isolated points.

Using this, we can make one more estimate on the dimension of the space of solutions to $\sum_{i=1}^{N_f}\alpha_i \beta_i^*=\eta$. Arbitrarily pick generic $\alpha_1,\dots, \alpha_{N_f-1}$ and $\beta_1^*,\dots,\beta_{N_f-1}^*$, which are just $(N_f-1)\left(h^0(\lambda)+h^1(\lambda)\right)$ arbitrary complex numbers. Then, the following equation:
\begin{equation}
\alpha_N \beta_N^*=\eta - \sum_{i=1}^{N_f-1}\alpha_i\beta_i^*,
\end{equation}
as an equation on $(\alpha_N, \beta_N^*)$, assuming that the right-hand side is non-zero, either has (after modding out $\C^*$) isolated solutions, in which case we say that $\dim_\C \CM_{N_f}=(N_f-1)\left(h^0(\lambda)+h^1(\lambda)\right)$, or has no solutions at all. In the latter case it is still possible that it has solutions for $\alpha_1,\dots, \alpha_{N_f-1}$ and $\beta_1^*,\dots,\beta_{N_f-1}^*$ not arbitrary but chosen from a certain subspace. In such a situation, it must be that $\dim_\C \CM_{N_f}<(N_f-1)\left(h^0(\lambda)+h^1(\lambda)\right)$. Combining this with the previous inequalities, we get:

\textbf{Proposition 2:} If $\CM_{N_f}$ is non-empty, it is true that:
\begin{equation}
{\rm VirDim}_\C\, \CM_{N_f}\leq {\rm RegDim}_\C\, \CM_{N_f}\leq \dim_\C \CM_{N_f}\leq (N_f-1)\left(h^0(\lambda)+h^2(\lambda)\right).
\end{equation}

This has an important

\textbf{Corollary:} if $\CM_{N_f}$ is non-empty and ${\rm VirDim}_\C\,\CM_1 \geq 0$, then in fact ${\rm VirDim}_\C\,\CM_1= h^1(\lambda)=0$, and all inequalities in Proposition 2 become equalities.

Indeed, Riemann-Roch estimate combined with ${\rm VirDim}_\C\,\CM_1 \geq 0$ implies $h^0(\lambda) + h^2(\lambda)\geq 1 + h^{2,0}$, with the equality only for ${\rm VirDim}_\C\,\CM_1 = h^1(\lambda)= 0$. But then ${\rm RegDim}_\C \CM_{N_f} = N_f(h^0(\lambda)+h^2(\lambda))-1-h^{2,0}\geq (N_f-1)(h^0(\lambda)+h^2(\lambda))$, and due to Proposition 2, this is only possible if $h^{2,0}+1=h^0(\lambda)+h^2(\lambda)$, so ${\rm VirDim}_\C\,\CM_1 = h^1(\lambda)= 0$ and all inequalities turn into equalities.

Recall from the previous subsection (see also Appendix~\ref{sec:twist}) that our equivariant multi-monopole invariants were determined by solutions to the 1-monopole SW equations. For K\"ahler manifolds, those exist only when $\CM_1$ is zero-dimensional. Here we have shown a bit more, but only for simply-connected $M_4$: moduli space $\CM_{N_f}$ is non-empty only if ${\rm VirDim}_\C\,\CM_1\leq 0$ and $h^1(\lambda)=0$.

\subsubsection*{Connectedness}

According to Proposition 1, for K\"ahler $M_4$, the 1-monopole moduli space consists of isolated points. In particular, it can be disconnected. The multi-monopole moduli space $\CM_{N_f}$ with $N_f>1$ and simply-connected $M_4$, however, is connected, or more precisely, its locus that contains the fixed points of the $U(1)^{N_f-1}\subset SU(N_f)$ action is definitely connected. Pick $\alpha^{(1)}, \beta^{(1)*}$ and $\alpha^{(2)}, \beta^{(2)*}$, two (possibly equal) solutions to $\alpha\beta^*=\eta$. Consider the following family of solutions:
\begin{align}
\alpha_1 &= \alpha^{(1)}\cos\theta,\quad \beta_1^* = \beta^{(1)*}\cos\theta,\cr
\alpha_2 &= \alpha^{(2)}\sin\theta,\quad \beta_2^* = \beta^{(2)*}\sin\theta,\cr
\alpha_i&=\beta_i^*=0, \text{ for } i>2.
\end{align}
This obviously solves $\sum_{i=1}^{N_f}\alpha_i\beta_i^*=\eta$. As we change $\theta$ from $0$ to $\pi/2$, it connects solution $(\alpha^{(1)},\beta^{(1)*},0,\dots,0)$ to $(0,0,\alpha^{(2)},\beta^{(2)*},0,\dots,0)$. It is clear, that using similar paths within $\CM_{N_f}$, we can connect all fixed points of the $U(1)^{N_f-1}$ action. In particular, they all lie in the same connected component.

As we discussed before, this means that all these fixed points contribute with the same sign to the $N_f$-monopole invariant, as well as to the 1-monopole SW invariant (to which the equivariant $N_f$-monopole problem reduces). This might look surprising, but actually it already follows from the description in \cite{Witten:1994cg} that for simply-connected K\"ahler $M_4$, all solutions to $\alpha\beta^*=\eta$ contribute with the same sign.\footnote{In \cite{Witten:1994cg}, for $c_1(L)\cdot[\omega]>0$, the sign of each solution was argued to be $(-1)^{\dim_\C H^0(M_4,R)}$, where $R$ is a sheaf which fits into the short exact sequence $0 \rightarrow \CO \rightarrow E \rightarrow R \rightarrow 0$, and the map $\CO \rightarrow E$ is a multiplication by $\alpha$. Taking long exact sequence in the cohomology and using $H^1(M_4, \CO)=0$, we see that $\dim_\C H^0(R)= h^0(\lambda)-1$, and hence all solutions contribute with the same sign. For $c_1(L)\cdot[\omega]<0$, one replaces $\alpha$ by $\beta^*$. }

Connectedness of $\CM_{N_f}$ might fail for a non-simply-connected $M_4$.

\subsection{Multi-monopole homology of 3-manifolds}
\label{sec:3manifolds}

Even though one of our main goals, starting with \eqref{cobordisms}, was to bring the categorification of
3-manifold and knot invariants closer to its roots, namely to the corresponding invariants of 4-manifolds,
we conclude this section by going back and consider 4-manifolds of the form
\be
M_4 \; = \; \R \times M_3 \; =
{\,\raisebox{-1.6cm}{\includegraphics[width=5.0cm]{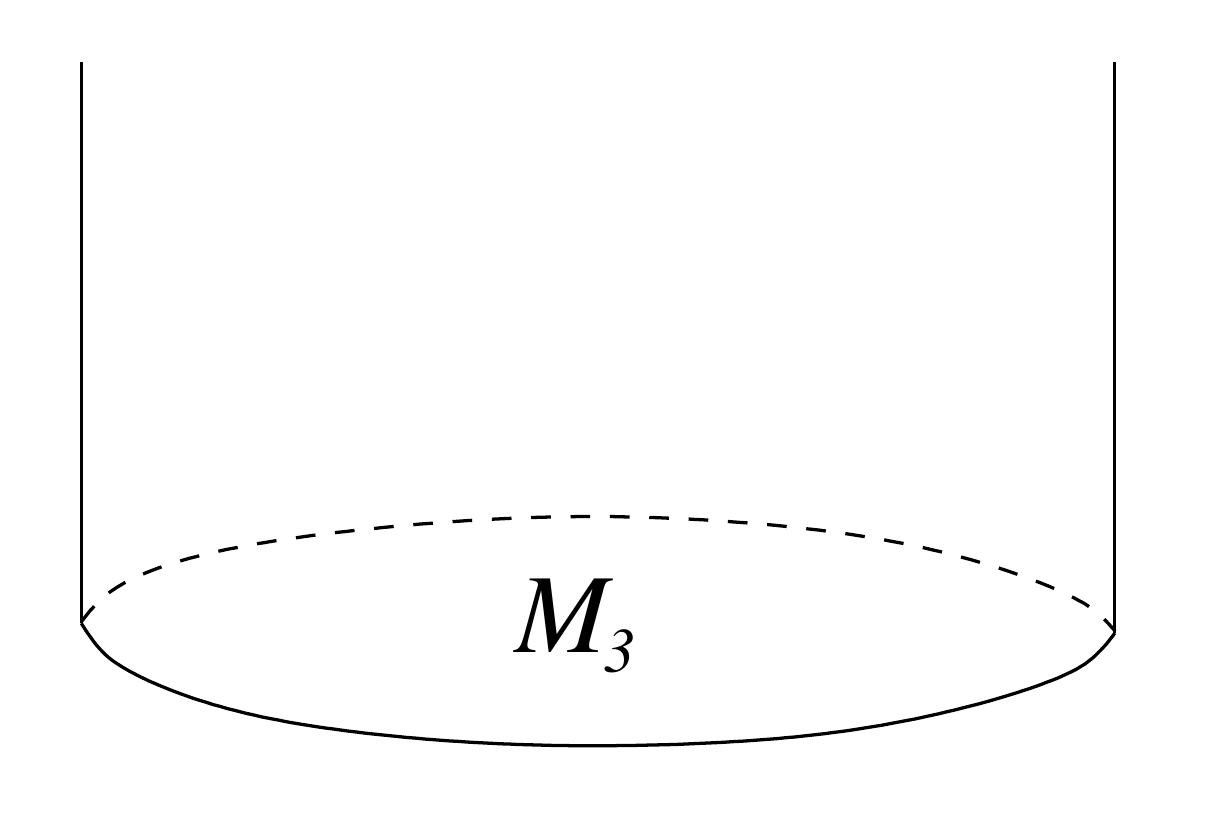}}\,}
\ee
This leads to a version of Floer homology $\CH_{N_f} (M_3)$ based on multi-monopole equations \eqref{NfSWeq},
equivariant with respect to the $SU(N_f)$ action.

Note, the $N_f = 1$ version is the familiar theory, the so-called monopole Floer homology,
$HM(M_3) \cong HF(M_3) \cong ECH (M_3)$, based on the ordinary Seiberg-Witten equations.
As a module over $\Z [U]$, this homology is naturally a part of the 4d TQFT that associates graded vector spaces
to 3-manifolds (equipped with a choice of Spin$^c$ structure) and, more importantly for us here,
it admits an equivariant interpretation, with respect to a circle action, such that $H^*_{S^1} (\text{pt}) \cong \Z [U]$.
Our equivariant multi-monopole homology $\CH_{N_f} (M_3)$ is a natural generalization of that, where the role of $U$
is played by $N_f$ equivariant parameters $z_i$, $i=1, \ldots, N_f$.

In general, there are various ways to compute $\CH_{N_f} (M_3)$ via different compactifications
of 6d fivebrane theory on $\R \times \Sigma \times M_3$ and, possibly, using additional dualities.
For example, first compactifying on $M_3$ one can compute $\CH_{N_f} (M_3)$ as a $Q$-cohomology
of 3d $\CN=2$ theory $T[M_3]$ on $\R \times \Sigma$ or, reversing the order of the compactification,
as $Q$-cohomology of a 4d TQFT on $M_4 = \R \times M_3$, {\it cf.} \eqref{6d4d2d}.
Furthermore, when $M_3$ is a Seifert manifold, both of these routes lead to a computation
of a certain partition function of 3d $\CN=2$ theory, which we illustrate below for a small sample of simple 3-manifolds.

A simple example of a Seifert manifold is the total space of a circle fibration (with no singular fibers)
over genus-$g$ Riemann surface $C_g$.
Further simplifications can be achieved by setting either the degree $p$ or the genus $g$ to zero.
The corresponding 3d theory $T[S^1 \xrightarrow[~]{~p~} C_g]$,
that describes the low-energy physics of $N$ fivebranes on $M_3$,
is $\CN=2$ level-$p$ super-Chern-Simons coupled to $2g+1$ adjoint chiral multiplets \cite{Gukov:2017kmk}.
In the case of a single fivebrane ($N=1$) that we are interested in here, all adjoint chiral
multiplets are neutral:
\be
\begin{array}{c@{\;}|@{\;}c@{\;}|@{\;}c@{\;}|@{\;}c}
& \; U(1)_{\text{gauge}} \; & \; \text{R-charge} \; & \; U(1)_{\beta} \; \\\hline
\text{chiral} & 0 & 2 & 1 \\
g~\text{chirals} & 0 & 0 & -1 \\
g~\text{chirals} & 0 & 0 & 0
\end{array}
\ee
The BPS spectrum of this theory in the presence of $N_f$ impurity operators $\CS_+ (z_i)$ on $\Sigma$
gives the desired multi-monopole homology $\CH_{N_f} (M_3)$.
Due to a phenomenon of ``homological-flavor locking'' that holds for Seifert 3-manifolds,
its Poincar\'e polynomial is equal to the ``refined index'' (= graded Euler characteristic)
of $\CH_{N_f} (M_3)$, graded by the extra flavor symmetry $U(1)_{\beta}$.
In \cite{Gukov:2016gkn}, this computation was performed for $N_f = 1$ and many simple 3-manifolds,
and we leave it to an interested reader to generalize it to $N_f > 1$.

Instead, we compute the refined index and, therefore, the Poincar\'e polynomial of $\CH_{N_f} (M_3)$,
by interpreting it as a fivebrane partition function on $S^1 \times \Sigma \times M_3$ and reducing on $S^1 \times \Sigma$ first.
This gives a 3d theory $T[S^1 \times \Sigma]$ topologically twisted on $M_3$.
In our case, $T[S^1 \times \Sigma]$ is a 3d $\CN=4$ vector multiplet coupled to $N_f$ charged hypermultiplets
or, in $\CN=2$ language,
\be
\begin{array}{c@{\;}|@{\;}c@{\;}|@{\;}c}
& \; U(1)_{\text{gauge}} \; & \; \text{R-charge} \; \\\hline
\text{chiral} & 0 & 0 \\
N_f~\text{chirals} & 1 & 1 \\
N_f~\text{chirals} & -1 & 1
\end{array}
\label{TSSigma}
\ee
Its topological partition function on $S^1 \xrightarrow[~]{~p~} C_g$
can be computed as in \cite{Gukov:2016gkn} or \cite{Benini:2016hjo,Closset:2016arn}.
For simplicity, let us set $p=0$ and start with $g=0$, that is $M_3 = S^1 \times S^2$.
Then, the topologically twisted index of 3d theory $T[S^1 \times \Sigma]$ on a 2-sphere is:
\be
\left(\frac{y^{-1}}{1-y^{-2}}\right) \sum_{h\in \mathbb{Z}} (-q)^{-h}
\int \frac{dx}{2\pi i x}\, \prod_{i=1}^{N_f}
\left(\frac{{x^{1/2} z_i^{1/2} y^{1/2}}}{1-x z_i y} \right)^{h}
\,\left(\frac{{x^{-1/2} z_i^{- 1/2} y^{1/2}}}{1-x^{-1} z_i^{-1} y} \right)^{-h},
\label{TSSigmaonSS}
\ee
where we follow the conventions of \cite[sec.3]{Gukov:2016gkn} and, in particular,
$h \in \Z$ labels the choice of Spin$^c$ structure on $M_3 = S^1 \times S^2$.
Using the Jeffrey-Kirwan residue prescription and picking up the residues of negatively-charged fields,
we obtain (with $h < 0$):
\be
\frac{1}{2\pi i} \sum_{j=1}^{N_f} \text{Res}_{x = y/z_j}
\left( \frac{1}{x} \, \prod_{i=1}^{N_f} \Big( \frac{x z_i - y}{1 - x z_i y} \Big)^h \right)
\; = \; (-1)^{h N_f} \left( y^{-h N_f} - y^{h N_f} \right).
\label{genuszeroHMNf}
\ee
Incorporating the factor $\frac{1}{y-y^{-1}}$ from \eqref{TSSigmaonSS}, we come to a prediction
that a multi-monopole analogue of the homology $HF^+ (M_3)$ has dimension $h N_f$ for $M_3 = S^1 \times S^2$
with Spin$^c$ structure $h$.

Similarly, the genus-$g$ topologically twisted index of 3d theory \eqref{TSSigma} can be expressed as a sum:
\be
Z_{S^1 \times C_g} \; = \;
\sum_{x = x_{(\alpha)}} Z_{\text{cl,1-loop}} \vert_{h=0}
\, \left( i \frac{\partial B}{\partial \log x} \right)^{g-1},
\ee
taken over solutions to the Bethe ansatz equation
$1 = e^{iB} := \exp \Big( \frac{\partial \log Z_{\text{cl,1-loop}}}{\partial h} \Big)$,
\be
1 \; = \; q \cdot \prod_{i=1}^{N_f} \left( \frac{x z_i - y}{1 - x z_i y} \right),
\ee
where:
\be
Z_{\text{cl,1-loop}} \; = \; q^h \, \prod_{i=1}^{N_f}
\left( \frac{x^{1/2} z_i^{1/2} y^{1/2}}{1 - x z_i y} \right)^h
\left( \frac{x^{-1/2} z_i^{-1/2} y^{1/2}}{1 - x^{-1} z_i^{-1} y} \right)^{-h}.
\ee
For example, when $N_f = 2$, the Bethe ansatz equation is quadratic in $x$ and, for $g=2$,
gives the following generating series for the equivariant multi-monopole invariants:
\be
\sum_{h} q^h P_{N_f = 2} (z_1, z_2) \; = \;
\frac{(y^2 + 1) \big( 4 z_1 z_2 (q - y^2) (q y^2 - 1) - (q-1)^2 y^2 (z_1 + z_2)^2 \big)}{q (y^2-1) (z_1 - y^2 z_2) (z_2 - y^2 z_1)}.
\label{PqyzNf2}
\ee
On the other hand, in the genus-0 case we recover the result \eqref{genuszeroHMNf} obtained
earlier by a different method.

Note, if we restore the overall factor $\frac{1}{y-y^{-1}}$ which was present in \eqref{TSSigmaonSS}
but for brevity omitted in \eqref{genuszeroHMNf} and \eqref{PqyzNf2},
both of these expressions have a well defined unrefined limit ($y \to 1$).
In this limit, the dependence on the equivariant parameters $z_i$ disappears,
as expected from the fact that, for $M_4 = M_3 \times S^1$,
the 2d theory $T[M_4]$ has enhanced $\CN=(2,2)$ supersymmetry
and the half-twisted correlator on $\Sigma$ is, in fact, topological.


\section{Non-abelian generalizations}
\label{sec:non-abelian}

One of the main goals in this paper was to explain how VOA$_G [M_4]$ or, equivalently,
the $\bar Q_+$-cohomology of 2d $\CN=(0,2)$ theory $T[M_4,G]$ knows about traditional 4-manifold invariants,
such as Seiberg-Witten invariants. In the process, we had to establish a dictionary between
gauge theory on $M_4$ and half-twisted correlators in $T[M_4,G]$.
This dictionary, then, can be used as a very effective tool to study the structure of
4-manifold invariants, old and new, when $G$ is abelian or non-abelian.

Examples of new invariants which are relatively simple but nonetheless non-trivial
are the equivariant multi-monopole invariants introduced in the previous section.
Based on $G=U(1)$, they can be viewed as a stepping stone toward a more powerful
invariant VOA$_G [M_4]$ for non-abelian $G$, which we expect to be at least as strong
as Donaldson invariants and, hopefully, even stronger.

\begin{figure}[ht]
\centering
\includegraphics[scale=1.1]{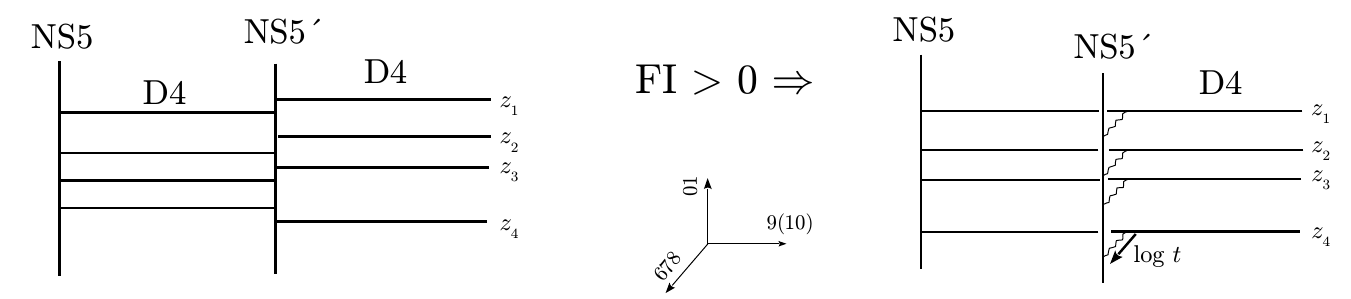}
\caption{Brane construction of $U(N)$ theory with $N_f = N$ fundamental hypermultiplets and its Higgs phase with non-zero FI parameters.}
\label{fig:branes-UN-Nhypers}
\end{figure}

Relegating a more thorough study of such non-abelian invariants to future work,
here we illustrate how the rules of section~\ref{sec:structure} can predict the structure
of these invariants thanks to the dictionary with 2d correlators.
Consider topological gauge theory on $M_4$ with gauge group $G = U(N)$
and $N_f = N$ fundamental hypermultiplets of mass $z_i$, $i = 1, \ldots, N$.
The corresponding brane setup is shown in Figure~\ref{fig:branes-UN-Nhypers}.
Turning on the FI parameter forces all $N$ D4 branes stretched between NS5 branes to align with semi-infinite D4 branes
leading to the ``color-flavor locking.''

In order to give a 2d dual formulation of this problem, as in section~\ref{sec:Kondo},
we lift it to a fivebrane configuration in M-theory and then reduce to two dimensions, {\it cf.} \eqref{6d4d2d}.
In particular, the NS5$'$ brane is lifted to a fivebrane on $M_4\times \Sigma$, with $\Sigma=\C$. From the viewpoint of
$T[M_4]$ on $\Sigma$, we have insertions of impurity operators $\CS(z_i)$,
the operators that correspond to fivebrane intersections at points $z_i$ (see Figure~\ref{fig:4mfldplane}):\footnote{The NS5 brane is also lifted to a fivebrane on $M_4\times\widetilde\Sigma$, with $\widetilde\Sigma=\C$ as well. Since there are D4 branes ending on NS5, the theory $T[M_4]$ on $\widetilde\Sigma$ could contribute another factor to the answer that would look like $\langle \CS_+(z_1)\ldots \CS_+(z_N)\rangle$. Whether such a contribution is really present will be studied in the future work.}
\begin{equation}
 \sum_{\lambda} \langle \CS(z_1)\ldots \CS(z_N) \, e^{- i \lambda \cdot X(\infty) } \rangle \,t^{\lambda}.
\label{UNN}
\end{equation}
It would be interesting to test this prediction by calculating the $SU(N_f)$-equivariant
invariants directly in 4d gauge theory, here with $N_f = N$.
Note, in derivation of \eqref{UNN} it was important that the gauge group is $U(N)$ and not $SU(N)$.

Similarly, one can consider the case with $N_f>N$ fundamental hypermultiplets.
Turning on the FI parameter then localizes the Coulomb branch (``$u$-plane'') integral
on $\binom{N_f}{N}$ configurations where $N$ finite D4 branes are distributed among $N_f$ semi-infinite D4 branes.
This leads to a Higgs phase of 4d gauge theory with the following ``color-flavor locking'' pattern:
\be
U(N)_{\text{gauge}} \times SU(N_f) \quad \to \quad S[U(N)_{\text{diag}} \times U(N_f - N)].
\ee
We leave it as an exercise to an interested reader to write down the corresponding 2d correlators.

Returning to the case $N_f = N$, let us make the proposed structure \eqref{UNN} a little bit more explicit
for minimal surfaces of general type (with $b_2^+ > 1$). Such $M_4$ have only one basic class (up to a sign):
\be
\lambda = \pm K,
\ee
where $K$ is the canonical line bundle. The corresponding Seiberg-Witten invariants are:
\be
\text{SW} (\lambda) \; = \; 1 ~\text{and}~ (-1)^{\chi_h},
\ee
where $\chi_h = \frac{1}{4} (\chi + \sigma)$ was introduced in \eqref{chihc}.
Note, for a simply-connected $M_4$, we have $\frac{\chi + \sigma}{4} = \frac{1 + b_2^+}{2}$,
which is an integer since the definition of Seiberg-Witten invariants requires $b_2^+ - b_1$ to be odd.
Therefore, for minimal surfaces of general type the impurity operator \eqref{SviaV} is in the same
$\bar Q_+$-cohomology class as the sum of two winding-momentum vertex operators of the form \eqref{Vlambda},
\be
V_{\lambda} (z) \; \sim \; e^{i k_L X_L (z) + k_R \sigma (z)},
\ee
where ``$\sim$'' means that we focus on the left-moving sector and ignore the right-moving sector.
These operators have the following chiral correlators:
\be
\langle \, V_{\lambda_1} (z_1) \cdots V_{\lambda_n} (z_n) \, \rangle_{\lambda}
\; = \; \prod_{i < j} (z_i - z_j)^{k_L^i k_L^j - k_R^i k_R^j}
\; = \; \prod_{i < j} (z_i - z_j)^{\lambda_i \lambda_j},
\label{chiralVVcorr}
\ee
which must satisfy the ``neutrality'' condition $\lambda_1 + \ldots + \lambda_n = - \lambda$.
Here, $\lambda$ is the ``background charge'' at infinity,
\be
\CO_{\text{bkgr}} \; = \; \lim_{z \to \infty} z^{\lambda^2} e^{-i \lambda X (z)}.
\ee
It corresponds to the choice of Spin$^c$ structure in the generating function \eqref{UNN},
which we can now explicitly evaluate for each given rank.

\noindent
$\underline{N = 2}$:
\be
\langle \, \CS (z_1) \, \CS (z_2) \, \rangle_{\lambda} \; = \;
\langle
( V_K (z_1) + (-1)^{\chi_h} V_{-K} (z_1) )
( V_K (z_2) + (-1)^{\chi_h} V_{-K} (z_2) )
\rangle_{\lambda},
\ee

\be
\langle \, \CS (z_1) \, \CS (z_2) \, \rangle_{\lambda} \; = \;
\begin{cases}
(z_1 - z_2)^{c}, & \text{if } \lambda = 2K, \\
2 (-1)^{\chi_h} (z_1 - z_2)^{-c}, & \text{if } \lambda =  0, \\
(z_1 - z_2)^{c}, & \text{if } \lambda = -2K.
\end{cases}
\ee
\newpage
$\underline{N = 3}$:
$$
\langle
( V_K (z_1) + (-1)^{\chi_h} V_{-K} (z_1) )
( V_K (z_2) + (-1)^{\chi_h} V_{-K} (z_2) )
( V_K (z_3) + (-1)^{\chi_h} V_{-K} (z_3) )
\rangle_{\lambda},
$$
$$
\; = \;
\begin{cases}
(-1)^{\chi_h} (z_1 - z_2)^{c} (z_2 - z_3)^{c} (z_1 - z_3)^{c} , & \text{if } \lambda = 3K, \\
(z_1 - z_2)^{c} (z_2 - z_3)^{- c} (z_1 - z_3)^{- c} + \text{permutations} , & \text{if } \lambda = K, \\
(-1)^{\chi_h} (z_1 - z_2)^{c} (z_2 - z_3)^{- c} (z_1 - z_3)^{- c} + \text{permutations} , & \text{if } \lambda = -K, \\
(z_1 - z_2)^{c} (z_2 - z_3)^{c} (z_1 - z_3)^{c} , & \text{if } \lambda = -3K.
\end{cases}
$$
Following \eqref{chihc}, here we use $c = K^2$.


\subsection{Coulomb-branch index and new 4-manifold invariants}

As noted in \cite{Gukov:2007ck}, the most interesting homological invariants point in the direction
of 4d gauge theories with non-anomalous R-symmetry $U(1)_R$.
Indeed, the Khovanov-Rozansky knot homology and its recent generalization to 3-manifolds are equipped
with a $\Z$-valued homological grading,\footnote{as opposed, say, to $\Z_8$ grading in the original Donaldson-Floer theory}
which in the physical realization \cite{Gukov:2004hz,Gukov:2016gkn} corresponds to the R-charge.
Therefore, formulating such homological invariants as Floer homologies in 4d TQFTs
(in the sense of Figure~\ref{fig:TQFT} and discussion around it) suggests that most powerful 4-manifold
invariants come from theories with $\Z$-valued $U(1)_R$ charge.

While fivebranes compactified on $\Sigma = \R^2_q$ (= ``cigar'') give rise to a rather exotic
4d $\CN=2$ theory with non-anomalous $U(1)_R$ symmetry, its simpler analogues can be obtained
by considering 4d $\CN=2$ superconformal theories (possibly, non-Lagrangian).
Their topological twists give rise to TQFTs with $\Z$-valued homological grading.
And, it is natural to ask whether any of these simpler variants produce 4-manifold invariants
which are stronger than Donaldson or Seiberg-Witten invariants.

As a practical way to answer this question, below we offer a quick ``detector'' based on a simple
and highly computable quantity. In any 4d $\CN=2$ theory with non-anomalous R-symmetry $U(1)_R$
one can define a topological $S^3 \times S^1$ index:
\be
\CI \; = \; \Tr_{\CH (S^3)} (-1)^F e^{- \beta H} {\frak t}^R.
\label{index}
\ee
It is basically a partition function of the topologically twisted 4d $\CN=2$ theory on $M_4 = S^3 \times S^1$,
refined by the $U(1)_R$ charge.\footnote{It is the same ``refinement'' as in the study of refined BPS invariants
of refined topological strings. This point is explained in many of the above mentioned papers and plays
a role in many concrete calculations.}
It can be viewed as a four-dimensional analogue of the 3d topologically twisted $S^2 \times S^1$ index \cite{Benini:2015noa,Closset:2015rna}
or, alternatively, as a topological cousin of various $S^3 \times S^1$ superconformal indices in four dimensions \cite{Gadde:2011uv}.
In fact, it is closely related to the Coulomb branch index, as we shall see next.

In a 4d $\CN=2$ theory with a Lagrangian description, it is easy to determine contributions
of vector and hypermultiplets to the index \eqref{index} or a closely related quantity: the corresponding Floer homology $\CH (S^3)$.
Among fields in a vector multiplet (summarized in Appendix~\ref{sec:twist}) only the scalar $\phi$ and the gluino $\rho$ have zero modes.
After the topological twist, all fields in a hypermultiplet transform as spinors and have no zero modes.
Therefore, only vector multiplets contribute to the topological $S^3 \times S^1$ index.
Furthermore, the contribution of vector multiplets can be evaluated using localization,
as in the case of 3d topological index \cite{Benini:2015noa,Closset:2015rna} or as in 4d superconformal index \cite{Gadde:2011uv};
schematically, the result looks like:
\be
\CI \; = \; \int_{\CM_{\text{BPS}}} \mathcal{D} \varphi_0 \, Z_{\text{1-loop}} \, e^{- S [\varphi_0]},
\label{indexint}
\ee
where:
\be
Z_{\text{1-loop}} \; = \; \frac{\det \CO_{\text{Fermi}}}{\det \CO_{\text{Bose}}}
\ee
is a ratio of one-loop determinants.
Concretely, for $G = SU(N)$, the contribution of a vector multiplet is given by the following expression:
\be
\CI_{\text{vector}} \; = \; \frac{1}{N! (1 - {\frak t})^{N-1}} \oint_{\mathbb{T}^{N-1}}
\prod_{i=1}^{N-1} \frac{da_i}{2\pi i a_i} \prod_{i \ne j}
\frac{1 - a_i / a_j}{1 - {\frak t} \, a_i / a_j},
\ee
in which the reader familiar with 4d superconformal index can recognize the Coulomb-branch index of
a 4d $\CN=2$ vector multiplet \cite{Gadde:2011uv}.
(See also \cite{Gukov:2016lki,Fredrickson:2017yka} for a connection between the Coulomb-branch index
and the equivariant Verlinde formula that counts graded dimensions of the Hilbert space in complex Chern-Simons theory.)

What can the index $\CI$ tell us about 4-manifold invariants?
It can give us some useful clues about behavior of the invariants computed by the topologically twisted 4d $\CN=2$ theory
(not necessarily Lagrangian) under the operation of connected sum:
\be
M_4 \; = \; M_4^+ \; \# \; M_4^-.
\label{connsum}
\ee
The standard by now neck-stretching argument \cite{MR1066174}
shows that both Donaldson and Seiberg-Witten \cite{Witten:1994cg} invariants of $M_4$ vanish when $b_2^+ (M_4^{\pm}) \ge 1$,
simply because there are no non-trivial solutions in the neck $\cong \R \times S^3$.
In fact, we have already touched upon this a few times, first when we computed the Floer homology of a 3-sphere \eqref{Tplus} from physics,
and then in the equivariant setting of section~\ref{sec:NfSW}, when we interpreted $\CT^+ \cong H^*_{S^1} (\text{pt})$
as the contribution of the (trivial) reducible solution, treated equivariantly.

Now, we can compare this to the index of $SU(2)$ vector multiplet (whose topological twist gives Donaldson-Witten TQFT)
to see that $\CI = \frac{1}{1 - {\frak t}^2} = 1 + {\frak t}^2 + \ldots$ is basically the graded dimension of $\CT^+$.
Therefore, factors like this will signal us that a 4d $\CN=2$ theory in question has trivial (reducible) solutions in the neck $\R \times S^3$,
and the more non-trivial the structure of $\CI$ the more likely we find new 4-manifold invariants.
In particular, the topological twist of a 4d $\CN=2$ theory that has non-trivial solutions on $\R \times S^3$
may lead to 4-manifold invariants that do not enjoy the above vanishing theorem and, therefore,
may be stronger than Donaldson or Seiberg-Witten invariants.\footnote{Of course, the index $\CI$ is a rather simple characteristic,
which has its advantages --- {\it e.g.}, it is relatively easy to compute, even in non-Lagrangian theories --- and disadvantages.
For example, even an interesting theory that leads to new 4-manifold invariants may have the index as simple as $\frac{1}{1 - {\frak t}^2}$.}

To summarize, in search of new 4-manifold invariants, we should focus special attention on theories with non-trivial structure of $\CI$.
Let us consider a few simple examples. Among 4d $\CN=2$ superconformal theories, one of the simplest yet interesting examples
is the original Argyres-Douglas theory, often denoted $(A_1,A_2)$, whose Coulomb branch index is:
\be
\CI_{(A_1,A_2)} \; = \; \frac{1}{1 - {\frak t}^{6/5}}.
\ee
The somewhat non-trivial structure of this expression suggests that a topological twist of the Argyres-Douglas theory
may be a good candidate for constructing new invariants. Note, this theory has no global flavor symmetry.
Similarly, the Coulomb branch index of $E_6$ SCFT can be obtained via Argyres-Seiberg duality \cite{Argyres:2007cn},
as in \cite{Gadde:2010te}:
\be
\CI_{E_6} \; = \; \frac{1}{1 - {\frak t}^3}.
\ee
Our next example is more advanced.

As we mentioned earlier, for applications to refined BPS states and homological invariants of knots and 3-manifolds,
one of the most interesting 4d $\CN=2$ theories is a 6d $(2,0)$ fivebrane theory ``compactified'' on the cigar $\Sigma = \R^2_q$.
Even though this exotic theory has all the features of the 4d $\CN=2$ superconformal theory,
strictly speaking, it is not four-dimensional (since $\R^2_q$ is non-compact). Nevertheless, its states can be graded
by $U(1)_q$ symmetry that acts as a rotation of $\R^2_q$ --- which explains the use of the subscript --- and each
graded component behaves much like a standard 4d $\CN=2$ SCFT.\footnote{Note, here we use the same cure for
non-compactness as in section~\ref{sec:NfSW}. This time, equivariant technique regulates non-compactness of space-time,
bringing us even closer to the origins of such methods~\cite{Moore:1997dj,Nekrasov:2002qd}.}
In particular, for a theory $T[\text{cigar}]$ of $N$ fivebranes on $\R^2_q$,
the following equivariant version of the topological index \eqref{index} is well-defined:
\be
\CI_{T[\text{cigar}]} \; = \; \Tr_{\CH (S^3)} (-1)^F e^{- \beta H} {\frak t}^R q^{n}.
\ee
As far as we know, the superconformal index of this theory (or its limits) was not studied in the literature.
Here, we propose the answer for the Coulomb-branch index, based on the same trick we used earlier,
in sections \ref{sec:Kondo} and \ref{sec:3manifolds}.
Namely, exchanging the order of the compactification, we can compute it as a supersymmetric partition function
of the 3d $\CN=2$ theory $T[S^3]$ on $S^1 \times \R^2_q$:
\begin{align}
\CI_{T[\text{cigar}]} & \; = \; Z_{S^1 \times \R^2_q} (T[S^3]) \cr
& \; = \; \prod_{i=1}^N \frac{1}{({\frak t}^i q^i; q)_{\infty}}
\; = \; 1 + {\frak t} q + (2 {\frak t}^2 + {\frak t}) q^2 + (3 {\frak t}^3 + 2 {\frak t}^2 + {\frak t}) q^3 + \ldots
\label{cigarindex}
\end{align}
To be more precise, this is the ``unreduced'' version of the partition function \cite{Gukov:2017kmk};
the ``reduced'' version $\prod_{i=1}^N \frac{({\frak t} q; q)_{\infty}}{({\frak t}^i q^i; q)_{\infty}}$
does not include the contribution from the Cartan of the adjoint chiral multiplet.
It would be interesting to find other independent checks of \eqref{cigarindex}.

Clearly, \eqref{cigarindex} is a good example of a non-trivial index, indicating that the corresponding
4-manifold invariants may not vanish under the connected sum operation \eqref{connsum} when $b_2^+ (M_4^{\pm}) \ge 1$.
Note, the choice $\Sigma = \R^2_q$ (= cigar) is precisely the one that gives 4d TQFT with surface operators
whose spaces of states are identified with $sl(N)$ Khovanov-Rozansky homology.


\acknowledgments{We would like to thank J.~Bryan, A.~Dabholkar, A.~Gadde, A.~Haydys, M.~Mari\~no, V.~Mikhaylov, J.W.~Morgan, H.~Nakajima,
H.~Ooguri, J.~Rasmussen, S.~Schafer-Nameki, A.~Soldatenkov, M.~Stosic,
E.~Verlinde, H.~Verlinde, J.~Wong, and K.~Ye for useful discussions and comments.
The work of M.D. and S.G. is supported in part by the U.S. Department of Energy, Office of Science,
Office of High Energy Physics, under Award Number DE-SC0011632.
In addition, the work of S.G. is supported in part by the ERC Starting Grant no. 335739 ``Quantum fields and knot homologies''
funded by the European Research Council under the European Union Seventh Framework Programme.
P.P. gratefully acknowledges the support from Marvin L. Goldberger Fellowship and the DOE Grant DE-SC0009988.
Opinions and conclusions expressed here are those of the authors and do not necessarily reflect the views of funding agencies.
}

%
\appendix

\section{Curvature of the canonical connection on the universal bundle}
\label{sec:CurvApp}

In this Appendix we are going to the compute curvature of the quotient connection $\nabla$ on the universal bundle $\CL \to \CM^*\times M_4$. The quotient connection was defined in \eqref{QuotNablaDef} using equivariant lift of sections and horizontal lift of vector fields. So we do the following calculation:
\begin{align}
\left(F^\nabla(X,Y)s\right)^\wedge&=\left(\big(\nabla_X \nabla_Y - \nabla_Y \nabla_X -\nabla_{[X,Y]}\big)s\right)^\wedge=(\widehat\nabla_{X^h} \widehat\nabla_{Y^h} - \widehat\nabla_{Y^h} \widehat\nabla_{X^h} -\widehat\nabla_{[X,Y]^h})\widehat{s}\cr
&=(\widehat\nabla_{X^h} \widehat\nabla_{Y^h} - \widehat\nabla_{Y^h} \widehat\nabla_{X^h} - \widehat\nabla_{[X^h,Y^h]} + \widehat\nabla_{[X^h,Y^h]} -\widehat\nabla_{[X,Y]^h})\widehat{s}\cr
&=F^{\widehat{\nabla}}(X^h,Y^h)\widehat{s} - \widehat{\nabla}_{[X,Y]^h - [X^h,Y^h]}\widehat{s}.
\end{align}
In this expression $F^{\widehat{\nabla}}=\widehat{\nabla}^2= d_xA + \delta A = \frac12 F_{\mu\nu}dx^\mu\wedge  dx^\nu + \delta A_\mu(x)\wedge dx^\mu$, and also we notice that:
\begin{equation}
[X,Y]^h - [X^h,Y^h] = hor([X^h,Y^h]) - [X^h,Y^h] = 2\Theta(X^h,Y^h),
\end{equation}
where $hor(.)$ denotes the horizontal projection of the vector field, $\Theta = (d_x+\delta)\theta$ is the curvature two-form on the principal $\CG$-bundle $\CC^*\times M_4$, and we have identified Lie algebra of $\CG$ (where $\Theta$ takes values) with the vertical vector fields (where $hor([X^h,Y^h]) - [X^h,Y^h]$ takes values). So taking derivative along the vertical vector field $\widehat{\nabla}_{[X,Y]^h - [X^h,Y^h]}$ is the same as acting with the Lie algebra element $2\Theta(X^h,Y^h)$, which acts simply by multiplication. So we can write:
\begin{align}
\left(F^\nabla(X,Y)s\right)^\wedge=F^{\widehat{\nabla}}(X^h,Y^h)\widehat{s} - 2\Theta(X^h,Y^h)\widehat{s}.
\end{align}
This gives the following expression for $F^\nabla$:
\begin{equation}
F^\nabla = d_xA + \delta A - 2 \Omega,
\end{equation}
where $\Omega$ is understood to be a two-form on $\CM^*\times M_4$ that corresponds to the form $\Theta=(d_x+\delta)\theta$ on $\CC^*\times M_4$, in a sense $\Omega(X,Y)=\Theta(X^h,Y^h)$.

To calculate $\delta\theta$ more explicitly, we first note that (as can be found by varying the defining equation for the Green's function):
\begin{equation}
\delta G(x,y) = -\int_{M_4} d^4 u\, G(x,u)\big( \overline{\Psi}(u) \delta \Psi(u) + \delta\overline\Psi(u) \Psi(u) \big)G(u,y),
\end{equation}
so we find:
\begin{align}
\label{deltheta}
&\delta\theta = \int_{M_4} d^4 y\, G(x,y)\delta\overline\Psi(y)\wedge \delta \Psi(y)\cr
&+i\int_{M_4\times M_4} d^4y\, d^4 u\, G(x,u)G(u,y)\Big[ \overline\Psi(u) \delta M(u) + \delta\overline\Psi(u) \Psi(u) \Big]\cr
&\wedge \Big[ d^*\delta A(y) + \frac{i}{2}\big(\overline\Psi(y)\delta \Psi(y) - \delta\overline\Psi(y) \Psi(y)\big) \Big].
\end{align}

When we substitute a pair of horizontal vector fields into $(d_x+\delta)\theta$, because $d^*_y\delta A(y) + \frac{i}{2}\big(\overline\Psi(y)\delta \Psi(y) - \delta\overline\Psi(y) \Psi(y)\big)$ vanishes on horizontal vectors by definition, the only part which survives is $\int_{M_4} d^4y\, G(x,y)\delta\overline\Psi(y)\wedge\delta \Psi(y)$. In particular, if we pick a point $x\in M_4$, consider the inclusion:
\begin{equation}
\CM^*\times \{x\} \subset \CM^*\times M_4,
\end{equation}
and pull back $\CL$ to $\CM^*\times {x}$, we get a line bundle on $\CM^*$, which we also call $\CL$. It has a pull-back connection, and its curvature $\Omega$ evaluated at point $m\in\CM^*$ on vectors $U, V\in T_m\CM^*$ is given by $\Theta(U^h,V^h)$ evaluated at some point $c\in \CC^*$ lying above $m$, where:
\begin{equation}
\Theta=\int_{M_4} d^4y\, G(x,y)\delta\overline\Psi(y)\wedge\delta \Psi(y) + (\text{terms that vanish on horizontal vectors}),
\end{equation}
and $U^h, V^h \in T_c \CC^*$ are horizontal lifts of $U, V$. This is the expression for $\Theta$ that was advertised in \eqref{ChernClassTh}.

By construction, $\Omega$ has to be closed, but it is slightly tricky to see it explicitly from the above equation. A way to check it is to use the formula:
\begin{align}
&(d\Omega)(U,V,W)=U\cdot \Omega(V,W) + c.p. - \Omega([U,V],W)-c.p.\cr
&=U^h\cdot \Theta(V^h,W^h) + c.p. - \Theta([U,V]^h,W^h)-c.p.\cr
&=U^h\cdot \Theta(V^h,W^h) + c.p. - \left(\Theta([U^h,V^h],W^h)+c.p.\right).- \left(\Theta([U,V]^h-[U^h,V^h],W^h)+c.p.\right)\cr
&=(\delta\Theta)(U,V,W) - 2\left(\Theta(\Theta(U^h,V^h),W^h)+c.p. \right),
\end{align}
where $c.p.$ stands for cyclic permutations. Since $\Theta$ is exact on $\CC^*$, the first term is zero. Then we look more closely at:
\begin{align}
\Theta(\Theta(U^h,V^h),W^h),
\end{align}
where $\Theta(U^h,V^h)$, as an element of the lie algebra of $\CG$, is identified with the vertical vector fields. Using expression \eqref{deltheta} for $\Theta$, one can explicitly check (through a two-step computation) that $\Theta(\Theta(U^h,V^h),W^h)=0$.
\subsection{$SU(N_f)$ invariance and the moment map}
To check that $\Omega$ is invariant with respect to the $SU(N_f)$ action given by the vector field $v_a$ from \eqref{SUNvectorField}, we need to show that $\CL_{d\pi(v_a)}\Omega=0$, where $\pi: \CC^*\times M_4 \to \CM^*\times M_4$ is our principal $\CG$-bundle. Since $d\Omega=0$, we have to prove that $d\iota_{d\pi(v_a)}\Omega=0$. According to our construction of $\Omega$, to find $\iota_{d\pi(v_a)}\Omega$, we have to compute $\iota_{V^h_a}\Theta$, where $V_a^h$ is a horizontal lift of $d\pi(v_a)$. By definition, $V^h_a$ is a vector field on $\CC^*$, and it differs from $v_a$ by a vertical vector field in such a way that $V^h_a$ is horizontal. Write $V^h_a$ as $v_a$ plus a general vertical vector field parametrized by $\chi(x)\in C^\infty(M_4)$:
\begin{align}
\label{horlift}
V^h_a = \int_{M_4}{\rm Vol}_{M_4}\,\left[ i(T^a)_i^{\,\, j}\left(\Psi_{\alpha j}\frac{\delta}{\delta \Psi_{\alpha i}} - \overline\Psi^{\alpha i}\frac{\delta}{\delta \overline\Psi^{\alpha j}}\right)\right.\cr
\left. + i\chi(x)\left(\Psi_{\alpha i}\frac{\delta}{\delta \Psi_{\alpha i}} - \overline\Psi^{\alpha i}\frac{\delta}{\delta \overline\Psi^{\alpha i}}\right) - \partial_\mu\chi(x)\frac{\delta}{\delta A_{\mu}(x)}\right].
\end{align}
The horizontality condition is that this vector field is in the kernel of one-form $d_x^*\delta A +\frac{i}{2}(\overline\Psi\delta \Psi - \delta\overline\Psi \Psi)$. This gives the equation on $\chi(x)$:
\begin{equation}
(d_x^* d_x + \overline\Psi(x)\Psi(x))\chi(x) = -\overline\Psi(x)T^a \Psi(x),
\end{equation}
which is solved by:
\begin{equation}
\chi(x) = -\int_{M_4}d^4y G(x,y)\overline\Psi(y)T^a \Psi(y).
\end{equation}
With this $\chi(x)$, equation \eqref{horlift} gives a horizontal lift of $v_a$. A computation shows:
\begin{align}
\iota_{V_a^h} \Theta &= -i\int_{M_4}d^4y\, \left[G(x,y)\delta(\overline\Psi(y) T^a \Psi(y))\right.\cr
&\left.-\int_{M_4} d^4u\, G(x,u)\delta(\overline\Psi(u)\Psi(u))G(u,y)\overline\Psi(y)T^a \Psi(y) \right]\cr
&=-\delta\int_{M_4}d^4y\, G(x,y)\overline\Psi(y)iT^a \Psi(y).
\end{align}
In particular, this implies that $\delta\iota_{V_a^h}\Theta=0$. To prove that $d\iota_{d\pi(v_a)}\Omega=0$, we do the following standard computation:
\begin{align}
(d\iota_{d\pi(v_a)}\Omega)(U,V)&= U\cdot (\iota_{d\pi(v_a)}\Omega)(V)-V\cdot (\iota_{d\pi(v_a)}\Omega)(U) - (\iota_{d\pi(v_a)}\Omega)([U,V])\cr
&=U^h\cdot (\iota_{V^h_a}\Theta)(V^h)-V^h\cdot (\iota_{V^h_a}\Theta)(U^h) - (\iota_{V^h_a}\Theta)([U,V]^h)\cr
&=(\delta\iota_{V^h_a}\Theta)(U^h,V^h) - (\iota_{V^h_a}\Theta)(hor[U^h,V^h]-[U^h,V^h])\cr
&=-2(\iota_{V^h_a}\Theta)(\Theta(U^h,V^h))=0,
\end{align}
Where the Lie algebra element $\Theta(U^h,V^h)$ is identified with the vertical vector field. Therefore, substituting the vertical vector field $\Theta(U^h,V^h)$ into the one-form $\iota_{V^h_a}\Theta$ gave zero in the last step. This proves that indeed, $d\iota_{d\pi(v_a)}\Omega=0$. Notice that actually we have shown more:
\begin{equation}
\iota_{d\pi(v_a)}\Omega = - i d H_a,
\end{equation}
where
\begin{equation}
H_a = \int_{M_4} d^4y\, G(x,y)\overline\Psi(y) T_a \Psi(y).
\end{equation}


\section{Topological twist of SQED}
\label{sec:twist}

It is possible to write a Lagrangian 4d TQFT that gives multi-monopole equations \eqref{NfSWeq}
as its localization equations. It was done in \cite{Labastida:1995bs} using the Mathai-Quillen formalism,
and the result was a topologically twisted 4d $\CN=2$ gauge theory with gauge group $U(1)$ and one charged hypermultiplet.
It is trivial to generalize it to $N_f$ hypers, and here we summarize such a theory.

Topological $\CN=2$ vector multiplet contains: a local connection one-form $A_\mu$, a complex bosonic scalar $\phi$, an auxiliary bosonic self-dual two form $D_{\mu\nu}$, a fermionic scalar $\rho$, a fermionic 1-form $\psi_\mu$ and a fermionic self-dual two-form $\chi_{\mu\nu}$. Topological $\CN=2$ hypermultiplet contains a complex bosonic spinor field of left (or positive) chirality $\Psi_\alpha$, an auxiliary complex bosonic spinor field of right (or negative) chirality $h_{\dot \alpha}$, a pair of left-handed fermion spinors $\mu_\alpha$, $\overline{\mu}^\alpha$, a pair of right-handed fermion spinors $\nu_{\dot{\alpha}}$, $\overline{\nu}^{\dot \alpha}$. (The latter four spinors are independent because we work in Euclidean space, in Minkowski space they would be related by complex conjugation which would flip chirality.) In the $N_f$-monopole case, all hypermultiplets are equipped with the additional index $i=1,\dots,N_f$. The $Q$-transformations of these fields are:\footnote{With apologies to QFT practitioners, we remind that $z_i$ denote mass parameters of the hypermultiplets, $i = 1, \ldots, N_f$.}
\begin{align}
[Q,A_\mu]&=\psi_\mu,\quad \{Q,\psi_\mu\} = \partial_\mu\phi,\quad \{Q,\phi\}=0,\cr
[Q,\overline{\phi}]&=\rho,\quad \{Q,\rho\}=0,\cr
\{Q,\chi_{\mu\nu}\}&=D_{\mu\nu},\quad [Q,D_{\mu\nu}]=0,\cr
[Q,\Psi_{\alpha i}]&=\mu_{\alpha i},\quad \{Q,\mu_{\alpha i}\}=-i(\phi+z_i)\Psi_{\alpha i},\cr
[Q,\overline\Psi^{\alpha i}]&=\overline\mu^{\alpha i},\quad \{Q,\overline\mu^{\alpha i}\}=-i(\overline\phi+\overline{z}_i)\overline\Psi^{\alpha i},\cr
\{Q,\nu_{\dot{\alpha} i}\}&=h_{\dot{\alpha} i},\quad [Q,h_{\dot{\alpha} i}]=-i(\phi+z_i)\nu_{\dot{\alpha} i},\cr
\{Q,\overline\nu^{\dot{\alpha} i}\}&=\overline{h}^{\dot{\alpha} i},\quad [Q,\overline{h}^{\dot{\alpha} i}]=-i(\overline\phi+\overline{z}_i)\overline\nu^{\dot{\alpha} i}.
\end{align}
The action is defined as:
\begin{equation}
S=\{Q,\Omega + \Omega_\eta\} + S_{\rm top},
\end{equation}
where $\Omega$ is given by:
\begin{align}
\Omega=\int_{M_4}{\rm Vol}_{M_4} \Big[ -\chi^{\alpha\beta}\left( \frac{i}{\sqrt{2}}(F^+_{\alpha\beta} - i \overline{\Psi}^i_{(\alpha}\Psi_{\beta)i})-\frac14 D_{\alpha\beta} \right)
- \frac{i}2 (\overline{\nu}^{i\dot{\alpha}}D_{\alpha\dot \alpha}\Psi_i^\alpha + \overline{\Psi}^{\alpha i}D_{\alpha\dot \alpha}\nu^{\dot \alpha}_i)\cr
+\frac18 (\overline{\nu}^{\dot \alpha i}h_{\dot \alpha i} - \overline{h}_{\dot \alpha i}\nu^{\dot \alpha i}) + \overline{\phi}d^*\psi -\frac{i}2 \overline{\mu}^{\alpha i}(\overline{\phi}+\overline{z}_i)\Psi_{\alpha i} +\frac{i}2 \overline{\Psi}^{\alpha i}(\overline{\phi}+\overline{z}_i)\mu_{\alpha i}
\Big],
\end{align}
$\Omega_\eta$ gives a topological FI term determined in terms of a fixed harmonic $\eta\in \CH^{2,+}(M_4)$:
\begin{equation}
\Omega_\eta = -\frac{i}{\sqrt{2}}\int_{M_4} \chi\wedge\eta,
\end{equation}
and $S_{\rm top}$ is a topological term that is $Q$-closed but not necessarily $Q$-exact. It might include $\theta$-term (which we choose not to add) and for our purposes has the following form:
\begin{equation}
S_{\rm top} = -\frac{1}{2\pi}\int_{M_4} F\wedge \log t,
\end{equation}
where $\log t \in \CH^{2,+}(M_4)$. From the topological theory point of view, this $t$ is simply a fugacity for the first Chern class of our ${\rm Spin}^c$ structure. However, if we think of our theory as obtained by twisting of the physical model (which we do, since we rely on brane constructions), then $t$ is not a free parameter but rather $t=\exp i\eta$.

Of course this $\eta$ is the same perturbation in multi-monopole equations that we introduced before.
For reference, we also provide the full action of the model with auxiliary fields integrated out:
\begin{align}
S=\int_{M_4}{\rm Vol}_{M_4} \Big[ \frac12\left( F^+_{\alpha\beta} - i\overline{\Psi}^i_{(\alpha}\Psi_{\beta)i} + \eta_{\alpha\beta} \right)^2+D^{\dot \alpha}_{\,\,\, \alpha}\overline{\Psi}^{\alpha i}D_{\beta\dot \alpha}\Psi^\beta_i+\overline\phi d^*d\phi +\rho d^*\psi\cr
+ \frac{i}{\sqrt{2}}\chi^{\alpha\beta}\left((d\psi)^+_{\alpha\beta} - i\overline{\Psi}^i_{(\alpha}\mu_{\beta)i}- i\overline\mu^i_{(\alpha}\Psi_{\beta)i}\right) + \frac{i}2 (\overline{\nu}^{i\dot{\alpha}}D_{\alpha\dot \alpha}\mu_i^\alpha - \overline\mu^{\alpha i}D_{\alpha\dot \alpha}\nu^{\dot \alpha}_i)\cr
-\frac{1}2 (\overline{\nu}^{i\dot{\alpha}}\psi_{\alpha\dot \alpha}\Psi_i^\alpha - \overline{\Psi}^{\alpha i}\psi_{\alpha\dot \alpha}\nu^{\dot \alpha}_i)-\frac{i}2 \rho\overline{\mu}^{\alpha i}\Psi_{\alpha i} +\frac{i}2\rho \overline{\Psi}^{\alpha i}\mu_{\alpha i}\cr
+\frac{i}{4} \overline\nu^{\dot\alpha i} (\phi+z_i)\nu_{\dot\alpha i} +\overline{\Psi}^{\alpha i}(\overline\phi + \overline{z}_i)(\phi+z_i) \Psi_{\alpha i} + i\overline\mu^{\alpha i}(\overline\phi + \overline{z}_i)\mu_{\alpha i}
\Big] + S_{\rm top}.
\end{align}

One can see that computations in this model go in parallel with our previous analysis based on the moduli space geometry and its universal bundle. Since the action is $Q$-exact, one multiplies it by $\k$ and takes the limit $\k\to\infty$ without affecting the answer for path integral. In this way, computation localizes to configurations which, in the case of zero masses $z_i=0$, are given by the perturbed multi-monopole equations and an additional condition:
\begin{equation}
\label{BranchEq}
\phi \Psi_i=0.
\end{equation}
This equation has two branches of solutions. One is the Coulomb branch with all $\Psi_i=0$ and $\phi$ arbitrary. In the case of generic metric and/or perturbation (and $b_2^+>1$), this branch is empty because one of the equations reduces to $F^+ + \eta=0$, which generically has no solutions. Another one is the Higgs branch, where
$\phi=0$ and $\Psi_i$ are determined from the multi-monopole equations. This branch contributes non-trivially, and the answer is expressed as the integral over the multi-monopole moduli space $\CM_{N_f}$. The integrand consists of determinants for fluctuations of various fields around the localization locus, as well as possible insertions of observables.

One can argue in a multitude of ways that in order to get a sensible answer, we actually have to insert an observable in the path integral, the simplest option being:
\begin{equation}
\label{Insertion}
\phi(x)^{\frac12 {\rm VirDim}\, \CM_{N_f}}.
\end{equation}
Of course, the most obvious way to argue this is using the R-symmetry anomaly (which was already used in section~\ref{sec:Kondo}). Alternatively (which is equivalent to anomaly considerations), one could, in the limit $\k\to \infty$, expand the fields as $\text{field} = \text{field}_0 + \frac1{\sqrt{\k}}\text{field}_1$, where $\text{field}_0$ is the localization locus value and $\text{field}_1$ is the fluctuation. By accurately counting powers of $\k$, one gets $\k^{\frac12 \dim\CM_{N_f}}$ from the path integral measure, $\k^{-\frac14 (\dim \CM_{N_f} + \dim {\rm coker}\, ds)}$ from fermion zero modes (here $ds$ is a map in the linearization complex of multi-monopole equations) and $\k^{-\frac14 {\rm VirDim}\, \CM_{N_f}}$ from the above insertion. These three contributions cancel.

In order to get a non-zero answer, we need to bring down $\frac12 {\rm VirDim}\, \CM_{N_f}$ powers of $\overline{\phi}$ from the action to contract it with \eqref{Insertion}. For contractions, we use the Green's function $\langle\phi(x)\overline{\phi}(y)\rangle_{free}$, which is given by:
\begin{equation}
G(x,y)=\langle x|\frac{1}{d^*d + \sum_i \overline{\Psi_i}\Psi_i}|y\rangle.
\end{equation}
This is the same Green's function as we had in \eqref{GreenF}. Each power of $\overline{\phi}$ comes from the interaction term $i\overline\mu^{\alpha i}\overline{\phi}\mu_{\alpha i}$ in the action, and after contracting $\phi$'s and leaving only zero modes for $\mu$'s, we end up with the following insertion in the action:
\begin{equation}
\left(\int_{M_4} d^4y\, G(x,y) i\overline\mu^{\alpha i}(y)\mu_{\alpha i}(y)\right)^{\frac12 {\rm VirDim}\, \CM_{N_f}}.
\end{equation}
We recognize expression in parentheses as the curvature form $\Theta$ representing the first Chern class of the universal bundle, as computed in \eqref{ChernClassTh}, if we identify $\delta\Psi_i$ with the zero mode of $\mu_i$. So indeed, as expected, the answer is given by an integral $\int_{\CM_{N_f}} (c_1(\CL))^{d/2}$, where $d={\rm VirDim}\, \CM_{N_f}$. As we said, this integral is in general divergent due to non-compactness of $\CM_{N_f}$.

This problem is resolved by switching on masses, which are nothing else but equivariant parameters
for the maximal torus of the flavor symmetry group $SU(N_f)$ \cite{Moore:1997dj,Nekrasov:2002qd}.
With masses, equation \eqref{BranchEq} gets replaces by:
\begin{equation}
(\phi + z_i)\Psi_i=0.
\end{equation}
Again, the Coulomb branch would have all $\Psi_i=0$ and arbitrary $\phi$, but it is generically empty. The Higgs branch in the presence of masses gets partially lifted, and we have only fixed points of the maximal torus $U(1)^{N_f-1}\subset SU(N_f)$ left on it. They occur at
\begin{align}
\phi&=-z_i,\cr
\Psi_j&=0,\quad j\ne i,
\end{align}
where $i=1,\dots, N_f$, and for each $i$, $\Psi_i$ and the connection satisfy the 1-monopole SW equations. We call the $i$-th component of the localization locus $F_i$, and this is precisely the equivariant localization discussed in the main text. One advantage of such TQFT approach is that reduction to the 1-monopole problem might be somewhat more transparent for some readers.

The contribution of the $i$-th fixed point set $F_i$ is simply given by the path integral of the 1-monopole problem (localized to the 1-monopole SW moduli space), coupled to the extra $N_f-1$ hypermultiplets, but this coupling vanishes in the $t\to\infty$ limit, as one can see by studying the topological action more carefully. Thus the final result factorizes into the 1-monopole answer and the determinant for $N_f-1$ topological hypers.

The 1-monopole answer is simply $\text{SW} (\lambda)$, the standard SW invariant. To compute the determinant of the $N_f-1$ hypers, we write the quadratic part of the action which describes their fluctuations:
\begin{align}
&\int_{M_4}{\rm Vol}_{M_4} \Big[ \sum_{j\ne i} D^{\dot \alpha}_{\,\,\, \alpha}\overline{\Psi}^{\alpha j} D_{\beta\dot \alpha}\Psi^\beta_j+ \frac{i}2\sum_{j\ne i} (\overline{\nu}^{j\dot{\alpha}}D_{\alpha\dot \alpha}\mu_j^\alpha - \overline\mu^{\alpha j}D_{\alpha\dot \alpha}\nu^{\dot \alpha}_j)\cr
&+ \sum_{j\ne i}|z_j-z_i|^2 \overline{\Psi}^{\alpha j}\Psi_{\alpha j} + \sum_{j\ne i} \frac{i}{4}\overline{\nu}^{\dot{\alpha} j}(z_j-z_i)\nu_{\dot \alpha j} + \sum_{j\ne i} i\overline{\mu}^{\alpha j}(\overline{z}_j - \overline{z}_i)\mu_{\alpha j}
\Big].
\end{align}
The Dirac operator matches its own non-zero modes of positive and negative chirality, so their eigenvalues coincide and cancel between bosonic and fermionic contributions. This is not true for the zero modes of $D \!\!\!\! \slash$,
and their contribution brings down factors of masses in the following way:
\begin{equation}
\frac{\prod_{j\ne i}(z_j-z_i)^{\# R_0}(\overline{z}_j - \overline{z}_i)^{\# L_0}}{\prod_{j\ne i}|z_j - z_i|^{2\# L_0}}=\frac{1}{\prod_{j\ne i}(z_j-z_i)^{\# L_0 - \# R_0}},
\end{equation}
where $\# L_0$ and $\# R_0$ are numbers of left and right zero modes of $D \!\!\!\! \slash$. Of course, $\# L_0 - \# R_0= {\rm Ind}_\C(D \!\!\!\! \slash)=\frac18 (\lambda^2-\sigma)$, and the above expression gives the determinant due to the $N_f-1$ hypers. Combining this with the 1-monopole answer $\text{SW} (\lambda)$ and summing over $i=1,\dots,N_f$ which labels localization loci in this case, we finally obtain:
\begin{equation}
\text{SW} (\lambda) \, \sum_{i=1}^{N_f} \frac{1}{\prod_{j\ne i}(z_j-z_i)^{\frac18 (\lambda^2-\sigma)}},
\end{equation}
which is the same answer as we got for $\text{ESW}_{M_4}(\lambda,z_i)$ in section~\ref{sec:simpletype} by performing equivariant integral over the moduli space. This is very satisfying, and also shows that the answer does not really depend on the simple type assumption that we had in section~\ref{sec:simpletype}.

The partition function of the topological theory is given by the sum over fluxes $\lambda$ and includes the fugacity $t=\exp i\eta$:
\begin{equation}
Z  \; = \; \sum_{\lambda}t^\lambda\, \text{ESW}_{M_4}(\lambda, z_i).
\end{equation}


\newpage

\bibliographystyle{JHEP}
\bibliography{classH}

\end{document}